\documentclass[10pt,reqno,a4paper]{amsart}
\usepackage[centering]{geometry}
\usepackage[british]{babel}
\usepackage{amsfonts,amssymb,amstext,amsthm,eucal,bbm,mathrsfs,amscd,bbold,pifont,tensor,dutchcal}
\usepackage{array}
\usepackage[svgnames,table]{xcolor}
\usepackage[T1]{fontenc}
\usepackage[utf8]{inputenc}
\usepackage{contour}
\usepackage{enumitem}
\usepackage[small]{eulervm}
\usepackage{microtype}
\usepackage{booktabs,caption}
\usepackage{verbatim}
\usepackage{braket}
\usepackage{float}
\usepackage{mdframed}
\usepackage{bm}
\usepackage{pdflscape}
\usepackage{tikz,pgfplots}
\usetikzlibrary{calc,arrows,cd}
\usepackage[unicode,pagebackref=true]{hyperref}
\renewcommand*{\backref}[1]{}
\renewcommand*{\backrefalt}[4]{%
  \ifcase #1%
  \or [Page~#2.]%
  \else [Pages~#2.]%
  \fi%
}

\hypersetup{%
  pdftitle   = {Carrollian and celestial spaces at infinity},
  pdfkeywords = {Lie algebras, kinematical, spacetime, homogeneous spaces, Poincaré, gauging, Cartan},
  pdfauthor  = {José Figueroa-O'Farrill, Emil Have, Stefan Prohazka, Jakob Salzer},
  pdfcreator = {\LaTeX\ with package \flqq hyperref\frqq},
  linkcolor=NavyBlue,
  citecolor=ForestGreen,
  urlcolor=OrangeRed,
  anchorcolor=OrangeRed,
  colorlinks=true
}
\theoremstyle{plain}

\theoremstyle{definition}

\newcommand{\g}{\mathfrak{g}}
\newcommand{\h}{\mathfrak{h}}
\renewcommand{\a}{\mathfrak{a}}
\renewcommand{\b}{\mathfrak{b}}

\renewcommand{\d}{\partial}

\newcommand{\gl}{\mathfrak{gl}}

\newcommand{\so}{\mathfrak{so}}
\newcommand{\iso}{\mathfrak{iso}}
\newcommand{\ISO}{\mathrm{ISO}}

\renewcommand{\k}{\mathfrak{k}}

\newcommand{\m}{\mathfrak{m}}

\newcommand{\eX}{\mathscr{X}}
\newcommand{\eO}{\mathscr{O}}

\newcommand{\x}{\boldsymbol{x}}
\newcommand{\y}{\boldsymbol{y}}
\newcommand{\zz}{\boldsymbol{z}}
\newcommand{\bv}{\boldsymbol{v}}

\ifdefined\C
\renewcommand{\C}{\boldsymbol{C}}
\else
\newcommand{\C}{\boldsymbol{C}}
\fi

\newcommand{\be}{\boldsymbol{e}}

\newcommand{\bzero}{\boldsymbol{0}}

\newcommand{\stab}{\mathfrak{stab}}
\newcommand{\Gr}{\mathrm{Gr}}
\newcommand{\Graff}{\mathrm{Graff}}
\newcommand{\RP}{\mathbb{RP}}
\newcommand{\ckv}{\mathfrak{ckv}}

\newcommand{\Tr}{\operatorname{Tr}}

\newcommand{\ad}{\operatorname{ad}}

\newcommand{\1}{\mathbb{1}}

\newcommand{\PP}{\mathbb{P}}

\newcommand{\EE}{\mathbb{E}}
\newcommand{\E}{\mathrm{E}}
\newcommand{\MM}{\mathbb{M}}

\newcommand{\RR}{\mathbb{R}}

\newcommand{\ZZ}{\mathbb{Z}}

\newcommand{\CO}{\operatorname{CO}}
\newcommand{\SO}{\operatorname{SO}}

\newcommand{\Ort}{\operatorname{O}}

\renewcommand{\div}{\operatorname{div}}

\newcommand{\eL}{\mathscr{L}}
\newcommand{\eM}{\mathscr{M}}
\newcommand{\eN}{\mathscr{N}}
\newcommand{\eQ}{\mathscr{Q}}
\newcommand{\gLC}{\mathbb{L}}
\newcommand{\zSpi}{\mathsf{Spi}}
\newcommand{\zTi}{\mathsf{Ti}}
\newcommand{\zNi}{\mathsf{Ni}} 
\newcommand{\zAdSC}{\mathsf{AdSC}}

\newcommand{\zLC}{\mathscr{L}}
\newcommand{\scri}{\mathscr{I}}
\newcommand{\eH}{\mathscr{H}}
\newcommand{\zdS}{\mathcal{dS}}

\newcommand{\zCS}{\mathbb{CS}}

\newcommand{\zdSC}{\mathsf{dSC}}
\newcommand{\zAdS}{\mathsf{AdS}}
\newcommand{\zC}{\mathsf{C}}

\definecolor{gris}{rgb}{0.5,0.5,0.5}
\definecolor{darkgreen}{rgb}{0.0,0.5,0.0}

\allowdisplaybreaks[0]
\numberwithin{equation}{section}
\usetikzlibrary{shapes.misc,shapes.symbols,snakes,decorations.text}
\tikzset{cross/.style={cross out, draw=black, thick, minimum size=2*(#1-\pgflinewidth), inner sep=0pt, outer sep=0pt},
cross/.default={3pt}}
\pgfdeclarelayer{nodelayer}
\pgfdeclarelayer{edgelayer}
\pgfsetlayers{edgelayer,nodelayer,main}
\tikzstyle{ghost}=[fill=none, draw=none, shape=circle]
\tikzstyle{dot}=[fill=black, draw=black, shape=circle, scale=0.5]
\tikzstyle{grey line}=[-, draw={rgb,255: red,128; green,128; blue,128}]
\tikzstyle{blue line}=[-, draw=blue,thick]
\tikzstyle{red line}=[-, draw=red,thick]
\tikzstyle{blue line2}=[-, draw=blue]
\tikzstyle{dash blue}=[-, draw=blue, dashed]
\tikzstyle{dash red}=[-, draw=red, dashed, thick]
\tikzstyle{dash black}=[-, draw=black, dashed, thick]
\tikzstyle{dash grey}=[-, draw={rgb,255: red,128; green,128; blue,128}, dashed]
\tikzstyle{thick black line}=[-, thick]
\tikzstyle{thick black line dashed}=[-, thick,dashed]
\tikzstyle{blue fill}=[-, draw=none, fill={rgb,255: red,126; green,214; blue,255}]
\tikzstyle{black line}=[-]
\tikzstyle{thick red}=[-,draw=red,thick]
\tikzstyle{red fill}=[-, fill={rgb,255: red,255; green,162; blue,164}, draw={rgb,255: red,255; green,0; blue,4}]
\tikzstyle{purple fill}=[-, fill={rgb,255: red,128; green,0; blue,255}, draw={rgb,255: red,100; green,15; blue,128}]
\tikzstyle{arrow}=[draw=black, <-]
\tikzstyle{green arrow}=[draw=ForestGreen, ->]
\definecolor{colour1}{RGB}{252,57,0}
\definecolor{colour2}{RGB}{252,115,0}
\definecolor{colour3}{RGB}{252,173,0}
\definecolor{colour4}{RGB}{252,202,0}
\definecolor{colour5}{RGB}{252,255,130}
\definecolor{contournoin}{RGB}{255,255,0}
\definecolor{contournoout}{RGB}{228,0,5}

\begin{document}

\title{Carrollian and celestial spaces at infinity}
\author[Figueroa-O'Farrill]{José Figueroa-O'Farrill}
\author[Have]{Emil Have}
\author[Prohazka]{Stefan Prohazka}
\author[Salzer]{Jakob Salzer}
\address[JMF,EH,SP]{Maxwell Institute and School of Mathematics, The University
  of Edinburgh, James Clerk Maxwell Building, Peter Guthrie Tait Road,
  Edinburgh EH9 3FD, Scotland, United Kingdom}
\address[JS]{Physique Théorique et Mathématique, Université libre de Bruxelles and
International Solvay Institutes, Campus Plaine C.P. 231, B-1050
Bruxelles, Belgium
}
\email[JMF]{\href{mailto:j.m.figueroa@ed.ac.uk}{j.m.figueroa@ed.ac.uk}, ORCID: \href{https://orcid.org/0000-0002-9308-9360}{0000-0002-9308-9360}}
\email[EH]{\href{mailto:emil.have@ed.ac.uk}{emil.have@ed.ac.uk}, ORCID: \href{https://orcid.org/0000-0001-8695-3838}{0000-0001-8695-3838}}
\email[SP]{\href{mailto:stefan.prohazka@ed.ac.uk}{stefan.prohazka@ed.ac.uk}, ORCID: \href{https://orcid.org/0000-0002-3925-3983}{0000-0002-3925-3983}}
\email[JS]{\href{mailto:jakob.salzer@ulb.be}{jakob.salzer@ulb.be}, ORCID: \href{https://orcid.org/0000-0002-9560-344X}{0000-0002-9560-344X}}
\begin{abstract}
  We show that the geometry of the asymptotic infinities of Minkowski
  spacetime (in $d+1$ dimensions) is captured by homogeneous spaces of
  the Poincaré group: the blow-ups of spatial ($\mathsf{Spi}$) and timelike
  ($\mathsf{Ti}$) infinities in the sense of Ashtekar--Hansen and a novel
  space $\mathsf{Ni}$ fibering over $\mathscr{I}$.  We embed these spaces à la
  Penrose--Rindler into a pseudo-euclidean space of signature
  $(d+1,2)$ as orbits of the same Poincaré subgroup of $\operatorname{O}(d+1,2)$.
  We describe the corresponding Klein pairs and determine their
  Poincaré-invariant structures: a carrollian structure on $\mathsf{Ti}$, a
  pseudo-carrollian structure on $\mathsf{Spi}$ and a ``doubly-carrollian''
  structure on $\mathsf{Ni}$.  We give additional geometric characterisations
  of these spaces as grassmannians of affine hyperplanes in Minkowski
  spacetime: $\mathsf{Spi}$ is the (double cover of the) grassmannian of
  affine lorentzian hyperplanes; $\mathsf{Ti}$ is the grassmannian of affine
  spacelike hyperplanes and $\mathsf{Ni}$ fibers over the grassmannian of
  affine null planes, which is $\mathscr{I}$.  We exhibit $\mathsf{Ni}$ as the
  fibred product of $\mathscr{I}$ and the lightcone over the celestial
  sphere.  We also show that $\mathsf{Ni}$ is the total space of the bundle
  of scales of the conformal carrollian structure on $\mathscr{I}$ and show
  that the symmetry algebra of its doubly-carrollian structure is
  isomorphic to the symmetry algebra of the conformal carrollian
  structure on $\mathscr{I}$; that is, the BMS algebra.  We show how to
  reconstruct Minkowski spacetime from any of its asymptotic
  geometries, by establishing that points in Minkowski spacetime
  parametrise certain lightcone cuts in the asymptotic geometries.  We
  include an appendix comparing with (A)dS and observe that the
  de~Sitter groups have no homogeneous spaces which could play the
  rôle that the celestial sphere plays in flat space
  holography.
\end{abstract}
\thanks{EMPG-21-15}
\maketitle
\tableofcontents

\section{Introduction}

In the search for a quantum theory of gravity it is by now widely
assumed that holography will act as our guide in this
endeavour~\cite{tHooft:1993dmi,Susskind:1994vu}. The benchmark result
to which all other instances of holography can be compared is
clearly the Anti-de Sitter/Conformal Field Theory (AdS/CFT)
correspondence~\cite{Maldacena:1997re,Gubser:1998bc,Witten:1998qj}.
This correspondence relates the dynamics of a gravitational theory on
$(d+1)$-dimensional AdS space to a $d$-dimensional CFT on its
conformal boundary.

Distilled down to its very essentials, one could argue that the
AdS/CFT correspondence is a consequence of the existence of two
natural spaces for the same symmetry group $\textrm{SO}(d,2)$: on the
one hand $(d+1)$-dimensional AdS space and on the other hand
$d$-dimensional Minkowski spacetime on which this group acts (locally) by
conformal transformations. Both of these are \emph{homogeneous spaces}
of $\textrm{SO}(d,2)$, i.e., they are of the form $\textrm{SO}(d,2)/H$
where $H$ is a closed subgroup of $\textrm{SO}(d,2)$. These spaces
differ in the choice of $H$ which is the stabiliser group of points in
the respective spacetime.

Of course this is an extreme simplification, for the AdS/CFT
correspondence is much more than the mere observation of the existence
of a lower-dimensional space with the same symmetry group as AdS.
Nevertheless, when trying to generalise the holographic principle to
asymptotically flat spacetimes even the simple observation in the last
paragraph becomes less obvious. In such a foggy situation, it is often
the use of symmetries which shines a light on the forward
path. Starting from $(d+1)$-dimensional Minkowski spacetime with the
Poincaré group as its 
symmetry group, which other (lower-dimensional) spaces share the same
symmetries? In analogy to the AdS case, these spaces would be
potential candidates on which to define the dual theory of an
asymptotically flat spacetime. This leads us to the question: What are
the homogeneous spaces of the Poincaré group and their geometric
properties?

The homogeneous spaces of the $(d+1)$-dimensional Poincaré group are
determined locally by a \emph{Klein pair} $(\iso(d,1),\mathfrak{h})$
consisting of the Poincaré Lie algebra $\iso(d,1)$ and a Lie
subalgebra $\mathfrak{h}$. The most obvious example is, of course,
Minkowski spacetime $\MM$ with Klein pair $(\iso(d,1),\so(d,1))$, with
$\so(d,1)$ the Lorentz subalgebra.  A slightly less obvious example is
obtained by instead considering the Klein pair
$(\iso(d,1),\iso(d-1,1))$, i.e., by replacing the Lorentz algebra by
the $d$-dimensional Poincaré algebra $\iso(d-1,1)$ which is clearly
also a subalgebra of $\iso(d,1)$. In contrast to Minkowski space, the
Poincaré group acts on the resulting space in a way that does not
allow for the construction of a nondegenerate invariant metric.
Instead, one finds a \emph{pseudo-carrollian structure} consisting of
a degenerate Lorentzian metric and a distinguished vector field. As we
will explain in more detail below, the resulting $(d+1)$-dimensional
spacetime fibers over $d$-dimensional de Sitter space $\zdS_d$ and the
degenerate metric is the pull-back by the projection of the constant
positive curvature metric on $\zdS_d$.  Although the physical
significance of this construction appears rather opaque at first
sight, it was observed by Gibbons in~\cite{Gibbons:2019zfs} that this
is precisely the universal structure at spatial infinity $\zSpi$ of
Ashtekar and Hansen's (AH)~\cite{Ashtekar:1978zz}. In a generic
asymptotically flat spacetime various physical fields acquire
direction-dependent limits at the point $i^0$. One therefore considers
a blow-up of $i^0$, such that fields at $i^0$ can be regarded as
smooth fields on the blow-up. The blow-up is constructed as the space
of space-like geodesics approaching $i^0$ with unit tangent vector.
The set of all such curves turns out to be parametrised by the
homogeneous space discussed above where the $\zdS_d$-slices
parametrise the choices of tangent vectors and the coordinate along
the fibre correspond to the tangential acceleration which is not fixed
by the construction of \cite{Ashtekar:1978zz}. We will therefore refer
to the homogeneous space of the Poincaré group with Klein pair
$(\iso(d,1),\iso(d-1,1))$ as $\zSpi$.

The above construction immediately suggests the existence of another
homogeneous space of the Poincaré group corresponding to the universal
structure at (either future or past) timelike infinity that we will
refer to as $\zTi$. In this case the subgroup is isomorphic to the
euclidean group in one lower dimension. The homogeneous space is now
equipped with a \emph{carrollian structure} and fibers over
$d$-dimensional hyperbolic space $\eH^d$ instead of de Sitter
space. In fact, the existence of this space was already revealed in
the classification of spatially isotropic homogeneous spacetimes
of~\cite{Figueroa-OFarrill:2018ilb} (see also~\cite{Morand:2018tke})
where it was called the anti-de~Sitter--Carroll spacetime (henceforth
$\zAdSC$)\footnote{In the seminal work \cite{Bacry:1968zf}, the
  corresponding kinematical Lie algebra was termed a ``para-Poincaré''
  algebra, but we will not use that terminology here.} and identified
with the carrollian limit of $\zAdS$.

\begin{figure}
    \centering
    \begin{tikzpicture}[scale=0.9]
      \foreach \Escala [count=\xi] in {1,0.8,...,0.4}
      \node[starburst, scale=\Escala, fill=colour\xi, minimum width=1cm, minimum height=1cm, line width=1.5pt]at (6.5,0) {};
      \foreach \Escala [count=\xi] in {1,0.8,...,0.4}
      \node[starburst, scale=\Escala, fill=colour\xi, minimum width=1cm, minimum height=1cm, line width=1.5pt]at (1.2,5.5) {};
    \begin{pgfonlayer}{nodelayer}
		\node [style=dot] (0) at (0, 5) {};
		\node [style=dot] (1) at (0, -5) {};
		\node [style=dot] (2) at (5, 0) {};
		\node [style=dot] (3) at (-5, 0) {};
		\node [style=ghost] (4) at (2.5, 2.5) {};
		\node [style=ghost] (5) at (-2.5, 2.5) {};
		\node [style=ghost] (6) at (-2.5, -2.5) {};
		\node [style=ghost] (7) at (2.5, -2.5) {};
		\node [style=ghost] (8) at (0, 0) {};
		\node [style=ghost] (9) at (3, 3) {$\scri^+_d$};
		\node [style=ghost] (10) at (-3, 3) {$\scri^+_d$};
		\node [style=ghost] (11) at (3, -3) {$\scri^-_d$};
		\node [style=ghost] (12) at (-3, -3) {$\scri^-_d$};
		\node [style=ghost] (13) at (0.1, 5.5) {$i^+$};
		\node [style=ghost] (14) at (5.5, 0) {$i^0$};
		\node [style=ghost] (15) at (-5.5, 0) {$i^0$};
		\node [style=ghost] (16) at (0.1, -5.5) {$i^-$};
 		\node [style=ghost] (17) at (6, 0.25) {};
		\node [style=ghost] (18) at (6, -0.25) {};
		\node [style=ghost] (19) at (7.5, 1) {};
		\node [style=ghost] (20) at (7.5, -1) {};
        \node [style=ghost] (21) at (0.75, 5.75) {};
		\node [style=ghost] (22) at (0.75, 5.25) {};
		\node [style=ghost] (23) at (2.25, 6.5) {};
		\node [style=ghost] (24) at (2.25, 4.5) {};
		\node [style=ghost] (25) at (3, 6.5) {$\zTi_{d+1}$};
		\node [style=ghost] (26) at (3, 4.5) {{\color{red}$\eH^d$}};
		\node [style=ghost] (27) at (3, 6) {};
		\node [style=ghost] (28) at (3, 5) {};
		\node [style=ghost] (29) at (8.4, 1) {$\zSpi_{d+1}$};
		\node [style=ghost] (30) at (8.25, -1) {{\color{blue}$\zdS_d$}};
		\node [style=ghost] (31) at (8.25, 0.5) {};
		\node [style=ghost] (32) at (8.25, -0.5) {};
		\node [style=ghost] (33) at (3.25, -2.5) {};
		\node [style=ghost] (34) at (4, -1.75) {};
		\node [style=ghost] (35) at (4.7, -1.2) {$\zNi_{d+1}$};
		\node [style=ghost] (36) at (3.6, -3) {};
		\node [style=ghost] (37) at (4.75, -1.7) {};
	\end{pgfonlayer}
	\begin{pgfonlayer}{edgelayer}
		\draw [style=thick black line] (0.center) to (3.center);
		\draw [style=thick black line] (0.center) to (2.center);
		\draw [style=thick black line] (3.center) to (1.center);
		\draw [style=thick black line] (1.center) to (2.center);
		\draw [style=thick black line] (5.center) to (7.center);
		\draw [style=thick black line] (4.center) to (6.center);
		\draw [style=red line, bend left, looseness=1.25] (6.center) to (7.center);
		\draw [style=red line, bend right, looseness=1.25] (6.center) to (7.center);
		\draw [style=red line, bend right=45, looseness=1.75] (6.center) to (7.center);
		\draw [style=red line, bend left=45, looseness=1.75] (6.center) to (7.center);
		\draw [style=red line] (6.center) to (7.center);
		\draw [style=red line, bend right=45] (5.center) to (4.center);
		\draw [style=red line, bend left, looseness=1.25] (5.center) to (4.center);
		\draw [style=red line, bend right=45, looseness=1.75] (5.center) to (4.center);
		\draw [style=red line, bend left=45, looseness=1.75] (5.center) to (4.center);
		\draw [style=red line] (5.center) to (4.center);
		\draw [style=blue line, bend right, looseness=1.25] (5.center) to (6.center);
		\draw [style=blue line, bend left, looseness=1.25] (5.center) to (6.center);
		\draw [style=blue line, bend right=45, looseness=1.75] (5.center) to (6.center);
		\draw [style=blue line, bend left=45, looseness=1.75] (5.center) to (6.center);
		\draw [style=blue line] (5.center) to (6.center);
		\draw [style=blue line, bend left, looseness=1.25] (4.center) to (7.center);
		\draw [style=blue line, bend right, looseness=1.25] (4.center) to (7.center);
		\draw [style=blue line, bend left=45, looseness=1.75] (4.center) to (7.center);
		\draw [style=blue line, bend right=45, looseness=1.75] (4.center) to (7.center);
		\draw [style=blue line] (4.center) to (7.center);
		\draw [style=thick black line, bend left, looseness=1.25] (22.center) to (24.center);
		\draw [style=thick black line, bend right] (21.center) to (23.center);
		\draw [style=thick black line, bend right] (17.center) to (19.center);
		\draw [style=thick black line, bend left, looseness=1.25] (18.center) to (20.center);
		\draw [style=arrow] (28.center) to (27.center);
		\draw [style=arrow] (32.center) to (31.center);
		\draw [style=arrow] (33.center) to (34.center);
		\draw [style=green arrow, postaction={decorate,decoration={raise=-2ex,text along path,text align=center,text={|\tiny| bundle of scales}}}, bend left=-45, looseness=1.25] (36.center) to (37.center);
	\end{pgfonlayer}
\end{tikzpicture}
    \caption{The Penrose diagram of Minkowski spacetime $\MM$ with its
      hyperbolic slicing. We have also illustrated how $\zTi$ and
      $\zSpi$ arise as the blow-ups of, respectively, timelike and
      spacelike infinities, while $\zNi$ fibers over $\scri$ and can
      be understood as the bundle of scales of the conformal
      carrollian structure of $\scri$.}
    \label{fig:Minkowski}
\end{figure}

Looking at the Penrose diagram (cf.\ Figure~\ref{fig:Minkowski}) of an
asymptotically flat spacetime, the appearance of the universal
structure at timelike and spacelike infinities as $(d+1)$-dimensional
homogeneous spaces of the Poincaré group further suggests the
existence of another homogeneous space related to the universal
structure at null infinity.  While the latter is indeed described by a
homogeneous space of the Poincaré group, namely $\scri$, it is only of
dimension $d$. The above picture is nevertheless completed by an
additional $(d+1)$-dimensional space\footnote{For the
  avoidance of doubt, let us emphasise that despite the spelling,
  $\zNi$, just like $\zTi$ and $\zSpi$, is pronounced to rhyme with
  $\scri$, and not with ``knee'' \cite{Knights_who_say_Ni}.} $\zNi$
fibering over $\scri$.  We will see that $\zNi$ also fibers over the
light-cone and that both the lightcone and $\scri$ fiber over the
celestial sphere, resulting in a commuting square of fibrations
displayed below together with all the other homogeneous spaces under
consideration:
\begin{equation}\label{eq:null-fibrations}
  \begin{tikzcd}
\MM & \zSpi \arrow[d]  & \zTi \arrow[d]       &                             & \zNi \arrow[dl] \arrow[dr] &                 & d+1 \\
& \zdS             & \eH                  & \scri \arrow[dr]            &                             & \zLC \arrow[dl] & d \\
&                 &                     &                             & \zCS                        &                 & d-1 \\
  \end{tikzcd}
\end{equation}
where $\zLC$ is either the future or past lightcone (without the apex)
and $\zCS$ is the celestial sphere. To the right of the square we have
denoted their dimensions, which shows that $\zSpi$ and $\zTi$ do not
have the conventional interpretation as a boundary of one lower
dimension. As we will see, all manifolds in \eqref{eq:null-fibrations}
admit transitive actions of the Poincaré group; although the action is
not effective for $\zLC$ and $\zCS$, where the translations act
trivially. While the Poincaré-invariant structures of $\zTi$ and
$\zSpi$ are (pseudo)carrollian, that of $\zNi$ is a novel
carrollian-like structure which involves two invariant vectors and a
corank-two degenerate metric. We tentatively dub this structure a
``doubly-carrollian'' structure\footnote{The doubly-carrollian
  structure bears a superficial resemblance to so-called stringy
  carrollian structures encountered in a ``string Carroll geometry''
  \cite{Cardona:2016ytk}, which is the (much less studied) carrollian
  counterpart of string Newton--Cartan geometry
  \cite{Bergshoeff:2018yvt}. We stress, however, that they are not the
  same.}, by analogy with the fibration $\zLC \to \zCS$. Concretely,
one observes that the carrollian structure on $\zLC$ arises naturally
from interpreting $\zLC$ as the total space of the bundle of scales of
the conformal structure of $\zCS$. In the same way, the
doubly-carrollian structure of $\zNi$ arises naturally from
interpreting $\zNi$ as the total space of the bundle of scales of the
conformal carrollian structure of $\scri$, as discussed in
Section~\ref{sec:znul}. Consistent with this interpretation is the
fact that the symmetries of the doubly-carrollian structure of $\zNi$,
determined in Appendix~\ref{sec:symmetries-znul}, agree with the BMS
symmetries~\cite{Bondi:1962px,Sachs:1962zza}, which are the symmetries
of the conformal carrollian structure on $\scri$. Indeed, we claim
that the symmetries of the Poincaré-invariant structures of $\zNi$,
$\zSpi$ and $\zTi$ capture precisely the expected asymptotic
symmetries of flat space. The explicit Klein pairs of all the
aforementioned homogeneous spaces and their symmetries are summarised in
Tables~\ref{tab:overview}, \ref{tab:overview-HdS}
and~\ref{tab:overview-scriLCCS}, which might provide useful
orientation.

Whereas $\zTi$, $\zSpi$ and $\zNi$ might seem to be rather abstract,
remarkably they, together with $\MM$, embed simultaneously into
a pseudo-euclidean space $\EE^{d+1,2}$ of signature $(d+1,2)$ as orbits
of the \emph{same} Poincaré subgroup of $\Ort(d+1,2)$. This extends
the well-known embedding of four-dimensional Minkowski spacetime into
$\EE^6$ described, for example, in \cite[Section~9.2]{MR838301}.
Furthermore, any (non-trivial) orbit of the Poincaré group in this
pseudo-Euclidean space takes the form of one of the above
$(d+1)$-dimensional homogeneous spaces. This embedding picture
provides what is arguably the simplest description of the spaces
$\MM_{d+1}$, $\zTi_{d+1}$, $\zSpi_{d+1}$ and $\zNi_{d+1}$ and also
shares intriguing similarities with the embedding
picture originally due to Dirac \cite{Dirac:1936fq}  and used recently
in the AdS/CFT correspondence. We will elaborate on this in the
conclusions.

As we will demonstrate, both $\zTi$ and (a $\ZZ_2$ quotient of)
$\zSpi$ can also be interpreted as the grassmannians of affine
spacelike and lorentzian hyperplanes in Minkowski spacetime,
respectively.  Mirroring the discussion around
\eqref{eq:null-fibrations}, the grassmannian of affine null
hyperplanes is $d$-dimensional and may be identified with $\scri$,
whereas the $(d+1)$-dimensional space $\zNi$ can instead be viewed as
the space of pairs of null vectors in Minkowski spacetime. Conversely,
the embedding picture allows us to show that $\MM$ parametrises
certain geometrical objects in these other spaces; in other words, we
may reconstruct Minkowski spacetime from any of its associated
homogeneous geometries. For instance, the embedding space picture
allows us to show how certain hypersurfaces in $\zTi$, $\zSpi$ and
$\zNi$ correspond to points in Minkowski spacetime.  This should be
compared to the so-called good cuts \cite{Kozameh:1983yu,Adamo:2009vu}
that allow to reconstruct Minkowski space, or more generally
asymptotically flat spacetimes, from certain codimension-one sections
of null infinity.

This paper is organised as follows. We start in
Section~\ref{sec:embeddings} with arguably the simplest description of
the ($d+1$)-dimensional homogeneous spaces $\MM$, $\zTi$, $\zSpi$ and
$\zNi$ of the Poincaré group in terms of their embedding in
$\EE^{d+1,2}$ as orbits of the same Poincaré subgroup of $O(d+1,2)$.
Moreover we show that they exhaust the types of nontrivial Poincaré
orbits in $\EE^{d+1,2}$.  In addition we relate $\zNi$ to $\scri$ via
the passage to the projective space $\PP^{d+2}$ of lines through the
origin in $\EE^{d+1,2}$.  Using these embeddings, we show in
Section~\ref{sec:reconstruction} that we may reconstruct Minkowski
spacetime from the spaces $\zTi$, $\zSpi$ and $\zNi$, as well as
$\scri$, by exhibiting a bijective correspondence between points
in Minkowski spacetime and certain hypersurfaces in these spaces.  In
Section~\ref{sec:klein-pairs} we proceed to a more algebraic
description of these homogeneous spaces in terms of Klein pairs
$(\g,\h)$, where $\g$ is in all cases the Poincaré Lie algebra and
$\h$ is the relevant stabiliser subalgebra.  This will allow us to
easily determine the Poincaré-invariant structures in the homogeneous
spaces.  We will see that the lorentzian structure of Minkowski
spacetime is replaced by a carrollian structure for $\zTi$, a
pseudo-carrollian structure for $\zSpi$ and a doubly-carrollian
structure for $\zNi$.  In Section~\ref{sec:geometries}, after a brief
review of the basic notions of (affine) grassmannians,  we give
natural geometric realisations of the Klein pairs for $\zTi$
(resp.\ $\zSpi$) in terms of grassmannians of spacelike
(resp.\ lorentzian) affine hyperplanes in Minkowski spacetime.  We then
show that $\zNi$ arises as the bundle of scales of $\scri$, which we
identify with the grassmannian of affine null hyperplanes.  We also
exhibit $\zNi$ as the fibred product of $\zLC$ and $\scri$ over the
celestial sphere $\zCS$.  In Section~\ref{sec:conclusion} we present
our conclusions and describe some potential applications of the
results presented here.  There are four appendices. In Appendix~\ref{sec:AH-construction}, we review briefly the
Ashtekar--Hansen construction of $\zSpi$ as a blow-up of $i^0$ and how
the analogous blow-up of $i^\pm$ gives rise to $\zTi$. In
Appendix~\ref{sec:homog-spac-poinc}, we give a survey of
low-dimensional homogeneous spaces of the four-dimensional Poincaré
and de Sitter groups.  In Appendix~\ref{sec:symmetries-znul}, we
determine the symmetry Lie algebra of the Poincaré invariant
doubly-carrollian structure on $\zNi$. In
Appendix~\ref{sec:another-choice-section}, we describe an alternative
approach to the reconstruction discussed in
Section~\ref{sec:reconstruction} that uses sections corresponding to
eigenfunctions of the second Casimir of the Lorentz algebra.
Finally, in Appendix~\ref{sec:other-signatures} we discuss how our
results extend to arbitrary signature, although we concentrate mostly
on the Klein space of signature $(2,2)$.

\section{Embeddings}
\label{sec:embeddings}

Although the spaces under consideration were motivated as homogeneous
spaces of the Poincaré group, their simplest description turns out to
involve their embedding as codimension-2 submanifolds in a
pseudo-euclidean space.  The use of an auxiliary six-dimensional
pseudo-euclidean space to study four-dimensional physics has a long
and illustrious pedigree.  It was perhaps first used by Dirac
\cite{Dirac:1936fq} in order to discuss conformally invariant wave
equations and later by Kasner \cite{MR1506435} and Fronsdal
\cite{Fronsdal:1959zza} in order to embed the Schwarzschild black
hole.  It appears in Penrose and Rindler \cite{MR838301} in a context
very similar to ours in their discussion of the projective model for
compactified Minkowski spacetime, and more recently it has become part
of the holographic toolkit (see, e.g., \cite{Costa:2011mg}).

We will work in general dimension and in this section we will set up
our conventions for the pseudo-euclidean space, identify a Poincaré
subgroup of isometries and discuss its orbits.

\subsection{A Poincaré subgroup of $\Ort(d+1,2)$}
\label{sec:poinc-subgr-ortd+1}

We start by describing the pseudo-euclidean space $\EE^{d+1,2}$.  We
will be working with global coordinates
$x^A = (x^0,x^1,\dots,x^d, x^+, x^-)$ for $\EE^{d+1,2}$, closely related
to the cartesian coordinates, where $x^0$ is timelike and
$x^\pm = \tfrac1{\sqrt2}(x^{d+1} \pm x^{d+2})$ are null ($x^{d+1}$ and
$x^{d+2}$ are spacelike and timelike, respectively).  Relative to
these coordinates, the metric on $\EE^{d+1,2}$ is expressed as
\begin{equation}
  \label{eq:pseudo-euclidean-metric}
  g_{\EE} = \eta_{AB} dx^A dx^B = -(dx^0)^2 + \sum_{a=1}^d (dx^a)^2 +
  2 dx^+ dx^-.
\end{equation}
It clearly has signature $(d+1,2)$.  We will let $\RR^{d,1}$ denote
the lorentzian vector space $(\RR^{d+1},\bar\eta)$, where $\bar\eta=
\text{diag}(-1,1,\dots,1)$.  A typical point in $\EE^{d+1,2}$ is denoted
by $(\x,x^+,x^-)$ with $x^\pm \in \RR$ and $\x \in \RR^{d,1}$.

We now introduce some algebraic subspaces of $\EE^{d+1,2}$. Let
$\epsilon \in \RR$ and let $\eQ_\epsilon$ denote the quadric
hypersurface cut out by the equation $\eta_{AB}x^A x^B = \epsilon$. In
particular, if $\epsilon=0$, we shall call $\eQ_0$ the null quadric.
These quadrics are preserved by a subgroup $\Ort(d+1,2)$ of the
isometries of $\EE^{d+1,2}$, which acts transitively on every
$\eQ_{\epsilon \neq 0}$. The null quadric contains a singular point
(namely, the origin in $\EE^{d+1,2}$) and $\Ort(d+1,2)$ acts
transitively on the complement.

If $\epsilon = - \rho^2 < 0$, then the induced metric on $\eQ_\epsilon$ is
lorentzian of constant negative curvature, making $\eQ_{\epsilon<0}$ into
the hyperboloid model of $\zAdS_{d+2}$ with radius of curvature
$\rho$.  If $\epsilon = \rho^2 > 0$, then the induced metric on
$\eQ_\epsilon$ has signature $(d,2)$ and has constant positive
curvature, so that $\eQ_{\epsilon>0}$ is a pseudo-sphere of radius of
curvature $\rho$, or, equivalently a signature-$(d,2)$ version of
de~Sitter space.

Let $\sigma \in \RR$ and let $\eN_\sigma$ denote the null hypersurface
with equation $x^- = \sigma$. For $\sigma \neq 0$, the subgroup of
$\Ort(d+1,2)$ which preserves $\eN_\sigma$ is isomorphic to the
Poincaré group $\Ort(d,1) \ltimes \RR^{d,1}$. It is given explicitly
by the following matrices
\begin{equation}\label{eq:poincare-subgroup-ort}
  \left\{
    \begin{pmatrix}
    A & \bzero & \bv \\
    -\bv^T \bar\eta A & 1 & -\tfrac12 \bar\eta(\bv,\bv) \\
    \bzero^T & 0 & 1
  \end{pmatrix}
  \, \middle |\,
  A^T \bar\eta A = \bar\eta
  \quad\text{and}\quad
  \bv \in \RR^{d,1}\right\} \subset \Ort(d+1,2).
\end{equation}
The subgroup of $\Ort(d+1,2)$ which preserves $\eN_0$ is larger and it
includes also ``dilatations''.  Every matrix in the Poincaré
group~\eqref{eq:poincare-subgroup-ort} decomposes into a product
\begin{equation}\label{eq:poin=lor+trans}
  \begin{pmatrix}
    A & \bzero & \bv \\
    -\bv^T \bar\eta A & 1 & -\tfrac12 \bar\eta(\bv,\bv) \\
    \bzero^T & 0 & 1
  \end{pmatrix} = 
  \begin{pmatrix}
    \1 & \bzero & \bv \\
    -\bv^T \bar\eta & 1 & -\tfrac12 \bar\eta(\bv,\bv) \\
    \bzero^T & 0 & 1
  \end{pmatrix}
  \begin{pmatrix}
    A & \bzero & \bzero \\
    \bzero^T & 1 & 0 \\
    \bzero^T & 0 & 1
  \end{pmatrix}
\end{equation}
of a Lorentz transformation $A$ fixing the points $(\bzero,x^+,x^-) \in
\EE^{d+1,2}$ and a translation $\bv$.

At the level of the Lie algebra, $\so(d+1,2)$ is spanned by the vector fields
\begin{equation}
  M_{AB} := \eta_{AC} x^C \d_B - \eta_{BC} x^C \d_A \in \eX(\EE^{d+1,2}),
\end{equation}
with Lie brackets
\begin{equation}
  [M_{AB}, M_{CD}] = \eta_{BC} M_{AD} - \eta_{AC} M_{BD}- \eta_{BD}
  M_{AC} + \eta_{AD} M_{BC}.
\end{equation}

The Poincaré algebra $\g$ is the subalgebra of $\so(d+1,2)$ whose
vector fields are tangent to the null hypersurfaces $\eN_\sigma$ for
any $\sigma$.  It is spanned by
\begin{equation}
  \label{eq:poincare-generators}
  \begin{aligned}\relax
    L_{ab} := M_{ab} &= x^a \d_b - x^b \d_a\\
    B_a := M_{0a} &= - x^0 \d_a - x^a \d_0
\end{aligned}
\qquad\qquad
  \begin{aligned}\relax
    P_a := M_{a+} &= x^- \d_a - x^a \d_+\\
    H := M_{0+} &= -x^0 \d_+ - x^- \d_0,
\end{aligned}
\end{equation}
where $a,b = 1,\dots,d$.  Its Lie brackets are
\begin{equation}\label{eq:pre-para-poincare}
  \begin{aligned}\relax
    [L_{ab},L_{cd}] &= \delta_{bc} L_{ad} - \delta_{ac} L_{bd} - \delta_{bd} L_{ac} + \delta_{ad} L_{bc}\\
    [L_{ab}, B_c] &= \delta_{bc} B_a - \delta_{ac} B_b\\
    [L_{ab}, P_c] &= \delta_{bc} P_a - \delta_{ac} P_b
  \end{aligned}
  \qquad\qquad
  \begin{aligned}\relax
    [B_a, B_b] &= L_{ab}\\
    [H,B_a] &= - P_a\\
    [B_a, P_b] &= \delta_{ab} H.
  \end{aligned}
\end{equation}
If $\sigma = 0$, there is an enhancement of symmetry and the
subalgebra of $\so(d+1,2)$ tangent to $\eN_0$ has an additional
generator: namely, $D:= M_{-+} = x^+ \d_+ - x^- \d_-$.  This enhances
the Poincaré group to the subgroup of $\Ort(d+1,2)$ consisting of matrices
of the form
\begin{equation}
    \begin{pmatrix}
    A & \bzero & \bv \\
    -a \bv^T \bar\eta A & a & -a \tfrac12 \bar\eta(\bv,\bv) \\
    \bzero^T & 0 & a^{-1}
  \end{pmatrix} =
  \begin{pmatrix}
    \1 & \bzero & \bzero \\
    \bzero^T & a & 0\\
    \bzero^T & 0 & a^{-1}
  \end{pmatrix}
  \begin{pmatrix}
    \1 & \bzero & \bv \\
    -\bv^T \bar\eta & 1 & -\tfrac12 \bar\eta(\bv,\bv) \\
    \bzero^T & 0 & 1
  \end{pmatrix}
  \begin{pmatrix}
    A & \bzero & \bzero \\
    \bzero^T & 1 & 0 \\
    \bzero^T & 0 & 1
  \end{pmatrix},
\end{equation}
where the additional symmetry is given by nonzero $a \in \RR$.

\subsection{Poincaré orbits in $\EE^{d+1,2}$}
\label{sec:poincare-orbits}

In discussing the orbits of the Poincaré group on $\EE^{d+1,2}$ we find
it convenient to restrict ourselves to the identity component of the
Poincaré group, denoted $G$, and given by
\begin{equation}\label{eq:connected-poincare-subgroup}
  G = \left\{
    \begin{pmatrix}
    A & \bzero & \bv \\
    -\bv^T \bar\eta A & 1 & -\tfrac12 \bar\eta(\bv,\bv) \\
    \bzero^T & 0 & 1
  \end{pmatrix} ~ \middle | ~  A \in \SO(d,1)_0,\quad \bv \in \RR^{d,1}\right\},
\end{equation}
where $\SO(d,1)_0$ is the identity component of the Lorentz group.

Since $G \subset \Ort(d+1,2)$, it preserves the quadrics $\eQ_\epsilon$
for any $\epsilon \in \RR$, and by definition it also preserves the
null hyperplanes $\eN_\sigma$ for any $\sigma \in \RR$.  Therefore it
preserves their intersections
\begin{equation}
  \eM_{\epsilon,\sigma} := \eQ_\epsilon  \cap \eN_\sigma.
\end{equation}

\subsubsection{Embedding Minkowski}
\label{sec:embedding-m}

\begin{figure}[h!]
\centering
\definecolor{cd20000}{RGB}{210,0,0}
\definecolor{c03468f}{RGB}{3,70,143}
\definecolor{cffffff}{RGB}{255,255,255}
\begin{tikzpicture}[overlay]
\begin{pgfonlayer}{nodelayer}
		\node [style=ghost] (0) at (6, 3.8) {{\color{cd20000}$\eN_1$}};
		\node [style=ghost] (1) at (3, 1.7) {$\eQ_0$};
		\node [style=ghost] (2) at (3.2, 7.6) {{\color{c03468f}$\MM_{d+1}$}};
\end{pgfonlayer}
\end{tikzpicture}
\includegraphics[width=0.6\textwidth]{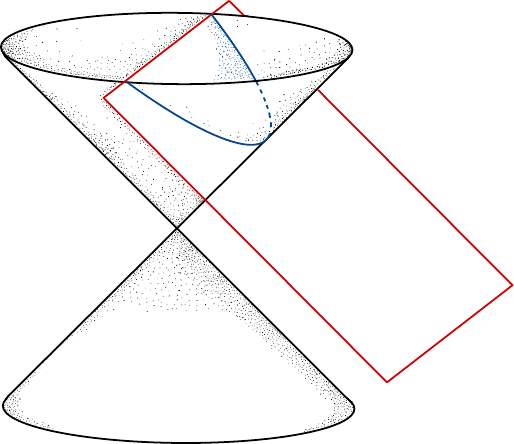}
\caption{An embedding of $(d+1)$-dimensional Minkowski spacetime
  $\MM_{d+1}$ as the intersection $\eQ_0 \cap \eN_1$ in the
  ambient space $\EE^{d+1,2}$}
    \label{fig:Embedding-Minkowski}
\end{figure}

Our first observation is that for any $\epsilon$, provided that
$\sigma \neq 0$, $\eM_{\epsilon,\sigma}$ is an embedding of Minkowski
spacetime $\MM_{d+1}$ in $\EE^{d+1,2}$.  Let us first show that
$\eM_{\epsilon,\sigma}$ is an orbit of $G$.  Suppose that
$(\x,x^+,\sigma)$ is a point in $\eM_{\epsilon,\sigma}$.  Because it
lies in the quadric $\eQ_\epsilon$ and $\sigma\neq 0$, we may solve
for $x^+$ in terms of $\x$:
\begin{equation}\label{eq:solving-x+}
  x^+(\x) = \frac{\epsilon - \bar\eta(\x,\x)}{2\sigma},
\end{equation}
so that $\eM_{\epsilon,\sigma}$ is the image of $\RR^{d,1}$ under the
embedding $\x \mapsto (\x, x^+(\x), \sigma)$.  The resulting
paraboloid is illustrated in Figure~\ref{fig:Embedding-Minkowski} for
$\epsilon = 0$ and $\sigma = 1$.

The action of the Poincaré group on $(\x,x^+,\sigma)$ can be read off
from equation~\eqref{eq:connected-poincare-subgroup} and we see that
it corresponds to $\x \mapsto A \x + \sigma \bv$. This action is
transitive on $\RR^{d,1}$ and hence transitive on
$\eM_{\epsilon,\sigma}$. Since $x^- = \sigma$ is a constant, the
pull-back to $\eM_{\epsilon,\sigma}$ of the pseudo-euclidean metric
$g_{\EE}$ in equation~\eqref{eq:pseudo-euclidean-metric} agrees with
the Minkowski metric, proving that for any $\epsilon\in \RR$ and
$\sigma \neq 0$, $\eM_{\epsilon,\sigma}$ is isometric to $\MM_{d+1}$.\footnote{We can make contact with the hyperboloid picture of $\zAdS_{d+2}$ in the following way. Setting $\epsilon=-\rho^2$ and parametrizing the hyperboloid as $x^0=y^0 \sigma \rho^{-1}, x^a=\sigma\rho^{-1} y^{a}, x^-=\sigma$ and $x^+$ as in \eqref{eq:solving-x+}, the induced metric on the hyperboloid becomes
  \begin{equation}
    \label{eq:poincarepatch}
d s^2=\rho^2\sigma^{-2}d \sigma^2+\sigma^2\rho^{-2}\left(-(d y^0)^2+(d y^a)(d y_a)\right).
\end{equation}
This is the usual Poincaré patch of $\zAdS_{d+2}$. For $\epsilon<0$ and $\sigma\neq 0$ the Minkowski spaces $\eM_{\epsilon,\sigma}$ described above therefore correspond to this slicing of $\zAdS_{d+2}$. The conformal boundary of $\zAdS_{d+2}$ is reached for $\sigma \rightarrow \infty$. The light-like hypersurface $\sigma=0$ is not covered by the Poincaré patch coordinates. As we will see below, the corresponding space $\eM_{-\rho^2,0}$ is not Minkowski space but $\zTi^\pm$.
}

We pick an origin $(\bzero,\frac{\epsilon}{2\sigma},\sigma) \in
\eM_{\epsilon,\sigma}$.  The subgroup $H \subset G$ fixing the origin
is the proper orthochronous Lorentz subgroup consisting of matrices of
the form
\begin{equation}\label{eq:lorentz-subgroup-ort}
  H = \left\{ 
    \begin{pmatrix}
      A & \bzero & \bzero \\
      \bzero^T & 1 & 0 \\
      \bzero^T & 0 & 1
    \end{pmatrix} ~\middle | ~ A \in \SO(d,1)_0  \right\}.
\end{equation}
Its Lie algebra $\h$ consists of the span of $L_{ab},B_a$, defined in
equation~\eqref{eq:poincare-generators}.  We write this as $\h =
\left<L_{ab},B_a\right>$.

If $\sigma=0$, the quadric condition does not fix $x^+$.  The Poincaré
orbits in $\eM_{\epsilon,0}$ depend on the sign of $\epsilon$, so we
must distinguish between three cases, depending on whether $\epsilon
>0$, $\epsilon<0$ or $\epsilon = 0$.

\subsubsection{Embedding $\zSpi$}
\label{sec:embedding-spi}

Let $\epsilon = \rho^2 > 0$.  Then $\eM_{\rho^2,0}$ consists of those
points $(\x,x^+,0)$ where $\bar\eta(\x,\x) = \rho^2$ and $x^+\in\RR$
is otherwise arbitrary.  The condition $\bar\eta(\x,\x) = \rho^2$ cuts
out a one-sheeted hyperboloid in $\RR^{d,1}$, i.e., a $d$-dimensional
de Sitter space $\zdS_d$. The proper orthochronous Lorentz group $\SO(d,1)_0$ acts transitively on this
hyperboloid.  The translation $\bv$ in $G$ acts via $(\x, x^+, 0)
\mapsto (\x, x^+ - \bar\eta(\bv,\x),0)$ and hence we see that $G$
acts transitively on $\eM_{\rho^2,0}$.  Let $\be_d := (0,0,\dots,0,1)
\in \RR^{d,1}$ be an elementary spacelike vector and let us choose an
origin $(\rho\be_d, 0, 0)$ for $\eM_{\rho^2,0}$.  The subgroup $H \subset
G$ which fixes the origin consists of matrices
\begin{equation}
H=  \begin{pmatrix}
    A & \bzero & \bv \\
    -\bv^T \bar\eta A & 1 & -\tfrac12 \bar\eta(\bv,\bv) \\
    \bzero^T & 0 & 1
  \end{pmatrix}
\end{equation}
where $A \in \SO(d,1)_0$ is such that $A \be_d = \be_d$ and $\bv  \in
\RR^{d,1}$ is such that $\bar\eta(\bv,\be_d) = 0$.  This subgroup $H$
is isomorphic to the Poincaré group $\SO(d-1,1)_0 \ltimes \RR^{d-1,1}$
in one lower dimension.  Its Lie algebra $\h$ can be determined as
those vector fields in \eqref{eq:poincare-generators} which vanish at
the origin and we can see that $\h = \left<L_{ij},B_i, P_i, H\right>$,
where $i,j = 1,\dots,d-1$.   For any $\epsilon >0$, $\eM_{\epsilon,0}$
is an embedding in $\EE^{d+1,2}$ of the blow-up $\zSpi_{d+1}$ of the
spatial infinity $i^0$ of Minkowski spacetime, as reviewed in
Appendix~\ref{sec:AH-construction}.

This embedding shows that $\zSpi_{d+1}$ fibers over $\zdS_d$,
identifying $\zdS_d$ with any one of the one-sheeted hyperboloids in
$\RR^{d,1}$. The projection $\zSpi_{d+1} \to \zdS_d$ sends
$(\x,x^+,0)\mapsto \x$.  This is a trivial bundle and hence
$\zSpi_{d+1} \cong \zdS_d \times \RR$.  Every smooth function $f$ on $\zdS_d$
defines a section $\zdS_d \to \zSpi_{d+1}$ by $\x \mapsto (\x, f(\x),
0)$.  These sections are in one-to-one correspondence with the
$\zSpi$-supertranslations, an infinite-dimensional abelian ideal of
the Lie symmetries of the pseudo-carrollian structure
\cite{Gibbons:2019zfs} on $\zSpi_{d+1}$. As we will see in
Section~\ref{sec:-m-from-SpiNiTi}, some of these supertranslations are
Poincaré translations and they will thus be associated to points in
Minkowski spacetime once we fix an origin and in this way we will
reconstruct Minkowski spacetime from its asymptotic geometry $\zSpi$.

\subsubsection{Embedding $\zTi^\pm$}
\label{sec:embedding-ti}

Let $\epsilon = - \rho^2 < 0$.  Now $\eM_{-\rho^2,0}$ consists of
points $(\x,x^+,0)$ where $\bar\eta(\x,\x) = - \rho^2$ and $x^+ \in
\RR$.  The condition $\bar\eta(\x,\x) = - \rho^2$ defines a
two-sheeted hyperboloid in $\RR^{d,1}$ which is acted on transitively by
$\SO(d,1)$.  Under the identity component $\SO(d,1)_0$, each sheet is
an orbit.  The translation $\bv$ in $G$ acts via $(\x,x^+,0) \mapsto
(\x, x^+ - \bar\eta(\bv,\x),0)$ and hence $G$ acts with two orbits on
$\eM_{-\rho^2,0}$:
\begin{equation}
  \eM_{-\rho^2,0} = \eM_{-\rho^2,0}^+ \cup \eM_{-\rho^2,0}^-,
\end{equation}
where
\begin{equation}
  \eM_{-\rho^2,0}^\pm = \left\{ \begin{pmatrix} \x \\ x^+ \\ 0
    \end{pmatrix} ~ \middle | ~ \bar\eta(\x,x) = - \rho^2,\quad \pm
    x^0 > 0,\quad\text{and}\quad x^+ \in \RR \right\}.
\end{equation}

Let $\be_0 = (1,0,\dots,0) \in \RR^{d,1}$ be an elementary timelike
vector and let us fix the origin $(\pm \rho \be_0,0,0) \in
\eM_{-\rho^2,0}^\pm$.  The subgroup $H \subset G$ which fixes the
origin is common to both orbits and consists of matrices
\begin{equation}
  H=    \begin{pmatrix}
    A & \bzero & \bv \\
    -\bv^T \bar\eta A & 1 & -\tfrac12 \bar\eta(\bv,\bv) \\
    \bzero^T & 0 & 1
  \end{pmatrix}
\end{equation}
where $A \in \SO(d,1)_0$ fixes $\be_0$ and $\bv \in \RR^{d,1}$ is
perpendicular to $\be_0$. This subgroup is isomorphic to the euclidean
group $\SO(d) \ltimes \RR^d$ in one lower dimension, corresponding to
the hyperplane of $\RR^{d,1}$ perpendicular to $\be_0$. Its Lie
algebra $\h$ is spanned by those vector fields in
equation~\eqref{eq:poincare-generators} which vanish at the origin:
namely, $\h = \left<L_{ab},P_a\right>$. As shown in
\cite{Figueroa-OFarrill:2018ilb} (see also \cite{Morand:2018tke}),
$\eM_{-\rho^2,0}^\pm$ define embeddings of $\zAdSC_{d+1}$, the
carrollian limit of $\zAdS_{d+1}$.  As discussed in
Appendix~\ref{sec:AH-construction}, $\zAdSC_{d+1}$ is isomorphic (as a
homogeneous space of the Poincaré group) to the blow-ups
$\zTi_{d+1}^\pm$ of the timelike infinities $i^\pm$ of Minkowski
spacetime.  We shall therefore refer to $\zAdSC_{d+1}$ simply as
$\zTi_{d+1}$.

Just as with $\zSpi_{d+1}$, this embedding shows that $\zTi_{d+1}$
fibers over hyperbolic space $\eH^d$, where we identify $\eH^d$ with
any one of the sheets of the two-sheeted hyperboloids $\bar\eta(\x,\x)
= -\rho^2$.  The fibration $\zTi_{d+1} \to \eH^d$ sends $(\x,x^+,0)
\mapsto \x$.  Again this is a trivial bundle and hence $\zTi_{d+1}
\cong \eH^d \times \RR$.  The smooth sections $\eH^d \to \zTi_{d+1}$
can be identified with the smooth functions on $\eH^d$ and correspond
to the $\zTi$-supertranslations, the infinite-dimensional abelian
ideal of the Lie symmetries of the carrollian structure of
$\zTi_{d+1} (= \zAdSC)$, which were determined in
\cite{Figueroa-OFarrill:2019sex}.  Again, some of the
supertranslations correspond to Poincaré translations and we will
revisit this in Section~\ref{sec:-m-from-SpiNiTi} when we discuss the reconstruction
of Minkowski spacetime from $\zTi$.

\subsubsection{Embedding $\zNi^\pm$}
\label{sec:embedding-ni}

Finally we let $\epsilon = 0$ and consider $\eM_{0,0} = \eQ_0
\cap \eN_0$.  The point $(\x,x^+,0)$ lies in $\eM_{0,0}$ if and only
if $\bar\eta(\x,\x) = 0$, so that $\x$ lies on the lightcone in
$\RR^{d,1}$.  Under $\SO(d,1)_0$, the lightcone $\zLC \subset
\RR^{d,1}$ breaks up into three orbits:
\begin{equation}
  \zLC = \zLC^- \cup \{\bzero\} \cup \zLC^+,
\end{equation}
where $\zLC^\pm$ are the future/past lightcones with the apex removed.
Provided that $\x \in \zLC^\pm$, the translations in $G$ can relate
any two values of $x^+$, but if $\x = \bzero$, then each of the points
$(\bzero,x^+,0)$ is fixed by $G$.  In summary,
$\eM_{0,0}$ breaks up into two $(d+1)$-dimensional orbits
$\eM_{0,0}^\pm$ and a line $\ell = \{(\bzero,x^+,0) \mid x^+ \in
\RR\}$ of fixed points under the Poincaré group $G$; that is,
\begin{equation}
  \eM_{0,0} = \eM_{0,0}^- \cup \ell \cup \eM_{0,0}^+
  \qquad\text{with}\qquad \ell = \bigcup_{x^+\in\RR} \{(\bzero, x^+, 0)\},
\end{equation}
where
\begin{equation}
  \eM_{0,0}^\pm = \left\{ \begin{pmatrix} \x \\ x^+ \\ 0
    \end{pmatrix} ~ \middle | ~ \bar\eta(\x,\x) = 0,\quad \pm
    x^0 > 0,\quad\text{and}\quad x^+ \in \RR \right\}.
\end{equation}

Let
$\be_- := \frac{1}{\sqrt2}(\be_d - \be_0) =
(-\frac1{\sqrt2},0,\dots,0,\frac1{\sqrt2}) \in \RR^{d,1}$ and let us
fix the origin $(\mp \be_-, x^+, 0) \in \eM_{0,0}^\pm$.  The subgroup
$H \subset G$ which fixes the origin is common to both $\eM_{0,0}^\pm$
and consists of matrices
\begin{equation}
 H=   \begin{pmatrix}
    A & \bzero & \bv \\
    -\bv^T \bar\eta A & 1 & -\tfrac12 \bar\eta(\bv,\bv) \\
    \bzero^T & 0 & 1
  \end{pmatrix}
\end{equation}
where $A \in \SO(d,1)_0$ fixes $\be_-$ and $\bv \in \RR^{d,1}$ is
perpendicular to $\be_-$.  This subgroup is isomorphic to the
$d$-dimensional Carroll group.  Its Lie algebra $\h$ is spanned by
those vector fields in equation~\eqref{eq:poincare-generators} which
vanish at the origin: namely, $\h = \left<L_{ij},P_i, L_{id} + B_i,
  H-P_d\right>$, for $i,j = 1,\dots,d-1$.  Again, since the stabiliser
subgroup is common to both $\zNi^\pm$, they are isomorphic as
homogeneous spaces of $G$.  We will therefore refer to either one of
these two spaces simply as $\zNi$.

As we will see in Section \ref{sec:znul}, we will identify
$\eM_{0,0}^\pm$ with $\zNi^\pm$, the bundle of scales of the conformal
carrollian structure of $\scri^\pm$.  This embedding of $\zNi^\pm$
shows that it fibers over the future/past lightcone $\zLC^\pm$, with
the fibration $\zNi^\pm \to \zLC^\pm$ given simply by $(\x,x^+,0)
\mapsto \x$.  Together with the identification of $\zNi^\pm$ as a
bundle over $\scri^\pm$, we can see that there is a double fibration
\begin{equation}\label{eq:double-fibration-1}
  \begin{tikzcd}
                     & \zNi^\pm_{d+1} \arrow[dl] \arrow[dr] &                 \\
    \scri^\pm_d \arrow[dr] &                             & \zLC^\pm_d \arrow[dl] \\
                     & \zCS^{d-1}                        &                 \\
  \end{tikzcd}
\end{equation}
which allows us to view $\zNi^\pm$ as sitting inside $\zLC^\pm \times
\scri^\pm$ as their fibred product over the celestial sphere $\zCS$.
Said differently, the fibration $\zNi^\pm \to \scri^\pm$ is the
pull-back fibre bundle of the fibration $\zLC^\pm \to \zCS$ via the
fibration $\scri \to \zCS$.

The fibration $\zNi^\pm \to \scri^\pm$ can also be understood from the
embedding picture.  Let $\PP^{d+2}$ be the projective space of
$\EE^{d+1,2}$.  It is the quotient of $\EE^{d+1,2}\setminus\{0\}$ by the
action of the nonzero reals $\RR^\times$ which rescales the nonzero
vectors: $x \mapsto \lambda x$ for $x \in \EE^{d+1,2}\setminus\{0\}$ and
$\lambda \in \RR^\times$.  As explained in
\cite[Section~9.2]{MR838301}, the image of the null quadric $\eQ_0$ in
$\PP^{d+2}$ is a conformal compactification $\MM_{d+1}^\#$ of
Minkowski spacetime $\MM_{d+1}$. The image of points in $\eQ_0$ with
$x^-\neq0$ correspond to the interior points of $\MM_{d+1}^\#$
(corresponding to Minkowski spacetime itself), whereas the image of
points with $x^- =0$ correspond to the conformal boundary of Minkowski
spacetime in this compactification.  The points in $\zNi^\pm_{d+1}$
map to $\scri$, which is the identification of $\scri^+$ and
$\scri^-$, which are after all indistinguishable as homogeneous spaces
of $G$, whereas the points in the singular line $\ell$ (except for the
origin) get mapped to the same point $I \in \PP^{d+2}$ which is the
identification of $i^0$, $i^+$ and $i^-$. Hence the fibration
$\zNi^\pm_{d+1} \to \scri_d$ to be discussed in
Section~\ref{sec:two-ways-celestial} is simply the restriction to
$\zNi^\pm_{d+1} \subset \EE^{d+1,2}$ of the projection
$\EE^{d+1,2}\setminus\{0\} \to \PP^{d+2}$.

\subsubsection{Summary}
\label{sec:summary}

We may summarise the above discussion by explicitly decomposing
$\EE^{d+1,2}$ in terms of orbits of the connected Poincaré group:
\begin{multline}
  \label{eq:orbit-decomposition}
    \EE^{d+1,2} = \left(
      \bigsqcup_{{\varepsilon,\sigma\in\RR}\atop{\sigma \neq 0}}
      \underbrace{\eM_{\varepsilon,\sigma}}_{\cong \MM} \right) \sqcup
    \left( \bigsqcup_{\varepsilon>0}
    \underbrace{\eM_{\varepsilon,0}}_{\cong\zSpi} \right) \sqcup
    \left( \bigsqcup_{\varepsilon<0}
    \underbrace{\eM^+_{\varepsilon,0}}_{\cong \zTi^+} \right)\sqcup
    \left( \bigsqcup_{\varepsilon<0}
      \underbrace{\eM^-_{\varepsilon,0}}_{\cong \zTi^-} \right)\\
    {} \sqcup
    \underbrace{\eM^+_{0,0}}_{\cong \zNi^+}\sqcup
    \underbrace{\eM^-_{0,0}}_{\cong \zNi^-} \sqcup \left( \bigsqcup_{x^+ \in
      \RR} \left\{ \begin{pmatrix}\bzero \\ x^+ \\ 0 \end{pmatrix}
    \right\} \right)
\end{multline}

We may now pass to the projective space $\PP^{d+2} =
(\EE^{d+1,2}\setminus\{0\})/\RR^\times$ to obtain
\begin{equation}
  \label{eq:projectivisation}
  \PP^{d+2} = \left( \bigsqcup_{\tau\in\RR} \MM \right) \sqcup \zSpi
  \sqcup \zTi \sqcup \scri \sqcup \{I\},
\end{equation}
where $\tau = \varepsilon/\sigma^2$ is a projective invariant.
Restricting to the projectivised null quadric we obtain the conformal
compactification
\begin{equation}
  \MM^\sharp = \MM \sqcup \scri \sqcup \{I\}
\end{equation}
of Penrose and Rindler \cite[Section 9.2]{MR838301}.  Although they
treat the four-dimensional case ($d=3$ here), their results are
dimension agnostic.  Here $\scri$ is the identification of $\scri^+$
and $\scri^-$ and $\{I\}$ is the singleton set obtain by identifying
$i^0$ and $i^\pm$.  However we see that if we do not restrict to the
null quadric, we actually obtain $\zSpi$ and $\zTi^\pm$ as limits of a
family of embedded Minkowski spacetimes.

\section{Fables of the reconstruction}
\label{sec:reconstruction}

The embedding formalism in Section~\ref{sec:embeddings} allows us to
explain how to reconstruct Minkowski spacetime $\MM_{d+1}$ from its
asymptotic geometries $\zSpi_{d+1}$, $\zTi_{d+1}$, $\zNi_{d+1}$ and
$\scri_d$.  In all cases, the idea is the same. Every point in
Minkowski spacetime is stabilised by a unique Lorentz subgroup of the
Poincaré group $G$.  Our strategy is to fix an origin in $\MM_{d+1}$
and consider the orbits of the corresponding (proper, orthochronous)
Lorentz subgroup of $G$ on $\zSpi_{d+1}$, $\zTi_{d+1}$, $\zNi_{d+1}$
and $\scri_d$. In all cases the orbits will be hypersurfaces, which
turn out to be cut out by a section of the (trivial) fibrations
$\zSpi_{d+1} \to \zdS_d$, $\zTi_{d+1} \to \eH^{d}$, $\zNi_{d+1} \to
\zLC_d$ and $\scri_d \to \zCS^{d-1}$. This means that we may
associate one such section to the origin in Minkowski spacetime.

Any other point in $\MM_{d+1}$ is obtained from the origin by a unique
translation.  Hence to see which sections correspond to points in
Minkowski spacetime, we can take the section corresponding to the
origin and apply a translation.  In this way we will obtain a family
of hypersurfaces in each of $\zSpi_{d+1}$, $\zTi_{d+1}$, $\zNi_{d+1}$
and $\scri_d$ or, equivalently, a family of sections of each of the
trivial fibrations $\zSpi_{d+1} \to \zdS_d$, $\zTi_{d+1} \to \eH^{d}$,
$\zNi_{d+1} \to \zLC_d$ and $\scri_d \to \zCS^{d-1}$.  Being trivial,
the sections can be identified with smooth functions on the base.  We
will see that the sections corresponding to the points in Minkowski
spacetime can be identified with (the restrictions to $\zdS_d$, $\eH^{d}$
and $\zLC_d$ of) affine functions on the ambient $\RR^{d,1}$ in the
first three cases, and from affine functions on the ambient $\RR^d$ in
the case of $\zCS^{d-1}$.\footnote{This is to be compared with the
reconstruction \cite{eastwood_tod_1982} of four-dimensional (complex)
Minkowski spacetime from $\scri$ as the space of certain hypersurfaces
of $\scri$ (the so-called ''good cuts''), which arise as sections of
the fibration $\scri \to \zCS$.  The space of good cuts is an affine
space modelled on the kernel of $\eth^2$, which for Minkowski
spacetime, consists of the spherical harmonics on the sphere with
$\ell = 0,1$.  But these are precisely the restriction to the sphere
of the affine functions on the ambient three-dimensional euclidean
space.} An additional freedom in the identification of points of Minkowski
space with sections of $\zSpi_{d+1}$, $\zTi_{d+1}$, $\zNi_{d+1}$
and $\scri_d$ is fixed by employing the (generalized) light-cone of the ambient space
$\mathbb{R}^{d+1,2}$.

The essence of our approach may be described as follows. Once we fix a
reference section through $\zSpi_{d+1} \to \zdS_d$,
$\zTi_{d+1} \to \eH^{d}$, $\zNi_{d+1} \to \zLC_d$ or
$\scri_d \to \zCS^{d-1}$, any other section is obtained from the
reference section via the action of a supertranslation in the
corresponding symmetry group. Some of the supertranslations are
Poincaré translations, and the sections obtained from the reference
section via Poincaré translations are in bijective correspondence with
points in Minkowski spacetime. This means we reconstruct points in
Minkowski spacetime using sections in the asymptotic geometries; something
we can interpret as a form of holographic reconstruction of Minkowski
spacetime. We will use the conventions introduced in
Section~\ref{sec:embeddings}.

\subsection{Reconstructing $\MM$ from $\zSpi$, $\zTi$ and $\zNi$}
\label{sec:-m-from-SpiNiTi}

Let us first treat the three cases: $\zSpi_{d+1}$, $\zTi_{d+1}$
and $\zNi_{d+1}$.   Let $\MM := \eM_{0,1}$ denote the embedded Minkowski
spacetime containing the point $(\bzero,0,1)$, which we shall think of
as the origin.  The origin is stabilised by the subgroup $\Ort(d,1)
\subset \Ort(d+1,2)$ consisting of matrices like those in
equation~\eqref{eq:lorentz-subgroup-ort}, whose identity component is
\begin{equation}
  H = \left\{   \begin{pmatrix}
    A & \bzero & \bzero \\
    \bzero^T & 1 & 0 \\
    \bzero^T & 0 & 1
  \end{pmatrix}
  \, \middle | \,
  A \in \SO(d,1)_0\right\},
\end{equation}
with $\SO(d,1)_0$ the identity component of $\Ort(d,1)$, i.e., they
are given by the proper, orthochronous Lorentz transformations
parametrised by $A$.

Let $\x_0 \in \RR^{d,1}$. As one can easily show, the action of $H$ is
given by $(\x_0,x^+,0) \mapsto (A \x_0,x^+,0)$. This means the
orbit of $(\x_0,x^+,0)$ under $H$ consists of the hypersurface with
points $(\x, x^+,0)$ where $\x$ is in the (proper, orthochronous)
Lorentz orbit of $\x_0$. For example, if $(\x_0,x^+,0)$ belongs to
$\zSpi_{d+1}$ or $\zTi_{d+1}$ or $\zNi_{d+1}$, its orbit consists of
the hypersurface in $\zSpi_{d+1}$ or $\zTi_{d+1}$ or $\zNi_{d+1}$ with
$x^+ =$ constant.
We can interpret them as sections of the fibrations
$\zSpi_{d+1} \to \zdS_d$, $\zTi_{d+1} \to \eH^{d}$ and
$\zNi_{d+1} \to \zLC_d$. Since these fibrations are trivial, sections
correspond to smooth functions on $\zdS_d$, $\eH^{d}$ and $\zLC_d$.
For example, if $f \in C^\infty(\zdS_d)$, then its graph defines a
section $\zdS_d \to \zSpi_{d+1}$ consisting of the points $(\x, f(\x),0)$.
It is then clear that if we take $f$ to be a constant function on
$\zdS_d$, $\eH^{d}$ and $\zLC_d$, its graph is precisely the section
of $\zSpi_{d+1} \to \zdS_d$, $\zTi_{d+1} \to \eH^{d}$ and
$\zNi_{d+1} \to \zLC_d$ corresponding to fixing $x^+$ to the constant
value of $f$.  We will now determine the functions giving rise
to sections whose corresponding hypersurfaces are parametrised by
the points in Minkowski spacetime.

Let us act on these hypersurfaces with the Poincaré translations
\begin{equation}\label{eq:translation-subgrp-ort}
  \begin{pmatrix}
    \1 & \bzero & \bv \\
    -\bv^T \bar \eta & 1 & -\tfrac12 \bar\eta(\bv,\bv) \\
    \bzero^T & 0 & 1
  \end{pmatrix}.
\end{equation}
Each translation, given by the vector $\bv$, is identified with a
unique point in Minkowski spacetime (once we choose an origin).
The components $(v^{0},v^{1},\ldots , v^{d})$ of $\bv$ are cartesian
coordinates for Minkowski spacetime centred at the origin, and hence
we can identify Minkowski spacetime with $\RR^{d,1}$.  The action of
the translation is then given by
\begin{equation}
  \begin{pmatrix}
    \x \\ x^+ \\ 0
  \end{pmatrix} \mapsto
  \begin{pmatrix}
    \x \\ x^+ - \bar\eta(\x,\bv)\\ 0
  \end{pmatrix}.
\end{equation}
Therefore the action on the hypersurfaces of $\zSpi_{d+1}$,
$\zTi_{d+1}$ and $\zNi_{d+1}$ corresponds to the sections of the
trivial fibrations $\zSpi_{d+1} \to \zdS_d$, $\zTi_{d+1} \to \eH^{d}$ and
$\zNi_{d+1} \to \zLC_d$ defined by the restriction to $\zdS_d$,
$\eH^{d}$ and $\zLC_d$ of the affine function $f : \RR^{d,1} \to \RR$
defined by $f(\x) = x^+ - \bar\eta(\bv,\x)$.

The space of such affine functions is ($d+2$)-dimensional:
parametrised by $x^+$ and $\bv \in \RR^{d,1}$.  Clearly, Minkowski
spacetime only knows about $\bv$, and hence to reconstruct it or,
equivalently, to put the hypersurfaces in bijective correspondence
with the points of Minkowski spacetime we would either fix $x^+$ or
else introduce an equivalence relation between hypersurfaces which are
related by a constant shift in $x^+$ and take equivalence classes.

We may choose a value of $x^+$ via the following geometric
construction, which is analogous to the one in \cite{Kozameh:1983yu}.
An alternate construction is presented in
Appendix~\ref{sec:another-choice-section}. In a nutshell, we will
draw a generalised lightcone $\gLC_p \subset \EE^{d+1,2}$ at every point
$p \in \MM \subset \EE^{d+1,2}$ and then study its intersection
with $\zTi$, $\zSpi$ and $\zNi$.  In this way we can associate with
every point $p \in \MM$ a hypersurface in $\zTi$, $\zSpi$ and 
$\zNi$ and reconstruct $\MM$ (and hence Minkowski spacetime) as
the parameter space of such hypersurfaces.  This is analogous to the
identification in \cite{Kozameh:1983yu} of lightcone cuts in $\scri^+$
with its intersections with the lightcone at a point in Minkowski
spacetime.

Choose a point
$p = (\bv,-\tfrac12 \bar\eta(\bv,\bv),1) \in \MM \subset \EE^{d+1,2}$
and let $\gLC_p$ denote the null quadric in $\EE^{d+1,2}$ centred at
$p$:
\begin{equation}
  \gLC_p = \left\{ (\x, x^+, x^-) \in \EE^{d+1,2} ~ \middle | ~
    \bar\eta(\x -\bv, \x-\bv) + 2 (x^++\tfrac12\bar\eta(\bv,\bv))(x^--1) =
  0\right\}.
\end{equation}
Notice that $\gLC_p$ intersects $\MM$ precisely at the lightcone
based at $p$.  Indeed, if $x^-=1$, then $\bar\eta(\x-\bv,\x-\bv) = 0$,
so that $\x$ lives in the lightcone of $\RR^{d,1}$ based at $\bv$.  The
value of $x^+$ is undetermined and we can always choose it to be
$-\tfrac12 \bar\eta(\x,\x)$ so that $(\x,x^+,1) \in \MM$.  Hence
we conclude that $\gLC_p \cap \MM$ is the Minkowski lightcone
based at the point $p$.  We call $\gLC_p$ the ``generalised
lightcone'' based at $p$.
\begin{figure}[h!]
\centering
\definecolor{cffffff}{RGB}{255,255,255}
\definecolor{cd20000}{RGB}{210,0,0}
\definecolor{c007aff}{RGB}{0,122,255}
\definecolor{c007355}{RGB}{0,115,85}
\begin{tikzpicture}[overlay]
\begin{pgfonlayer}{nodelayer}
		\node [style=ghost] (0) at (5, 3.8) {{\color{cd20000}$\MM$}};
		\node [style=ghost] (1) at (3.5, 1.0) {$\gLC_p$};
		\node [style=ghost] (2) at (3.2, 7.6) {{\color{c007aff}$\zLC$}};
		\node [style=ghost] (3) at (5.1, 4.7) {{\color{c007355} $p$}};
\end{pgfonlayer}
\end{tikzpicture}
\includegraphics[width=0.45\textwidth]{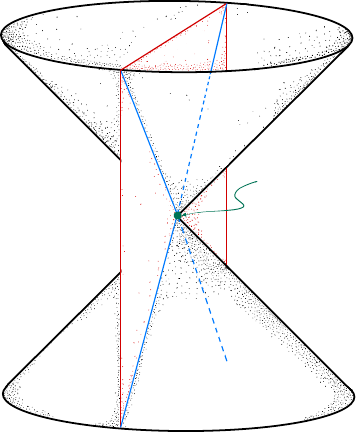}
\caption{The lightcone $\zLC$ at $p \in \MM$ as the
  intersection $\gLC_p \cap \MM$}
    \label{fig:lightcone-intersection}
\end{figure}

The ``light rays'' of $\gLC_p$ intersect the asymptotic geometries
$\zTi$, $\zSpi$ and $\zNi$, which are also embedded in $\EE^{d+1,2}$.
These intersections are easy to determine and we find the following:
\begin{equation}
  \label{eq:intersec}
  \begin{split}
    \gLC_p \cap \zSpi &= \left\{ (\x, x^+, 0)~\middle |~
      \bar\eta(\x,\x)=1~\text{and}~ x^+ = \tfrac12 -
      \bar\eta(\bv,\x)\right\}\\
    \gLC_p \cap \zTi &= \left\{ (\x, x^+, 0)~\middle |~
      \bar\eta(\x,\x)=-1~\text{and}~ x^+ = -\tfrac12 -
      \bar\eta(\bv,\x)\right\}\\
    \gLC_p \cap \zNi &= \left\{ (\x, x^+, 0)~\middle |~
      \bar\eta(\x,\x)=0,~ x^0>0~\text{and}~ x^+ = -
      \bar\eta(\bv,\x)\right\}.
  \end{split}
\end{equation}

The intersections in~\eqref{eq:intersec} relate a point $x \in
\RR^{d,1}$ in Minkowski spacetime to smooth functions $x^{+}(\bv)$, the
last equation in each line in~\eqref{eq:intersec}, on the respective
asymptotic geometries.  Let us consider the case of $\zSpi$, for
definiteness.  The origin of Minkowski spacetime corresponds to the
constant function $x^+(\bv) = \tfrac12$, which also shows that we have
indeed fixed the ambiguity of $x^+$ to $\tfrac12$.  A generic
point $(\bv,-\tfrac12 \bar\eta(\bv,\bv),1)$ in Minkowski spacetime leads then to
the affine function $x^+(\bv) = \tfrac12 - \bar\eta(\x,\bv)$, where
$\bar\eta(\bv,\bv)=1$, so that $x^+$ is a function on $\zdS$.  In this
way we have assigned to every point of Minkowski spacetime a function
in $\zSpi$, which is the restriction of an affine function on
$\RR^{d,1}$.  Conversely, given such a function it is clear we can
read off $\bv$ and therefore find the corresponding point in Minkowski
spacetime, establishing a bijection.  Similar arguments apply to
$\zTi$ and $\zNi$.

One final remark. Whereas the value of $x^+=0$ for $\zNi$ is
independent of choices, the values $x^+ = \pm \tfrac12$ for $\zSpi$
and $\zTi$, respectively, depend on the precise embeddings of
$\MM_{d+1}$, $\zSpi_{d+1}$ and $\zTi_{d+1}$ in $\EE^{d+1,2}$.
Re-embedding $\MM_{d+1}$ as the intersection $\eQ_0 \cap \eN_\sigma$,
for $\sigma \neq 0$, and similarly $\zTi_{d+1}$ as
$\eQ_{-\epsilon} \cap \eN_0$ and $\zSpi_{d+1}$ as
$\eQ_{\epsilon} \cap \eN_0$, for $\epsilon>0$, we can change $x^+$ to
$\pm \frac{\epsilon}{2\sigma}$, which can be any nonzero real
number.\footnote{Note that, in the case of $\zSpi$ and $\zTi$,
  Poincaré transformations cannot shift the value of $x^+$ by a
  constant, so that this statement is indeed Poincaré invariant.} The
important thing is that once a choice of $x^+$ has been made and hence
one hypersurface chosen, the other translates of that initial
hypersurface are in bijective correspondence with the points in
Minkowski spacetime. As noted previously in Sections
\ref{sec:embedding-spi} and \ref{sec:embedding-ti}, the more general
transformations in which $x^+$ is shifted by an arbitrary function
$f(\x)$ instead of the linear function $-\eta(\x,\bv)$ correspond to
$\zSpi,\zTi,\zNi$-supertranslations, respectively.

\subsection{Reconstructing $\MM$ from $\scri$}
\label{sec:m-from-scri}

Now we discuss the reconstruction of $\MM_{d+1}$ from $\scri_d$ along
the lines explained above.  We recall that $G$ denotes the identity
component of the Poincaré group.  We shall refer to $(A,\bv) \in G$ as
the Poincaré transformation consisting of a proper, orthochronous
Lorentz transformation $A$ and a translation $\bv$.  Let us consider
\begin{equation}
  \zNi^+ = \left\{ \begin{pmatrix} \x \\ x^+ \\ 0 \end{pmatrix}~
    \middle | ~ \x \in \zLC^+\quad\text{and}\quad x^+ \in \RR\right\},
\end{equation}
where $\zLC^+ \in \RR^{d,1}$ is the future lightcone.  The action of
$(A,\bv) \in G$ on $\zNi^+$ is given by
\begin{equation}
  (A,\bv) \cdot \begin{pmatrix} \x \\ x^+ \\ 0 \end{pmatrix}=
  \begin{pmatrix} A\x \\ x^+ - \bar\eta(\bv,A\x)\\ 0 \end{pmatrix}.
\end{equation}
Since the proper orthochronous Lorentz group acts transitively on
$\zLC^+$, it follows that $G$ acts transitively on $\zNi^+$ and hence
also on $\scri^+ = \PP_+\zNi^+$, where $\PP_+$ denotes the ray
projectivisation.  $\PP_+\zNi^+$ is the space of orbits of the $\RR^+$
action on $\zNi^+$ given by rescaling:
\begin{equation}
  \begin{pmatrix} \x \\ x^+ \\ 0 \end{pmatrix} \mapsto \begin{pmatrix}
    \lambda \x \\ \lambda x^+ \\ 0 \end{pmatrix}
\end{equation}
for $\lambda \in \RR^+$ a positive real number.

\subsubsection{Orbit decomposition of $\scri^+$ under the Lorentz group}
\label{sec:orbit-decomp-scri}

This section is somewhat outside the main narrative in the paper, but
we think it is interesting to point out the fact that although
$\scri^+$ is a homogeneous space of the Poincaré group, it is not far
from a homogeneous space of a Lorentz subgroup.  Indeed, pick the
origin in Minkowski spacetime which is stabilised by a subgroup
$\SO(d,1)_0$ of $G$.  What are the orbits of $\SO(d,1)_0$ on
$\scri^+$?  We claim that there are three orbits corresponding to
those $\left[ \begin{pmatrix} \x \\ x^+ \\ 0 \end{pmatrix}  \right]
\in \scri^+$ with $x^+=0$, $x^+>0$ and $x^+<0$.  The latter two cases
are open orbits of the same dimension as $\scri^+$, whereas the orbit
with $x^+=0$ is the desired hypersurface.

Let $x^+ > 0$.  We claim that
\begin{equation}
  \eO^+ := \left\{ \left[ \begin{pmatrix} \x \\ x^+ \\ 0 \end{pmatrix}  \right]
    ~\middle | ~ \x \in \zLC^+\quad\text{and}\quad x^+ > 0\right\}
  \subset \scri^+
\end{equation}
is an orbit of $\SO(d,1)_0$.  It is enough to show that any points in
$\eO^+$ are related by a proper orthochronous Lorentz transformation;
that is, that given any two points
\begin{equation}
  \left[ \begin{pmatrix} \x \\ x^+ \\ 0 \end{pmatrix}  \right]
  \qquad\text{and}\qquad \left[ \begin{pmatrix} \y \\ y^+ \\ 0 \end{pmatrix}  \right]
\end{equation}
with $\x,\y \in \zLC^+$ and $x^+,y^+>0$, there exists $A \in \SO(d,1)_0$
and $\lambda \in \RR^+$ such that $A \x = \lambda \y$ and $x^+ =
\lambda y^+$.  The second relation clearly sets $\lambda = x^+/y^+$
and the first equation says that $A \x = x^+/y^+ \y$.  But $x^+/y^+ \y
\in \zLC^+$ and $\SO(d,1)_0$ acts transitively, so that there exists some
$A \in \SO(d,1)_0$ sending $\x$ to $x^+/y^+ \y$.  A similar argument shows
that $\eO^-$, defined as $\eO^+$ but with $x^+<0$, is an orbit of
$\SO(d,1)_0$.

Suppose now that $x^+=0$.  Then to every $\left[ \begin{pmatrix} \x \\
    0 \\ 0   \end{pmatrix}  \right] \in \zNi^+$ there corresponds
    {\mbox{$[\x] \in \PP_+\zLC^+ \cong \zCS$}}.  Since $\SO(d,1)_0$ acts transitively on
$\zLC^+$, it acts transitively on $\zCS$ and, indeed, does so via
conformal transformations.  This orbit is thus a section of the
trivial fibration $\scri_d \to \zCS^{d-1}$.  The action of the
Poincaré translations in this hypersurface is given by
\begin{equation}
\left[ \begin{pmatrix} \x\\ 0 \\ 0 \end{pmatrix}  \right] \mapsto
\left[ \begin{pmatrix} \x\\ -\bar\eta(\bv,\x) \\ 0 \end{pmatrix} \right],
\end{equation}
which we may think of as a section of $\scri^+_d \to \zCS^{d-1}$
associated to a linear function in the ambient euclidean space $\RR^d$
into which the sphere embeds.

We think it is curious that to make $\scri^+$ into a homogeneous space
we need to extend the Lorentz group (which acts with three orbits) to
the full Poincaré group.

\section{Klein geometries}
\label{sec:klein-pairs}

In this section we will describe the homogeneous spaces of the
Poincaré group studied in the previous section as Klein geometries.

Recall that a Klein geometry of a Lie group $G$ is a homogeneous space
of $G$; that is, a smooth manifold $M$ on which $G$ acts smoothly and
transitively.  The intuition is that every point of $M$ ``looks the
same'' through the optics of $G$.  Pick an arbitrary point $o \in M$
and call it the \emph{origin}.  Let $H$ be the subgroup of $G$
consisting of elements which fix the origin. Then $H$ is a closed
subgroup of $G$ and $M$ is $G$-equivariantly diffeomorphic to the
space $G/H$ of left cosets $gH$, for $g \in G$, where the Lie group
$G$ acts on $G/H$ via left multiplication.  Let $\g$ and $\h$ denote
the Lie algebras of $G$ and $H$, respectively.  Then to a homogeneous
space of $G$ we may assign a Klein pair $(\g,\h)$.  Conversely a Klein
pair $(\g,\h)$ is said to be \emph{geometrically realisable} if there
exists a Lie group $G$ with Lie algebra $\g$ such that the connected
subgroup $H$ of $G$ corresponding to $\h$ is closed.  Not every Lie
pair is geometrically realisable, but it is possible to show that
there is a one-to-one correspondence between simply-connected
homogeneous spaces and geometrically realisable (effective) Klein
pairs. Said differently, such Klein pairs describe homogeneous spaces
up to coverings.

In the previous section we described a Poincaré subgroup of
$\Ort(d+1,2)$ acting linearly in a pseudo-euclidean space
$\EE^{d+1,2}$, which we then decomposed into orbits of the identity
component $G$ of the Poincaré group.  Not counting a line of
point-like orbits, all other orbits are $(d+1)$-dimensional and
$G$-equivariantly diffeomorphic to one of several homogeneous spaces
of $G$: Minkowski spacetime $\MM$ and three other spaces associated to
the asymptotic geometry of Minkowski space: $\zSpi$, $\zTi^\pm$ and
$\zNi^\pm$.  As homogeneous spaces of $G$ there is no distinction
between $\zTi^+$ and $\zTi^-$ nor between $\zNi^+$ and $\zNi^-$, and
we will therefore refer to the homogeneous spaces as $\zTi$ and
$\zNi$.

Already in Section~\ref{sec:embeddings} we chose some origins in the
homogeneous spaces and determined the corresponding stabiliser
subgroups and their Lie algebras as subalgebras of the Lie algebra $\g$ of
the Poincaré group, which are listed in
equation~\eqref{eq:poincare-generators} as vector fields in
$\EE^{d+1,2}$.  We collect those results here in order to further study
the Klein geometries and to determine, in particular, their invariant
geometrical structures.

To help to orient the reader let us provide a short overview.  In
Table~\ref{tab:overview} we list the Klein pairs $(\g,\h)$ in all
cases and identify the subalgebra $\h$ in the standard basis
$L_{\mu\nu} = - L_{\nu\mu}, P_\mu$, $\mu,\nu = 0,1,\dots,d$ for the
Poincaré Lie algebra
\begin{equation}\label{eq:poincare-1}
  \begin{split}
    [L_{\mu\nu}, L_{\rho\sigma}] &= \eta_{\nu\rho} L_{\mu\sigma} -
    \eta_{\mu\rho} L_{\nu\sigma} - \eta_{\nu\sigma} L_{\mu\rho} + \eta_{\mu\sigma} L_{\nu\rho} = - 4 \eta_{[\rho|[\mu}L_{\nu]|\sigma]}\\
    [L_{\mu\nu},P_\rho] &= \eta_{\nu\rho} P_\mu - \eta_{\mu\rho} P_\nu = - 2 \eta_{\rho[\mu}P_{\nu]}\\
    [P_\mu, P_\nu] &= 0,
  \end{split}
\end{equation}
related to the one in equation~\eqref{eq:poincare-generators} by $B_a
= L_{0a}$ and $H=P_0$.

The subalgebras of Poincaré which play in $\zTi$, $\zSpi$ or $\zNi$
the rôle of the Lorentz subalgebra in $\MM$ admit a uniform
description. If we think of Minkowski spacetime as an affine space
modelled on a lorentzian vector space $(V,\eta)$, where $\eta$ is the
Minkowski metric in mostly positive signature, then the subalgebras
$\h$ in the Klein pairs for $\zTi$, $\zSpi$ and $\zNi$ may be
described as follows: pick a vector $v \in V$ which is, respectively,
timelike, spacelike and null and let $P = v^\mu P_\mu$ be the
corresponding momentum generator. The subalgebra $\h$ in each of the
Klein pairs of $\zTi$, $\zSpi$ and $\zNi$ is a semidirect product
\begin{equation}
  \label{eq:uniP}
  \h = \stab(P) \ltimes P^\perp
\end{equation}
of the subalgebra $\stab(P)$ of $\so(V)$ which fixes $P$ and the
translations perpendicular to $P$. 

We now provide a brief description of this construction for each of the
spaces $\zTi$, $\zSpi$ and $\zNi$, while a more thorough analysis follows
below. For $\zTi$ we take $P = P_0$ so that $P^\perp =
\left<P_a\right>\cong \RR^d$, where the indices $a,b,\dots = 1,\dots,d$
run over the spatial directions. The stabiliser $\stab(P)$ preserving the
timelike momentum is $\so(d) \cong \left< L_{ab}\right>$. For $\zSpi$, we
pick $P = P_d$, which leaves $P^\perp = \left< P_\alpha \right>\cong
\RR^{d-1,1}$, where the $d$-dimensional lorentzian indices
$\alpha,\beta,\dots = 0,\dots,d-1$ run over all directions except for the
$d$-direction. The stabiliser of $P_d$ is the $d$-dimensional Lorentz
group $\so(d-1,1) \cong \left< L_{\alpha\beta} \right>$. Finally, for $\zNi$,
we choose the null momentum $P = P_- :=
\frac{1}{\sqrt{2}}(P_d - P_0)$, and hence $P^\perp = \left<P_-,P_i
\right>\cong \RR^d$, where $i,j,\dots = 1,\dots,d-1$ are
$(d-1)$-dimensional spatial indices. The stabiliser of $P_-$ is
$\iso(d-1) \cong \left<L_{ij},L_{-i}\right>$. These 
constructions are summarised in~Table~\ref{tab:overview}.

We will at various points change basis. In the primed basis the
carrollian nature of $\zTi$, $\zSpi$ and $\zNi$ is more manifest, with
$B'$ denoting in all cases the (generalised) carrollian boosts and
$P'$ (generalised) carrollian translations. The explicit relations
between the primed and unprimed bases for each of the spaces are given in
Table~\ref{tab:overview}. The invariants can also be given more
uniformly, schematically, by a carrollian metric $\pi'^{2}$ and 
carrollian vector field(s) $H'$, and we refer to Table~\ref{tab:overview}
for an overview of these.

\begin{landscape}
\begin{table}
  \centering
  \caption{Overview of the $(d+1)$-dimensional homogeneous spaces of the
    Poincaré group that are covered in this work}
    \rowcolors{2}{blue!10}{white}
   \resizebox{\linewidth}{!}{
  \begin{tabular}{l | >{$}l<{$}  >{$}l<{$}  >{$}l<{$}  >{$}l<{$} }
    \toprule
  $(\g,\h)$                                        & \MM_{d+1}                                                       & \zTi_{d+1} = \zAdSC_{d+1}                                          & \zSpi_{d+1}                                                                                         & \zNi_{d+1}                                                                                          \\ 
  Embedding                                        & \eQ_\epsilon \cap \eN_{\sigma \neq 0}                           & \left(\eQ_{\epsilon < 0} \cap \eN_{0}\right)^+                                    & \eQ_{\epsilon > 0} \cap \eN_{0}                                                                     & \left(\eQ_{0} \cap \eN_{0}\right)^+                                                                              \\ \midrule
    $\g^{\mathrm{orig}} \cong \mathfrak{iso}(d,1)$ & \left< L_{\mu\nu}, P_\mu\right>_{\mu,\nu = 0,\ldots,d}          & \left< L_{ab}, B_a = L_{0a}, P_a, H= P_0\right>_{a,b = 1,\ldots,d} & \left< L_{\alpha\beta}, B_\alpha = L_{\alpha d}, P_\alpha, P_d\right>_{\alpha,\beta = 0,\ldots,d-1} & \left< L_{ij}, L_{+i}, L_{-i}, L_{+-}, P_i, P_+, P_-\right>_{i,j = 1, \ldots, d-1}                  \\
    $\h^{\mathrm{orig}}$                           & \left<L_{\mu\nu}\right>                                         & \left<L_{ab},P_a\right>                                            & \left<L_{\alpha\beta},P_\alpha\right>                                                               & \left<L_{ij}, L_{-i}, P_i, P_-\right>                                                               \\
    $\h^{\mathrm{orig}} \cong$                     & \mathfrak{so}(d,1)                                              & \mathfrak{iso}(d) \cong \so(d)\ltimes \RR^d                        & \mathfrak{iso}(d-1,1) \cong \so(d-1,1) \ltimes \RR^{d-1,1}                                          & \mathfrak{iso}(d-1)\ltimes \RR^d \cong \textrm{Carroll}(d)                                          \\
                                                   & \eqref{eq:poincare-1}                                           & \eqref{eq:pre-para-poincare}                                       & \eqref{eq:poincare-2}                                                                               & \eqref{eq:poincare-witt}                                                                            \\ \midrule
    Redef                                          &                                                                 & B'_a = P_a, P'_a = B_a, H'= - H                                    & B'_\alpha = P_\alpha, P'_\alpha = B_\alpha                                                          & L'_{i} = L_{-i}, B'_i = P_i, B'_- = P_- , P'_i = L_{+i}, P'_-= L_{+-}                               \\
    $\h$                                           &                                                                 & \left<L'_{ab},B'_a\right>                                          & \left<L'_{\alpha\beta},B'_\alpha\right>                                                             & \left< L'_{ij} , L'_{i} , B'_i , B'_- \right>                                                       \\
                                                   &                                                                 & \eqref{eq:para-poincare}                                           & \eqref{eq:pseudo-para-poincare}                                                                     & \eqref{eq:semi-para-poincare}                                                                       \\
    Invariants                                     & \eta_{\mu \nu} \pi^{\mu}\pi^{\nu}, \eta^{\mu\nu} P_{\mu}P_{\nu} & \delta_{ab}\pi^{'a}\pi^{'b}, H'                                    & \eta_{\alpha\beta}\pi^{'\alpha}\pi^{'\beta}, P^{'}_{d}                                              & \delta_{ij}\pi'^{i}\pi'^{j}, P'_\pm \mod \h                                                         \\
    Class                                          &
                                                     \textrm{lorentzian}                                             & \textrm{carrollian}                                                & \textrm{pseudo-carrollian}                                                                          & \textrm{doubly-carrollian}                                                                          \\
    Symmetries                                     & \mathfrak{iso}(d,1)                                             & \so(d,1) \ltimes C^\infty(\eH^d)                                   & \so(d,1) \ltimes C^\infty(\zdS_{d})                                                                 & \so(d,1) \ltimes C^\infty(\zCS^{d-1}) \cong BMS_{d+1}\quad (d\geq 3)                                \\
    \rowcolor{white!10}                             &                                                                 &                                                                    &                                                                                                     & \eX(\zCS^1) \ltimes C^\infty(\zCS^1) \cong BMS_{3} \quad (d=2)                                      \\
    \bottomrule
  \end{tabular}
  }
  \caption*{The four columns correspond to the Klein
    pairs $(\g,\h)$ of the homogeneous spaces of the Poincaré group
    that we cover in this work. In particular, 
    $\g \cong \mathfrak{iso}(d,1)$ in all cases, but the
    subalgebra $\h$ differs between the Klein pairs. Additionally, 
    we recall from Section~\ref{sec:embeddings} the embeddings of these spaces 
    into $\EE^{d+1,2}$ as intersections of the quadric $\eQ_{\epsilon}$ with 
    null planes of the form $\eN_{\sigma}$, and in the case of $\zTi$
    and $\zNi$ with one of the components as explained in
    Section~\ref{sec:embeddings}.

    In the first section of the Table we provide the
    decomposition of $\g$ and the subalgebra $\h$ in terms the common
    Poincaré basis and describe the abstract Lie algebra structure of
    $\h$.

    In the main text of this work we change basis to make their
    carrollian nature more manifest. In the second section of the
    Table we provide this change of basis, listing only those elements
    where the change of basis amounts to more than simply adding a
    dash to the symbol. These changes are such that the subalgebra is
    now spanned by (generalised) rotations $L'$ and (generalised)
    boosts $B'$. Additionally we list the invariants of low rank which
    characterise the class of the geometry. It is understood that
    $\pi'$ is dual to $P'$. Finally, we list the Lie subalgebra of
    vector fields which preserve the respective invariants.}
  \label{tab:overview}
\end{table}
\end{landscape}
\begin{table}[H]
  \centering
  \caption{Overview of the $d$-dimensional homogeneous spaces of
    the Poincaré group that descend from $\zTi_{d+1} = \zAdSC_{d+1}$
    and $\zSpi_{d+1}$}
    \rowcolors{2}{blue!10}{white}
  \begin{tabular}{l | >{$}l<{$}  >{$}l<{$}  >{$}l<{$}  >{$}l<{$} >{$}l<{$} >{$}l<{$} }
    \toprule
    $(\g,\h)$                  & \eH^{d}                                                            & \zdS_d                                                                                              \\ \midrule
    $\g^{\mathrm{orig}} \cong \mathfrak{iso}(d,1)$       & \left< L_{ab}, B_a = L_{0a}, P_a, H= P_0\right>_{a,b = 1,\ldots,d} & \left< L_{\alpha\beta}, B_\alpha = L_{\alpha d}, P_\alpha, P_d\right>_{\alpha,\beta = 0,\ldots,d-1} \\
    $\h^{\mathrm{orig}}$       & \left<L_{ab},P_a, H\right>                                         & \left<L_{\alpha\beta},P_\alpha, P_d\right>                                                          \\
    $\h^{\mathrm{orig}} \cong$ & \so(d)\ltimes \RR^{d,1}                                            & \so(d-1,1) \ltimes \RR^{d,1}                                                                        \\
                               & \eqref{eq:pre-para-poincare}                                       & \eqref{eq:poincare-2}                                                                               \\ 
    Effective                  & No                                                                 & No                                                                                                  \\ \midrule
    Redef                      & B'_a = P_a, P'_a = B_a, H'= - H                                    & B'_\alpha = P_\alpha, P'_\alpha = B_\alpha                                                          \\
    $\h$                       & \left<L'_{ab},B'_a, H'\right>                                      & \left<L'_{\alpha\beta},B'_\alpha, P'_d\right>                                                       \\
    Noneff $\to$ Eff           & \eqref{eq:para-poincare} \to \eqref{eq:asyTi}                      & \eqref{eq:pseudo-para-poincare} \to \eqref{eq:asySpi}                                               \\
    Invariants                 & \delta_{ab}\pi^{'a}\pi^{'b}, \delta^{ab} P'_{a} P'_{b}             & \eta_{\alpha\beta}\pi^{'\alpha}\pi^{'\beta}, \delta^{\alpha\beta} P'_{\alpha} P'_{\beta}            \\
    Class                      & \textrm{riemannian}                                                & \textrm{lorentzian}                                                                                 \\
    Symmetries                 & \so(d,1)                                                           & \so(d,1)                                                                                            \\ \bottomrule
  \end{tabular}
  \caption*{\small This table summarises the $(d-1)$-dimensional homogeneous
    spaces of the Poincaré group that descend from
    $\zTi_{d+1} = \zAdSC_{d+1}$ and $\zSpi_{d+1}$, given by $\eH^{d}$
    and $\zdS_{d}$, respectively. Neither is effective and
    their invariants of low rank are given by nondegenerate
    metrics, i.e., they are (pseudo-)lorentzian and share the same
    symmetry algebra $\so(d,1)$. In the ``Noneff $\to$ Eff'' column we
    link the noneffective and effective Lie pairs.}
  \label{tab:overview-HdS}
\end{table}
\begin{table}[H]
  \centering
  \caption{Overview of $d$ and $(d-1)$-dimensional homogeneous spaces
    of the Poincaré group that descend from $\zNi_{d+1}$}
    \rowcolors{2}{blue!10}{white}
   \resizebox{\linewidth}{!}{
 \begin{tabular}{l | >{$}l<{$}  >{$}l<{$}  >{$}l<{$}  >{$}l<{$} >{$}l<{$} >{$}l<{$} }
    \toprule
    $(\g,\h)$                                     & \scri_{d}                                                            & \zLC_{d}                                                    & \zCS^{d-1}                                                  \\ \midrule
    $\g^{\mathrm{orig}}\cong \mathfrak{iso}(d,1)$ & \left< L_{ij}, L_{+i}, L_{-i}, L_{+-}, P_i, P_+, P_-\right>          & \left< L_{ij}, L_{+i}, L_{-i}, L_{+-}, P_i, P_+, P_-\right> & \left< L_{ij}, L_{+i}, L_{-i}, L_{+-}, P_i, P_+, P_-\right> \\
    $\h^{\mathrm{orig}}$                          & \left<L_{ij}, L_{-i}, P_i, P_-, L_{+-}\right>                        & \left<L_{ij}, L_{-i}, P_i, P_-, P_+ \right>                 & \left<L_{ij}, L_{-i}, P_i, P_-, P_+, L_{+-}\right>          \\
    $\h^{\mathrm{orig}} \cong$                    & (\mathfrak{iso}(d-1)\ltimes \RR^d ) \rtimes \RR                      & \mathfrak{iso}(d-1)\ltimes \RR^{d,1}                        & (\mathfrak{iso}(d-1)\ltimes \RR^{d,1}) \rtimes \RR          \\
                                                  & \eqref{eq:poincare-witt}                                             & \eqref{eq:poincare-witt}                                    & \eqref{eq:poincare-witt}                                    \\ 
    Effective                                     & Yes                                                                  & No                                                          & No                                                          \\ \midrule
    Redef                                         & \multicolumn{3}{c}{$L'_{i} = L_{-i},  \,\, B'_i = P_i, \,\, B'_- = P_- , \,\, P'_i = L_{+i}, \,\, P'_-= L_{+-}, \,\, P'_+=P_+$}                                                                  \\
    $\h$                                          & \left< L'_{ij} , L'_{i} , B'_i , B'_-, P'_- \right>                  & \left< L'_{ij} , L'_{i} , B'_i , B'_-, P'_+ \right>         & \left< L'_{ij} , L'_{i} , B'_i , B'_-, P'_-, P'_+ \right>   \\
    Noneff $\to$ Eff                              & \eqref{eq:semi-para-poincare}                                        & \eqref{eq:semi-para-poincare} \to \eqref{eq:LCeff}          & \eqref{eq:semi-para-poincare}  \to \eqref{eq:CS}            \\
    Invariants                                    & \delta_{ij}\pi'^{i}\pi'^{j}, P'_+ \mod \h                            & \delta_{ij}\pi'^{i}\pi'^{j}, P'_- \mod \h                   & \delta_{ij}\pi'^{i}\pi'^{j}, \delta^{ij} P'_i P'_j \mod \h  \\
\rowcolor{white!10}                               & \textrm{(up to
                                                    scale)}
                                                                                                                         &                                                             & \textrm{(up to scale)}                                        \\
\rowcolor{blue!10}    Class                       & \textrm{conformal carrollian}                                        & \textrm{carrollian}                                         & \textrm{conformal riemannian}                               \\
\rowcolor{white!10}    (Conformal)                & \so(d,1) \ltimes C^\infty(\zCS^{d-1}) \cong BMS_{d+1}\quad (d\geq 3) & \so(d,1) \quad (d\geq 3)                                    & \so(d,1) \quad (d\geq 3)                                    \\
\rowcolor{white!10}  symmetries                   & \eX(\zCS^1) \ltimes C^\infty(\zCS^1) \cong BMS_{3} \quad (d=2)       & \eX(\zCS^1) \quad (d=2)                                     & \eX(\zCS^1) \quad (d=2)                                     \\ \bottomrule
  \end{tabular}
}
\caption*{\small This table summarises the $d$ and $(d-1)$-dimensional
  homogeneous spaces of the Poincaré group that descend from
  $\zNi_{d+1}$, where $i,j = 1, \ldots, d-2$. Notably $\scri$ is
  effective, while the lightcone $\zLC$ and the celestial sphere
  $\zCS$ are not. The lightcone is the only case in possession of
  low-rank invariants. For the other two cases, the dilatation-like
  action of $L_{+-}=P'_{-}$ only allows for invariants up to scale:
  conformal carrollian and conformal riemannian, respectively.  In the
  last row we provide for $\zLC$ the symmetries of these invariants,
  and for the other two cases the conformal symmetries of their
  respective conformal invariants. In the ``Noneff $\to$ Eff'' column
  we link from the noneffective to the effective Lie pair or just link
  to the effective one.}
  \label{tab:overview-scriLCCS}
\end{table}

\subsection{Minkowski spacetime $\MM$}
\label{sec:minkowski-spacetime}

The Klein pair $(\g,\h_{\MM})$ for Minkowski spacetime has
\begin{align}
  \h_{\MM} = \langle L_{\mu\nu} \rangle,
\end{align}
which is a Lorentz subalgebra and hence spanned by the rotations and
Lorentz boosts.  We complete this basis for $\h_{\MM}$ to a basis for
$\g$ by the addition of translations $P_\mu$.  The Poincaré Lie algebra
in this basis takes the standard form given in
equation~\eqref{eq:poincare-1}.  We observe that the split
$\g = \h_{\MM} \oplus \m_{\MM}$, where $\m_{\MM}$ is the span of the $P_\mu$, is both
reductive $([\h_{\MM},\m_{\MM}] \subset \m_{\MM})$ and symmetric
$([\m_{\MM},\m_{\MM}] \subset \h_{\MM})$. The Poincaré-invariant
tensor fields correspond to the Lorentz-invariant tensors of the
linear isotropy representation $\m_{\MM}$: namely, $\eta_{\mu \nu}
\pi^{\mu}\pi^{\nu}$, corresponding to the Minkowski metric, and
$\eta^{\mu\nu} P_{\mu}P_{\nu}$, corresponding to its inverse. Here and in
what follows, $\pi^\mu$ is the basis of $\m_{\MM}^*$
canonically dual to $P_\mu$; that is,
$\langle \pi^{\mu}, P_{\nu}\rangle = \delta^{\mu}_{\nu}$.

The vector fields that preserve the invariant structure $\eta_{\mu \nu}
\pi^{\mu}\pi^{\nu}$ clearly generate the symmetry algebra of Minkowski
spacetime, i.e., the Poincaré algebra.

\subsection{Spatial infinity $\zSpi$}
\label{sec:spi}

As already discussed in Section~\ref{sec:embedding-spi}, the Klein
pair for $\zSpi$ is $(\g,\h_{\zSpi})$ where
\begin{equation}
  \h_{\zSpi} = \left<L_{ij}, B_i, P_i, H\right>,
\end{equation}
where $i=1,\dots,d-1$. In terms of the semidirect product
decomposition~\eqref{eq:uniP}, the subalgebra $\h_{\zSpi}$ is comprised
of the stabiliser of the spacelike momentum $P = P_d$, which is $\stab(P)
\cong \so(d-1,1)$, and the perpendicular translations $P^\perp = \left<
P_\alpha \right>\cong\RR^{d-1,1}$, where $\alpha,\beta,\dots =
0,1,\dots,d-1$. It is convenient to restore some of the symmetry by
breaking the manifest Lorentz symmetry in
\eqref{eq:poincare-1} via the basis $L_{\alpha\beta}$, $P_\alpha$,
$B_\alpha := L_{\alpha d}$ and $P_d$ with (nonzero) Lie brackets
\begin{equation}
  \label{eq:poincare-2}
  \begin{split}
    [L_{\alpha\beta},L_{\gamma\delta}] &= \eta_{\beta\gamma} L_{\alpha\delta} - \eta_{\alpha\gamma} L_{\beta\delta}- \eta_{\beta\delta} L_{\alpha\gamma} + \eta_{\alpha\delta} L_{\beta\gamma}\\
    [L_{\alpha\beta},B_\gamma] &= \eta_{\beta\gamma} B_\alpha - \eta_{\alpha\gamma} B_\beta\\
    [L_{\alpha\beta},P_\gamma] &= \eta_{\beta\gamma} P_\alpha - \eta_{\alpha\gamma} P_\beta\\
    [B_\alpha, B_\beta] &= - L_{\alpha\beta}\\
    [B_\alpha,P_\beta] &= - \eta_{\alpha\beta} P_d\\
    [B_\alpha, P_d] &= P_\alpha,
  \end{split}
\end{equation}
where $\eta_{\alpha\beta}$ is the $d$-dimensional lorentzian inner
product with mostly plus signature.  In this notation, the Klein pair
$(\g,\h_{\zSpi})$ is such that
\begin{equation}
  \h_{\zSpi} = \left<L_{\alpha\beta}, P_\alpha\right>.
\end{equation}
It is convenient to relabel $L'_{\alpha\beta} = L_{\alpha\beta}$,
$B'_\alpha = P_\alpha$, $P'_\alpha = B_\alpha$ and $P'_d = P_d$ to
arrive at
\begin{equation}
  \label{eq:pseudo-para-poincare}
  \begin{split}
    [L'_{\alpha\beta},L'_{\gamma\delta}] &= \eta_{\beta\gamma} L'_{\alpha\delta} - \eta_{\alpha\gamma} L'_{\beta\delta}- \eta_{\beta\delta} L'_{\alpha\gamma} + \eta_{\alpha\delta} L'_{\beta\gamma}\\
    [L'_{\alpha\beta},B'_\gamma] &= \eta_{\beta\gamma} B'_\alpha - \eta_{\alpha\gamma} B'_\beta\\
    [L'_{\alpha\beta},P'_\gamma] &= \eta_{\beta\gamma} P'_\alpha - \eta_{\alpha\gamma} P'_\beta\\
    [B'_\alpha,P'_\beta] &= \eta_{\alpha\beta} P'_d\\
    [P'_\alpha, P'_d] &= B'_\alpha\\
    [P'_\alpha, P'_\beta] &= - L'_{\alpha\beta}.
  \end{split}
\end{equation}
As in the case of Minkowski spacetime, the split $\g = \h_{\zSpi}
\oplus \m_{\zSpi}$, where 
\begin{align}
  \label{eq:spih}
  \h_{\zSpi}=\left<L'_{\alpha\beta},B'_\alpha\right> 
\end{align}
and $\m_{\zSpi}$ is now the span of $P'_\alpha$ and $P'_d$, is both
reductive and symmetric.

The computation of the low-rank invariants is formally identical to
those of $\zTi$ ($\zAdSC$ in \cite{Figueroa-OFarrill:2018ilb}),
changing $\delta \mapsto \eta$.  They result in a pseudo-carrollian
structure consisting of a nowhere-vanishing vector field corresponding
to $P'_d$ and a degenerate lorentzian metric corresponding to
$\eta_{\alpha\beta}\pi'^\alpha\pi'^\beta$, allowing us to conclude
that $\zSpi$ is a pseudo-carrollian symmetric
space~\cite{Gibbons:2019zfs}.

In parallel to the symmetries of carrollian structure of $\zTi$, which
were worked out in \cite{Figueroa-OFarrill:2019sex}, one
finds that the vector fields preserving the pseudo-carrollian
structure of $\zSpi$ generate the algebra
$\so(d,1) \ltimes C^\infty(\zdS_{d})$, i.e., a semi-direct product of
the Lorentz group, the Killing vectors of $\zdS_{d}$ and
Spi-supertranslations.   As in the case of $\zAdSC$, the
action of the Killing vectors of $\zdS_d$ on $C^\infty(\zdS_{d})$ is
not as functions, but as sections on the density line bundle.  These
results essentially go back to \cite{Ashtekar:1978zz}.

We can uncover the asymptotic geometry of $\zSpi$. We add $P'_{d}$ to
$\h_{\zSpi}$ to obtain
\begin{equation}
  \label{eq:dSh}
  \h_{\zdS} = \left<L_{\alpha\beta},P_\alpha, P_{d}\right> =
  \left<L'_{\alpha\beta},B'_\alpha, P'_{d}\right>.
\end{equation}
We observe that the Klein pair $(\g,\h_{\zdS})$ is not
effective because $\k = \langle P_{\mu} \rangle =\langle B'_{\alpha}, P'_{d} \rangle$ is an
ideal of $\g$ contained in $\h_{\zdS}$.   Quotienting by $\k$, we
arrive at an effective Klein pair $(\g/ \k,\h / \k)$.  The quotient
Lie algebra $\g/\k$ is spanned by the image of
$L'_{\alpha\beta},P'_\alpha$ in $\g/\k$.  If we again use the same
notation for their images, we obtain
\begin{equation}
  \label{eq:asySpi}
  \begin{split}
    [L'_{\alpha\beta},L'_{\gamma\delta}] &= \eta_{\beta\gamma} L'_{\alpha\delta} - \eta_{\alpha\gamma} L'_{\beta\delta}- \eta_{\beta\delta} L'_{\alpha\gamma} + \eta_{\alpha\delta} L'_{\beta\gamma}\\
    [L'_{\alpha\beta},P'_\gamma] &= \eta_{\beta\gamma} P'_\alpha - \eta_{\alpha\gamma} P'_\beta\\
    [P'_\alpha, P'_\beta] &= - L'_{\alpha\beta},
  \end{split}
\end{equation}
from where we see that
$(\g/\k, \h_{\zdS}/\k) \cong (\so(d,1),\so(d-1,1))$ is the Klein pair
of de~Sitter spacetime $\zdS_d$. In terms of the original basis,
$L'_{\alpha\beta}=L_{\alpha\beta}$ and $P'_{\alpha} = L_{\alpha d}$,
so that the Minkowski translations act trivially on the asymptotic
geometry.

\subsection{Timelike infinity $\zTi \cong \zAdSC$}
\label{sec:ads-carroll}

In Section~\ref{sec:embedding-ti}, we determined that the Klein pair
$(\g,\h_{\zTi})$ is such that
\begin{equation}
  \label{eq:h-ti}
  \h_{\zTi} = \left<L_{ab},P_a\right>.
\end{equation}
This subalgebra consists of the stabiliser of the timelike momentum
$P = P_0$, which is isomorphic to $\so(d)$, and the perpendicular momenta
$P^\perp = \left< P_a \right>\cong \RR^d$, where the indices $a,b,\dots
=1,\dots,d$ run over the spatial directions. To emphasise that $P_a$
should actually be understood as Carroll boosts we relabel the basis so
that the Carroll boosts are $B'_a = P_a$.  We relabel the rest of the
generators as $L'_{ab} = L_{ab}$, $P'_a = B_a$ and $H'= - H$.  In the new
basis, the nonzero Lie brackets then become:
\begin{equation}\label{eq:para-poincare}
  \begin{split}
    [L'_{ab},L'_{cd}] &= \delta_{bc} L'_{ad} - \delta_{ac} L'_{bd} - \delta_{bd} L'_{ac} + \delta_{ad} L'_{bc}\\
    [L'_{ab},B'_c] &= \delta_{bc} B'_a - \delta_{ac}B'_b\\
    [L'_{ab},P'_c] &= \delta_{bc} P'_a - \delta_{ac}P'_b\\
    [B'_a, P'_b] &= \delta_{ab} H'\\
    [H',P'_a] &= B'_a\\
    [P'_a, P'_b] &= L'_{ab}.
  \end{split}
\end{equation}
This means that the isotropy subalgebra is now
\begin{align}
  \h_{\zTi} = \left<L'_{ab},B'_a\right>
\end{align}
and $\m_{\zTi}$ is the span of $P'_a,H'$. We see that the split
$\g = \h_{\zTi} \oplus \m_{\zTi}$ is again both reductive and
symmetric, consistent with the fact that $\zTi$ is a carrollian
symmetric space.

One can now calculate the Poincaré-invariant tensor fields of this
spacetime.  This has been done in, e.g.,
\cite{Figueroa-OFarrill:2018ilb}, where it was shown that they are
given by a nowhere-vanishing vector field corresponding to $H' \in \m_{\zTi}$
and a degenerate metric corresponding to the $\h_{\zTi}$-invariant
bilinear form $\delta_{ab}\pi'^a\pi'^b \in \odot^2 \m_{\zTi}^*$, where
$\langle \pi'^a,P'_{b}\rangle = \delta^{a}_{b}$.

As in the case of $\zSpi$, the symmetry algebra of $\zTi$ is determined
by the algebra of vector fields preserving the above invariant structure.
In~\cite{Figueroa-OFarrill:2019sex} the symmetry algebra was determined to
be the infinite-dimensional algebra $\so(d,1) \ltimes
C^\infty(\eH^{d})$, where $C^\infty(\eH^{d})$ are the smooth functions on
$d$-dimensional hyperbolic space, although the action of $\so(d,1)$ on
$C^\infty(\eH^{d})$ is that of Killing vectors on $\eH^{d}$ not
on functions but on sections of the density line bundle.  The symmetry
algebra consists of Lorentz transformations and
$\zTi$-supertranslations.

The decomposition above suggests another Klein pair where one adds $H'$
to $\h_{\zTi}$ leading to
\begin{align}
\label{eq:stabiliser-H}
  \h_{\eH} = \left<L_{ab},P_a,H\right>  = \left<L'_{ab},B'_a,H'\right>  .
\end{align}
We observe that the Klein pair $(\g,\h_{\eH})$ is not effective, as
$\k=\langle P_\mu\rangle=\langle B'_{a}, H '\rangle$ is an ideal of
$\g$ contained in $\h_{\eH}$.  We may quotient by $\k$ to arrive at an
effective Klein pair $(\g/ \k,\h_{\eH} / \k)$, where the quotient Lie
algebra $\g/\k$ is spanned by the images of $L'_{ab},P'_a$.  Letting
them stand for their images in $\g/\k$, the brackets of $\g/\k$ are
given by
\begin{equation}\label{eq:asyTi}
  \begin{split}
    [L'_{ab},L'_{cd}] &= \delta_{bc} L'_{ad} - \delta_{ac} L'_{bd} - \delta_{bd} L'_{ac} + \delta_{ad} L'_{bc}\\
    [L'_{ab},P'_c] &= \delta_{bc} P'_a - \delta_{ac}P'_b\\
    [P'_a, P'_b] &= L'_{ab} .
  \end{split}
\end{equation}
The Klein pair $(\g/ \k,\h_{\eH} / \k) \cong (\so(d,1),\so(d))$ is the
infinitesimal description of $d$-dimensional hyperbolic space
$\eH^{d}$, whose invariant riemannian metric can then be understood as
the asymptotic geometry at $\zTi$ (in analogy to the one of $\zSpi$ as
described, e.g., in~\cite{Geroch:1977jn}).  The Killing symmetries of
this metric are again $\so(d,1)$ which is isomorphic to a Lorentz
subalgebra of the original Poincaré algebra.  This algebra is spanned
by the Minkowski rotations $L_{ab}$ and Minkowski boosts $B_{a}$, so
that the translations of Minkowski spacetime (corresponding to the ideal
$\k$) act trivially on the asymptotic geometry of $\zTi$.

\subsection{The doubly-carrollian manifold $\zNi$}
\label{sec:null}

In Section~\ref{sec:embedding-ni}, after choosing a suitable origin
for $\zNi$, we determined the stabiliser subgroup.  The corresponding
Lie pair is $(\g,\h_{\zNi})$, where
\begin{equation}
  \h_{\zNi} = \left< L_{ij},P_i, L_{id} + B_i,
  P_d + H\right>,
\end{equation}
for $i,j = 1,\dots,d-1$.  It is convenient to change basis to $P_i$
and $P_\pm := \frac{1}{\sqrt{2}}(P_d \pm P_0)$, and where $i=1,\ldots
,d-1$ now, with $\eta_{ij}= \delta_{ij}$ and $\eta_{+-}= \eta_{-+} =
1$ with other components zero. The Lorentz generators break up as
$L_{ij}$, $L_{+i}$, $L_{-i}$ and $L_{+-}$.  In this basis,
\begin{equation}
  \h_{\zNi} = \left<L_{ij}, P_i, L_{-i}, P_- \right>,
\end{equation}
and the (nonzero) Lie brackets~\eqref{eq:poincare-1} of the Poincaré Lie algebra are given by
\begin{equation}
  \label{eq:poincare-witt}
  \begin{aligned}\relax
   [L_{ij},L_{k\ell}] &= \delta_{jk} L_{i\ell} - \delta_{ik}L_{j\ell} - \delta_{j\ell} L_{ik} + \delta_{i\ell} L_{jk} \\
    [L_{ij},L_{\pm k}] &= \delta_{jk} L_{\pm i} - \delta_{ik} L_{\pm j}\\
    [L_{+i},L_{-j}] &= - L_{ij} - \delta_{ij} L_{+-}\\
    [L_{+-},L_{\pm i}] &= \pm L_{\pm i}\\
  \end{aligned}
  \qquad\text{and}\qquad
  \begin{aligned}\relax
    [L_{ij}, P_k] &= \delta_{jk} P_i - \delta_{ik} P_j\\
    [L_{\pm i}, P_j] &= \delta_{ij} P_\pm\\
    [L_{\pm i}, P_{\mp}] &= - P_i\\
    [L_{+-}, P_\pm] &= \pm P_\pm.
  \end{aligned}
\end{equation}
The subalgebra $\h_{\zNi}$ is a semidirect product (cf.,~\eqref{eq:uniP}) of the stabiliser of $P = P_-$, which is isomorphic to $\iso(d-1)$, and the perpendicular translations $P^\perp = \left< P_-,P_i \right>\cong \RR^d$.

Let us already make some observations.  Firstly, in contrast to $\MM$,
$\zTi$ and $\zSpi$, we keep only manifest symmetry under
$\mathfrak{so}(d-1)$ rather than the larger $\mathfrak{so}(d)$ or
$\mathfrak{so}(d-1,1)$.  This means that $\zNi$ has three
$\mathfrak{so}(d-1)$ vectors ($L_{\pm i}, P_{i}$) and three
$\mathfrak{so}(d-1)$ scalars ($L_{+-}, P_{\pm}$) rather than two
vectors and one scalar, as is the case in $\zTi$ and $\zSpi$.

Secondly, the $\mathfrak{so}(d,1)$ Lorentz subalgebra spanned by
$\langle L_{ij},L_{\pm i}, L_{+-} \rangle$, i.e., the first four
brackets of~\eqref{eq:poincare-witt}, are written in such
a way that the relation to euclidean conformal field theory in $d-1$
dimensions is manifest.  Indeed, both have $\mathfrak{so}(d-1)$
symmetries, however acting differently on the underlying manifold. The
dilatations are given by $L_{+-}$, and the translations and special
conformal transformations are given by $L_{\pm i}$. This observation
is one of the motivations to study flat space holography from the
point of view of ($d-1$)-dimensional celestial conformal field theory,
see, e.g.,~\cite{Strominger:2017zoo, Raclariu:2021zjz} for a review.

The Klein pair $(\g,\h_{\zNi})$ is not reductive.  In other words, we
cannot find a complement $\m_{\zNi}$ to $\h_{\zNi}$ in $\g$ for which
$[\h,\m_{\zNi}] \subset \m_{\zNi}$.   This means that we must make a
choice.  We choose $\m_{\zNi}$ to be the subspace of $\g$ spanned by
$P_+, L_{+i}, L_{+-}$.  Let us relabel the basis so that $P'_+ = P_+$,
$P'_-= L_{+-}$ and $P'_i = L_{+i}$, $L'_{ij}= L_{ij}$, $L'_{i} =
L_{-i}$, $B'_i = P_i$ and $B'_- = P_-$, in terms of which the
(nonzero) Poincaré Lie brackets~\eqref{eq:poincare-witt} are given by
 \begin{equation}
   \label{eq:semi-para-poincare}
   \begin{aligned}\relax
     [L'_{ij},L'_{k\ell}] &= \delta_{jk} L'_{i\ell} - \delta_{ik}    L'_{j\ell} - \delta_{j\ell} L'_{ik} + \delta_{i\ell} L'_{jk} \\
     [L'_{ij},L'_{k}] &= \delta_{jk} L'_{i} - \delta_{ik} L'_{j}\\    
     [L'_{ij},B'_k] &= \delta_{jk} B'_i - \delta_{ik} B'_j\\
     [L'_{i},B'_j] &= \delta_{ij} B'_-\\
     [L'_{ij},P'_k] &= \delta_{jk} P'_i - \delta_{ik} P'_j\\
     [L'_{i},P'_j] &= -L'_{ij} + \delta_{ij} P'_-\\
   \end{aligned}
   \qquad\text{and}\qquad
    \begin{aligned}\relax
     [B'_i,P'_j] &= -\delta_{ij} P'_+\\
     [B'_-,P'_i] &= B'_i\\
     [L'_{i},P'_-] &= L'_{i}\\
     [B'_-,P'_-] &= B'_-\\
     [L'_{i},P'_+] &= -B'_i\\
     [P'_-, P'_i] &= P'_i\\
     [P'_-, P'_+] &= P'_+.
   \end{aligned}
\end{equation}
In this basis, we define
\begin{equation}
  \label{eq:Nullh}
  \h_{\zNi}= \left<L_{ij}, L_{-i}, P_i, P_-\right> = \left< L'_{ij}, L'_{i}, B'_{i}, B'_{-} \right> \, .
\end{equation}
The subalgebra $\h_{\zNi}$ can be seen to be isomorphic to the
$d$-dimensional Carroll algebra, when $L'_{i}$ are interpreted as
spatial translations and $B'_{-}$ as time translations. This has a
simple geometric explanation: the orbits in Minkowski spacetime of the
subgroup of the Poincaré group generated by $\h_{\zNi}$ are null
hyperplanes, which as shown, e.g., in~\cite[Section
4.2.5.]{Figueroa-OFarrill:2018ilb}, are homogeneous spacetimes of the
Carroll group; namely, copies of the $d$-dimensional carrollian
spacetime $\zC_d$.

Since the Klein pair $(\g,\h_{\zNi})$ is non-reductive, the linear
isotropy representation is the quotient $\g/\h_{\zNi}$ and its dual is
the annihilator of $\h_{\zNi}$ in $\g^*$. This space is spanned by
$\left<\pi'^i,\pi'^+,\pi'^- \right>$ and the action of $\h_{\zNi}$ is
given by \begin{align} \begin{split}
  L'_{ij} \cdot \pi'^{k} &= \delta^{k}_{j} \delta_{im} \pi'^{m} - \delta^{k}_{i} \delta_{jm} \pi'^{m} \\
  L'_{i} \cdot \pi'^{-} &= - \delta_{ij} \pi'^{j}  \\
  B'_{i} \cdot \pi'^{+} &= \delta_{ij} \pi'^{j}.
\end{split}
\end{align}
We illustrate how these are obtained for the last of the expressions
above.  The action of $\h_{\zNi}$ on its annihilator is the
restriction of the coadjoint action of $\g$.  This is a linear map $\g
\to \gl(\g^*)$, which, when restricted to $\h_{\zNi}$, leaves invariant
the annihilator of $\h_{\zNi}$.  This action is defined as follows:
if $\alpha \in \g^*$ and $X \in \h_{\zNi}$, $X \cdot \alpha = -
\alpha \circ \ad_X$.  Applying this to $X=B_i'$ and $\alpha =
\pi'^{+}$, we find that for $X \in \g$, $B_i'\cdot \pi'^{+}(X) = -
\braket{\pi'^{+},[B_i',X]}$, where $\langle -,- \rangle$ denotes the
dual pairing.  The only contributing bracket is $[B'_i,P'_j] =
-\delta_{ij} P'_+$, which gives rise to $B_i'\cdot \pi'^{+}(P'_j) =
\delta_{ij}$, allowing us to conclude that $B'_{i} \cdot \pi'^{+} =
\delta_{ij} \pi'^{j}$.

It follows that the Poincaré-invariant tensor fields up to second rank
on $\zNi$ are generated by two nowhere-vanishing vector fields
corresponding to the $\h_{\zNi}$-invariant vectors $\overline{P'}_\pm
\in \g/\h_{\zNi}$ and the $\h_{\zNi}$-invariant symmetric bilinear
form $\delta_{ij} \pi'^i \pi'^j$.  In addition, let us highlight the
existence of the invariant volume form
\begin{align}
  \epsilon_{a_{1} \cdots a_{d-1}} \pi'^{a_{1}} \cdots \pi'^{a_{d-1}} \, .
\end{align}

An invariant structure of the above type, i.e., two nowhere-vanishing
vector fields together with a doubly degenerate metric, differs from
previously known structures such as carrollian, galilean, aristotelian
or their stringy versions. We will refer to such a structure
tentatively as \emph{doubly-carrollian}.  The justification is that
carrollian structures arise naturally in the bundle of scales of a
conformal manifold and this structure arises naturally in the bundle
of scales of a conformal carrollian manifold.

The Lie algebra of symmetries of the Poincaré-invariant tensors in
$\zNi$ is calculated in Appendix~\ref{sec:symmetries-znul}.
Summarising the results, we find that for $d\geq 3$, the symmetry
algebra is $\so(d,1) \ltimes C^\infty(\zCS^{d-1})$, whereas for $d=2$
it is isomorphic to $\eX(\zCS^1) \times C^\infty(\zCS^1)$, where
$\eX(\zCS^1)$ is the Lie algebra of smooth vector fields on the
(celestial) circle. In all cases, the abelian ideals of smooth
functions on $\zCS^{d-1}$ and the action of the Lie algebra of
conformal Killing vectors ($\so(d,1)$ or $\eX(\zCS^1)$) on them suggest
that they should be interpreted as sections of the density line
bundle, as explained in \cite[§10]{Figueroa-OFarrill:2019sex}.

These symmetry algebras are precisely $\textrm{BMS}_{d+1}$ which
suggests that $\zNi$ is closely related to null infinity. We will see
in the next section that this is indeed the case.

\subsection{$\scri$ and $\zLC$: two ways to the celestial sphere $\zCS$}
\label{sec:two-ways-celestial}

In Sections~\ref{sec:spi} and \ref{sec:ads-carroll}, we saw that the
respective Klein pairs $(\g,\h)$ of $\zSpi$ and $\zTi$ allowed for
enlargements of the subalgebra $\mathfrak{h}$ that led to the
naturally related lower-dimensional homogeneous
spaces~\eqref{eq:asyTi} and~\eqref{eq:asySpi}. Similarly, we find in
the following that there are several lower-dimensional Klein
geometries $(\g,\h)$ which can be obtained from $\zNi$ by enlarging
the stabiliser subalgebra $\h_{\zNi} \subset \h$ while keeping $\g$
fixed as the Poincaré algebra. Geometrically this corresponds to
viewing $\zNi$ as the total space of a principal bundle over some
lower-dimensional homogeneous spaces.

\subsubsection{Null infinity $\scri$}

Still working in the basis~\eqref{eq:semi-para-poincare} for the
Poincaré algebra, we may add $P'_-$ to $\h_{\zNi}$ to arrive at
\begin{equation}
  \label{eq:ANHh}
 \h_{\scri} = \left<L_{ij}, L_{-i}, P_i, P_-, L_{+-}\right> = \langle L'_{ij}, L'_{i}, B'_{i}, B'_{-}, P'_{-} \rangle.
\end{equation}
Since $\h_\scri$ does not contain a nonzero ideal of $\g$, 
the resulting Klein pair $(\g,\h_{\scri})$ is effective.  The
simply-connected homogeneous space based on this Klein pair
\begin{equation}
  \label{eq:scrihom}
  \mathrm{ISO(d,1)_0}/\left((\mathrm{ISO}(d-1)\ltimes \mathbb{R}^{d})\rtimes \mathbb{R}\right)
\end{equation}
can be identified with $\scri$, i.e., the null boundary of
$(d+1)$-dimensional Minkowski spacetime as discussed, e.g., in
\cite{Herfray:2020rvq}. As we will see in
Section~\ref{sec:nul-geometry} we can also view it as the grassmannian
of affine null hyperplanes in Minkowski spacetime.

The dual of the linear isotropy representation of $\h_{\scri}$
is given relative to the basis $\langle \pi'^{i}, \pi'^{+} \rangle$ by
\begin{align}
\begin{split}
  L'_{ij} \cdot \pi'^{k} &= \delta^{k}_{j} \delta_{im} \pi'^{m} - \delta^{k}_{i} \delta_{jm} \pi'^{m} \\
  B'_{i} \cdot \pi'^{+} &= \delta_{ij} \pi'^{j} \\
  P'_{-} \cdot \pi'^{i} &= - \pi'^{ i} \\
  P'_{-} \cdot \pi'^{+} &= -\pi'^{+}.
\end{split}
\end{align}
Since $P'_{-}$ is now in $\h_{\scri}$ and it acts like a dilatation,
there are no invariant tensors of low rank. However, it is natural
to look for invariant conformal classes of tensors. They turn out to
be a vector field $P'_{+}$ with conformal weight $-1$ and a degenerate
metric $\pi^{2} = \delta_{ij}\pi^{i}\pi^{j}$ with conformal weight
$2$. This space thus admits a conformal carrollian structure. The
symmetries of this structure give rise to the BMS group in $(d+1)$
dimensions~\cite{Geroch:1977jn,Duval:2014uva}
\begin{align}
\begin{split}
  \so(d,1) \ltimes C^\infty(\zCS^{d-1}) &\cong BMS_{d+1}\quad (d\geq 3)\\
  \eX(\zCS^1) \ltimes C^\infty(\zCS^1) &\cong BMS_{2}
\end{split}
\end{align}
as is expected from the identification of this homogeneous space with $\scri$.

\subsubsection{Lightcone $\zLC$}

Starting again from the Klein pair of $\zNi$, we may alternatively
add $P'_+$ to $\h_{\zNi}$ to arrive at
\begin{equation}
  \label{eq:LCh}
 \h_{\zLC} = \left<L_{ij}, L_{-i}, P_i, P_-, P_{+}\right> = \langle L'_{ij}, L'_{i}, B'_{i}, B'_{-}, P'_{+} \rangle .
\end{equation}
Since $\h_\zLC$ now contains the ideal $\k = \langle P_\mu\rangle = \langle B'_i,B'_-,P'_+ \rangle$,
the Klein pair is not effective as we may quotient both $\g$ and $\h_{\zLC}$ by this ideal. 
Therefore $(\g,\h_{\zLC})$ reduces to $(\g/\k, \h_{\zLC}/\k)$, where $\g/\k$ is
spanned by the residue classes modulo $\k$ of
$L'_{ij}, L'_i,P'_i,P'_-$ subject to the following Lie brackets, which
we obtain from those in $\g$ simply by dropping any terms in $\k$:
\begin{equation}
  \label{eq:LCeff}
  [L'_i,P'_j] = - L'_{ij} + \delta_{ij} P'_-, \qquad [L'_i,P'_-] =
  L'_i \qquad\text{and}\qquad [P'_-,P'_i] = P'_i,
\end{equation}
in addition to those brackets involving $L'_{ij}$ which simply say
that $L'_{ij}$ spans a subalgebra isomorphic to $\so(d-1)$ relative to
which $L'_i,P'_i$ are vectors and $P'_-$ is a scalar.  We recognise
this Lie algebra as the Lorentz algebra $\so(d,1)$ and $\h_{\zLC}/\k$ is
isomorphic to a euclidean algebra $\iso(d-1)$.  We thus recognise the
resulting Klein pair as the homogeneous space of the (proper,
orthochronous) Lorentz group describing the $d$-dimensional lightcone
\begin{equation}
  \label{eq:LChom}
\textrm{SO}(d,1)_0/\textrm{ISO}(d-1)
\end{equation}
in ($d+1$)-dimensional Minkowski
spacetime~\cite{Figueroa-OFarrill:2018ilb}.

The dual of the linear isotropy representation of $\h_{\zLC}/\k$ is
given in the basis $\langle \pi'^{i}, \pi'^{-} \rangle$ by
\begin{align}
\begin{split}
  L'_{ij} \cdot \pi'^{k} &= \delta^{k}_{j} \delta_{im} \pi'^{m} - \delta^{k}_{i} \delta_{jm} \pi'^{m} \\
  L'_{i} \cdot \pi'^{-} &= - \delta_{ij} \pi'^{j}.
\end{split}
\end{align}

We see that the invariant tensors on $\zLC$, as already discussed in
\cite{Figueroa-OFarrill:2018ilb}, correspond to the invariant
carrollian structure: a nowhere vanishing vector field corresponding
to the $\h_{\zLC}$-invariant vector $\overline{P'}_-$ (where the overbar 
denotes the image of $P'_-$ in $\g/\k$) and the
symmetric tensor $\delta_{ij} \pi'^i \pi'^j$. The algebra of vector
fields leaving this structure invariant was determined
in~\cite{Figueroa-OFarrill:2018ilb}. It yields the $(d+1)$-dimensional
Lorentz algebra $\mathfrak{so}(d,1)$.

\subsubsection{Celestial sphere $\zCS$}

Returning again to the Klein pair of $\zNi$, we find that the two
cases discussed above are the only $d$-dimensional homogeneous spaces
that can be obtained from an enlargement of the subalgebra
$\h_{\zNi}$. In contrast to $\zTi$ and $\zSpi$ one finds however,
that we can construct a $(d-2)$-dimensional homogeneous space by
adding both $P'_\pm$ to $\h_{\zNi}$. We arrive at
\begin{equation}
  \label{eq:CSh}
 \h_\zCS= \left<L_{ij}, L_{-i}, P_i, P_-, L_{+-}, P_{+}\right> = \langle L'_{ij}, L'_{i}, B'_{i}, B'_{-}, P'_{+}, P'_{-} \rangle   .
\end{equation}
Since both $\h_\scri$ and $\h_\zLC$ are contained in $\h_\zCS$, it contains once more the
ideal $\k$ spanned by $B'_i,B'_-,P'_+$, resulting in a non-effective Klein pair
$(\g,\h_\zCS)$.  The reduced Klein pair $(\g/\k,\h_\zCS/\k)$ is
effective, as can be seen can be seen from the following brackets:
\begin{equation}
   \label{eq:CS}
   \begin{split}
   [L'_{ij},L'_{k\ell}] &= \delta_{jk} L'_{i\ell} - \delta_{ik}    L'_{j\ell} - \delta_{j\ell} L'_{ik} + \delta_{i\ell} L'_{jk} \\
     [L'_{ij},L'_{k}] &= \delta_{jk} L'_{i} - \delta_{ik} L'_{j}\\    
     [L'_{ij},P'_k] &= \delta_{jk} P'_i - \delta_{ik} P'_j\\
     [L'_{i},P'_j] &= -L'_{ij} + \delta_{ij} P'_-\\
     [P'_-,L'_{i}] &= - L'_{i}\\
     [P'_-, P'_i] &= P'_i  .
   \end{split}
\end{equation}
The Lie algebra $\g/\k$ is again $\so(d,1)$, whereas now $\h_\zCS/\k$
is the parabolic subalgebra spanned by $L'_{ij}, L'_i,P'_-$. We
recognise the resulting Klein pair as that corresponding to the sphere
as a flat conformal geometry. In the present context, it is more
appropriate to call it the celestial sphere $\zCS^{d-1}$.  In summary,
the Lorentz group acts transitively on the celestial sphere via
conformal transformations.

The dual of the linear isotropy representation is given by
\begin{align}
\begin{split}
  L'_{ij} \cdot \pi'^{k} &= \delta^{k}_{j} \delta_{im} \pi'^{m} - \delta^{k}_{i} \delta_{jm} \pi'^{m} \\
  P'_{-} \cdot \pi'^{i} &= - \pi'^{i} .
\end{split}
\end{align}

There are again no invariant tensors due to the fact that $-P'_-$ acts
as a dilatation relative to which the linear isotropy representation
has weight $-1$.  Nevertheless, there is an invariant conformal class
of metric (with conformal weight $2$) associated to $\delta_{ij} \pi'^i
\pi'^j$ and its inverse (with conformal weight $-2$) $\delta^{ij}
\overline{P'}_i \overline{P'}_j$. As is well known, the vector fields
preserving this conformal structure, i.e., the conformal Killing vectors of
the $(d-1)$-sphere, generate the $(d+1)$-dimensional Lorentz algebra
$\mathfrak{ckv}(\zCS^{d-1})\cong \mathfrak{so}(d,1)$ for $d\ge 3$,
whereas for $d=2$ every vector field on the circle is conformal
Killing.

\subsection{Summary}

All $(d+1)$-dimensional homogeneous spaces, $\zTi,\zSpi,\zNi$, discussed in this section
allow for the construction of a related lower-dimensional homogeneous space
that is obtained by adding one scalar to the respective subalgebra $\h$. Interestingly,
the space $\zNi$ allows for a richer structure of lower-dimensional spaces
summarised in this diagram:
\begin{equation}\label{diag:relNul}
  \begin{tikzcd}
\zNi_{d+1} \arrow[dr] \arrow[r] & \zLC_{d}  \arrow[dr] &            & \\
  ?_{d+1}                        & \scri_{d} \arrow[r]  & \zCS^{d-1} & \\
  \end{tikzcd}
\end{equation}
where horizontal/diagonal arrows represent additions of elements of
$P_{\mu}$ and $L_{\mu\nu}$, respectively, to the subalgebra
$\mathfrak{h}$. One notices that $\scri$ constructed in this way is
somewhat special when compared to the way $\zLC/\zdS/\eH$ arise from
$\zNi/\zSpi/\zTi$. In the case of $\scri$ the additional generator is
an element of the Lorentz transformations $\mathfrak{so}(d,1)$,
whereas in all the other cases the additional generator is a
light-/space-/timelike element of the translations. One might wonder
if there exists a $(d+1)$-dimensional space ``$?_{d+1}$'' such that
$\scri$ can be constructed in analogy with $\zLC/\zdS/\eH$, i.e., by
adding a generator of $P_\mu$ to the subalgebra $\mathfrak{h}$ of this
putative space. Inspection of~\eqref{eq:poincare-witt}, however,
reveals that this is not possible.

\section{Geometric realisations}
\label{sec:geometries}

Now we will describe explicit geometric realisations of the Klein
pairs discussed above in terms of Minkowski spacetime.  We will see
that $\zTi$ and (a $\ZZ_2$-quotient of) $\zSpi$ can be described as
grassmannians of affine spacelike and lorentzian hyperplanes in
Minkowski spacetime, respectively; whereas $\zNi$ fibers over the
grassmannian of affine null hyperplanes, which is $\scri$. This
provides, after we recall some basic notions about grassmannians, a
comparably simple geometric and coordinate-independent realisation of
these spaces.

\subsection{Grassmannians and affine grassmannians}
\label{sec:grassm-affine-grassm}

The classical \textbf{grassmannians} are the spaces $\Gr(p,n)$ of
$p$-dimensional vector subspaces of $\RR^n$; equivalently,
$p$-dimensional planes through the origin.  We shall simply call
them $p$-planes from now on.   Every point in $\Gr(p,n)$
corresponds to such a $p$-plane.  Duality allows us to
identify $\Gr(p,n)$ and $\Gr(n-p,n)$, so that we can also think of
every point in $\Gr(p,n)$ as an ($n-p$)-plane.  If
we put a euclidean inner product on $\RR^n$ we can visualise this
duality as simply taking perpendicular complements.  Notice that
$\Gr(n-1,n)$ is the space of hyperplanes in $\RR^n$
and since every hyperplane has a perpendicular line, we see that
$\Gr(n-1,n)$ can be identified with $\Gr(1,n)$, which is the
projective space $\RP^{n-1}$.

Let us again put a euclidean inner product on $\RR^n$. Then if
$\Pi \subset \RR^n$ is a $p$-plane, we can choose an orthonormal basis
for $\Pi$ and complete it to an orthonormal basis for $\RR^n$.
Conversely the first $p$ vectors in any orthonormal basis of $\RR^n$
span a $p$-plane. Since $\Ort(n)$ acts transitively on orthonormal
basis, it acts transitively on $p$-planes. The subgroup of $\Ort(n)$
which stabilises a given $p$-plane $\Pi$ is isomorphic to
$\Ort(p) \times \Ort(n-p)$, which are the changes of orthonormal bases
for $\Pi$ and for its perpendicular complement $\Pi^\perp$. In other
words, $\Ort(p)$ is the subgroup of orthogonal transformations which
map the $p$-plane $\Pi$ into itself and $\Ort(n-p)$ is the subgroup of
orthogonal transformations which act trivially on $\Pi$. In summary,
$\Gr(p,n)$ can be thought of as a homogeneous space
\begin{equation}
  \Gr(p,n) \cong \Ort(n)/(\Ort(p) \times \Ort(n-p)),
\end{equation}
from where it follows that $\dim \Gr(p,n) = p (n-p)$.  Notice that
putting $p=1$ we get that $\dim \Gr(1,n) = n-1$, consistent with
$\Gr(1,n) \cong \RP^{n-1}$.

If we instead put a lorentzian inner product on $\RR^n$, we may refine
the notion of grassmannian by keeping track of the causal character
of the planes.  We will be solely interested in hyperplanes below.  If
$\mathcal{H} \subset \RR^n$ is a hyperplane (through the origin), then
its perpendicular $\mathcal{H}^\perp$ relative to a lorentzian inner
product is a line which can be either timelike, spacelike or null.  In
the former two cases, $\RR^n = \mathcal{H} \oplus \mathcal{H}^\perp$,
whereas in the null case $\mathcal{H}^\perp \subset \mathcal{H}$.
We may therefore partition the grassmannian of hyperplanes
$\Gr(n-1,n)$ into three sub-grassmannians: the grassmannians of
spacelike, timelike and null hyperplanes; i.e.,
\begin{equation}
  \Gr(n-1,n) = \Gr(n-1,n)^{\text{spacelike}} \cup
  \Gr(n-1,n)^{\text{timelike}} \cup \Gr(n-1,n)^{\text{null}}.
\end{equation}
Equivalently, this corresponds to partitioning the projective space
$\RP^{n-1}$ of lines in $\RR^n$ into three: the projective spaces of
timelike, spacelike or null lines, respectively. We will discuss the
three grassmannians of hyperplanes in more detail in Sections
\ref{sec:adsc-as-grassm}, \ref{sec:spi-as-grassmannian}, and
\ref{sec:nul-geometry}. Illustrations of the grassmannians of timelike
and spacelike hypersurfaces can be found in Figures
\ref{fig:adsc-hyper} and \ref{fig:spi}, respectively.

Now let's go back to the general discussion of grassmannians, not
necessarily in a lorentzian vector space.  A closely related notion is
the \textbf{grassmannian of affine $p$-planes} in $\RR^n$, denoted
$\Graff(p,n)$, which are translates of the $p$-planes which pass
through the origin. If we again put a euclidean inner product on
$\RR^n$, then it is clear that the euclidean group $\E(n)$ (with Lie
algebra $\iso(n) = \so(n) \ltimes \RR^{n}$) acts transitively: we can use the
translations to bring an affine $p$-plane to the origin, on which
orthogonal transformations act transitively as we saw above. This
allows us to describe $\Graff(p,n)$ as a homogeneous space of the
euclidean group, a first hint that the lorentzian generalisation could
be related to some of the spaces we have already discussed. The stabiliser of
a given affine $p$-plane $\Pi$ are the longitudinal translations along
$\Pi$, the orthogonal transformations of $\Pi$ and the orthogonal
transformations of $\Pi^\perp$; in other words,
\begin{equation}
  \Graff(p,n) \cong \E(n)/(\E(p) \times \Ort(n-p)),
\end{equation}
from where we see that $\dim \Graff(p,n) = (n-p)(p+1)$.  In the case
of hyperplanes (or dually lines), $\dim \Graff(n-1,n) = n$, which is
one dimension higher than $\Gr(n-1,n)$.  This is easy to visualise.
Since translating by a vector tangent to a hyperplane does not move
the hyperplane, we need only translate along the perpendicular line.
This exhibits $\Graff(n-1,n)$ as a principal $\RR$-bundle over
$\Gr(n-1,n)$: the fibre at a hyperplane $\mathcal{H}$ is precisely
$\mathcal{H}^\perp$.  If we identify $\Gr(n-1,n)$ with the projective
space $\RP^{n-1}$, then $\Graff(n-1,n)$ is the total space of the
tautological line bundle of $\RP^{n-1}$, so called because the fibre
at a point in $\RP^{n-1}$ is precisely the line to which that point
corresponds.

Now let us see what happens when we put a lorentzian inner product on
$\RR^{n}$ and let us restrict to hyperplanes. We may now partition the
affine grassmannian by the causal type of the affine hyperplane. Since
translations do not alter the causal type, we see that all translates
of a timelike (resp.  spacelike or null) hyperplane will be timelike
(resp. spacelike or null) affine hyperplanes. Translating an affine
hyperplane back to the origin gives a hyperplane of the same causal
type and hence the tautological fibration
$\Graff(n-1,n) \to \Gr(n-1,n)$ restricts to the submanifolds of
timelike, spacelike or null hyperplanes in $\Gr(n-1,n)$ to given three
principal $\RR$-bundles:
\begin{equation}
  \begin{split}
    & \Graff(n-1,n)^{\text{timelike}} \to \Gr(n-1,n)^{\text{timelike}}\\
    & \Graff(n-1,n)^{\text{spacelike}} \to \Gr(n-1,n)^{\text{spacelike}}\\
    & \Graff(n-1,n)^{\text{null}} \to \Gr(n-1,n)^{\text{null}}.
  \end{split}
\end{equation}
The first two are easy to visualise: they correspond to translating a
given timelike (resp. spacelike) hyperplane along its perpendicular
spacelike (resp. timelike) line.  In the null case, the perpendicular
lies on the plane and this description is not accurate.  We can of
course, translate by a line not on the null plane, which we could
choose to be a null line without loss of generality.

We will now proceed to describe $\zTi$, $\zSpi$ and $\zNi$
geometrically in this language.

\subsection{$\zTi \cong \zAdSC$ as a grassmannian}
\label{sec:adsc-as-grassm}

An explicit geometric realisation of the Klein pair $(\g,\h)$ for
$\zTi \cong \zAdSC$ is provided by the grassmannian of spacelike affine
hyperplanes in Minkowski spacetime $\MM$. The purpose of this section
is to prove this. We will also see that this description explains the
structure of the BMS-like algebra of symmetries of $\zAdSC$ determined
in~\cite{Figueroa-OFarrill:2019sex}.

Recall that Minkowski spacetime $\MM$ is an affine space modelled on a
lorentzian vector space $V$.  Affine hyperplanes are codimension-one
affine subspaces of $\MM$. They are all of the form
$p + W$, where $p$ is a point on $\MM$ and $W\subset V$ is a spacelike
hyperplane of $V$. Equivalently, the perpendicular line $W^\perp$ is
timelike, so it is contained in the interior of the lightcone of $V$.
The Lorentz group $\Ort(V)$ acts transitively on timelike lines and
hence it acts transitively on spacelike hyperplanes of $V$. The
translation subgroup acts transitively on Minkowski spacetime, hence
any two spacelike affine hyperplanes $p + W$ and $p’ + W’$ are related
by a Poincaré transformation. This shows that the Poincaré group acts
transitively on the space of spacelike affine hyperplanes on $\MM$.

Now, fix one such affine spacelike hyperplane: $p + W$. What is its
stabiliser subgroup?  The translations which preserve $p + W$ are
precisely translations by vectors in $W$.  And the Lorentz transformations
preserving $W$ are the ones which fix the timelike line $W^\perp$, which
is a subgroup isomorphic to the group $\Ort(W)$ of orthogonal
transformations of $W$.  In other words, the stabiliser subgroup is
the group of euclidean transformations of $W$, which is isomorphic to
the euclidean group $\Ort(d)\ltimes \RR^d$ for $\zTi_{d+1}$.

\begin{figure}[h!]
\centering
\definecolor{cfcfcfc}{RGB}{252,252,252}
\definecolor{cd20000}{RGB}{210,0,0}
\definecolor{c9b9b9b}{RGB}{155,155,155}
\definecolor{c03468f}{RGB}{3,70,143}
\definecolor{cffffff}{RGB}{255,255,255}
\begin{tikzpicture}[overlay]
\begin{pgfonlayer}{nodelayer}
		\node [style=ghost] (0) at (6, 3.4) {{\color{cd20000}$W$}};
		\node [style=ghost] (1) at (2.7, 0.9) {$W^\perp$};
		\node [style=ghost] (2) at (4, 7.9) {{\color{c03468f}$\eH^{d}$}};
\end{pgfonlayer}
\end{tikzpicture}
\includegraphics[width=0.55\textwidth]{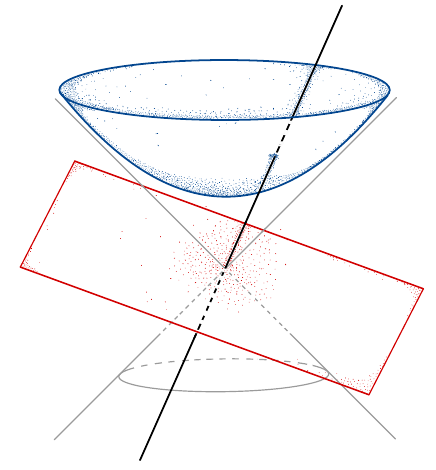}
    \caption{$\zTi \cong \zAdSC$ fibering over hyperbolic space.}
    \label{fig:adsc-hyper}
\end{figure}

This description shows that $\zTi_{d+1}$ fibers over $d$-dimensional
hyperbolic space $\eH^{d}$. This is nothing but the natural fibration of
the grassmannian of affine spacelike hyperplanes over the grassmannian
of spacelike hyperplanes, which admits a dual description as the
projective space of timelike lines.  Indeed, as we now explain, the
space of timelike lines in $V$ is naturally identified with $\eH^{d}$.
Recall that one model for hyperbolic space is given by any one of the
two sheets of the hyperboloid $\eta(x,x) = - \ell^2$, where $\eta$ is
the lorentzian inner product on $V$ and $\ell$ is the radius of
curvature of hyperbolic space. A timelike line will hit any one of
those hyperbolic spaces at exactly one point, as illustrated in
Figure~\ref{fig:adsc-hyper}. Hence $\zTi$ fibers over hyperbolic
space: the map sends the hyperplane $p + W$ to the point in hyperbolic
space which the line $W^\perp$ hits. The fibre is identified with
$W^\perp$ itself, since these are all the translates of $W$. So we
conclude that $\zTi$ is the total space of a line bundle over
hyperbolic space, which is tautological when we view hyperbolic space
as the space of timelike lines.  Furthermore, the carrollian degenerate
metric is the pullback via the projection of the hyperbolic metric on
$\eH^{d}$.

This allows us to understand the structure of the symmetry algebra of
$\zTi$, namely those vector fields which preserve the carrollian
structure. As shown in~\cite{Figueroa-OFarrill:2019sex} and discussed in Section~\ref{sec:ads-carroll}, the
Lie algebra of carrollian Killing vector fields on $\zTi \cong \zAdSC$ is
isomorphic to the semidirect product $\so(d,1) \ltimes
C^\infty(\eH^{d})$ of the Lorentz Lie algebra with the smooth functions on
hyperbolic space.  Some of the functions on $\eH^{d}$ correspond to the
Poincaré translations, but the rest are the so-called
``supertranslations''.  The emergence of the supertranslations is
clear in this geometric realisation.  Neither the carrollian vector
field nor the degenerate metric depend on the fibre coordinate, which
explains the symmetry of moving along each fibre at will. These are the
$\zTi$ supertranslations: they are sections of a line bundle and hence they
are abelian.  This line bundle is trivialisable and hence its sections
can be identified with the smooth functions on the base; although this
description hides the geometry.
\newpage
\subsection{$\zSpi/\ZZ_2$ as a grassmannian}
\label{sec:spi-as-grassmannian}

\begin{figure}[h!]
    \centering
\begin{tikzpicture}[overlay]
\definecolor{c9b9b9b}{RGB}{155,155,155}
\definecolor{c03468f}{RGB}{3,70,143}
\definecolor{cffffff}{RGB}{255,255,255}
\definecolor{cd20000}{RGB}{210,0,0}
\begin{pgfonlayer}{nodelayer}
		\node [style=ghost] (0) at (3.3, 7.7) {{\color{cd20000}$W$}};
		\node [style=ghost] (1) at (1, 2) {$W^\perp$};
		\node [style=ghost] (2) at (7, 4) {{\color{c03468f}$\zdS_{d}$}};
\end{pgfonlayer}
\end{tikzpicture}
\includegraphics[width=0.65\textwidth]{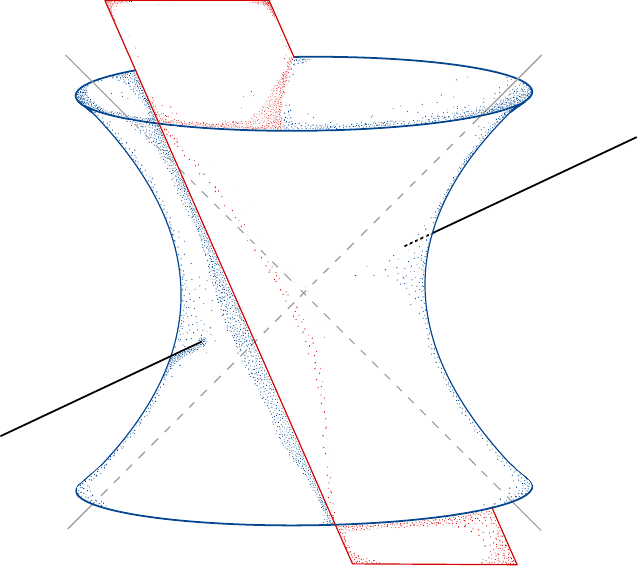}
    \caption{$\zSpi/\ZZ_2$ fibering over elliptic de~Sitter spacetime.}
    \label{fig:spi}
\end{figure}

A very similar picture exists for $\zSpi$, except that now it is a
double cover of the grassmannian of affine lorentzian hyperplanes in
Minkowski spacetime.  The discussion mimics that of $\zTi$, so we
will be brief.  An affine lorentzian hyperplane is again of the form $p
+ W$, but where now $W \subset V$ is a lorentzian hyperplane. Its
perpendicular line $W^\perp$ is now spacelike and shown in
Figure~\ref{fig:spi} such a line intersects any one of the one-sheeted
hyperboloids $\eta(x,x) = \ell^2$ at precisely two points. The induced
metric on the hyperboloid $\eta(x,x)= \ell^2$ is now that of
$d$-dimensional de~Sitter spacetime with radius of curvature $\ell$.
The space of spacelike lines in $V$ is then diffeomorphic to a $\ZZ_2$
quotient of de~Sitter spacetime $\zdS_d$, known as elliptic de~Sitter
spacetime~\cite{Schrodinger:1956jnw} and which we denote by
$\zdS_d/\ZZ_2$ in this paper. The action of $\ZZ_2$ is easy to
understand in the ambient vector space $V$: it changes the sign of all
coordinates simultaneously. That is clearly an isometry of the ambient
metric and since it preserves the hyperboloid it is also an isometry
of $\zdS_d$. Hence $\zdS_d/\ZZ_2$ inherits a metric from $\zdS_d$, making it
locally isometric to $\zdS_d$.

As argued for $\zTi$, the fact that the Lorentz group $\Ort(V)$ acts
transitively on spacelike lines shows that the Poincaré group acts
transitively on the space of affine lorentzian hyperplanes and the
Poincaré transformations which preserve such an affine hyperplane $p +
W$ are the translations along $W$ and the Lorentz transformations
$\Ort(W)$; in other words, the $d$-dimensional Poincaré group of $W$
with the induced lorentzian inner product.

Similarly to $\zTi$, also $\zSpi/\ZZ_2$ is seen to fiber over the 
projective space of spacelike lines, which is elliptic de~Sitter
spacetime, and hence it is the total space of a tautological line
bundle whose fibre at a point in the projective space of spacelike
lines is the corresponding spacelike line. The pseudo-carrollian
degenerate metric is the pull-back via the projection of the elliptic
de~Sitter metric on $\zdS_d/\ZZ_2$. As discussed in Section \ref{sec:spi},
the symmetry algebra of
$\zSpi/\ZZ_2$, consisting of those vector fields which preserve the
pseudo-carrollian structure is then isomorphic to
$\so(d,1) \ltimes C^\infty(\zdS_d/\ZZ_2)$, where $C^\infty(\zdS_d/\ZZ_2)$ are
the $\ZZ_2$-invariant functions in $C^\infty(\zdS_d)$. The
``supertranslations'' are again more properly interpreted as sections
of the tautological line bundle, which is trivialisable and hence can
be identified with the functions on $\zdS_d/\ZZ_2$.

\subsection{Geometric description of $\zNi$, $\scri$, $\zLC$ and $\zCS$}
\label{sec:nul-geometry}

Given the interpretations of $\zTi$ and $\zSpi/\ZZ_2$ as the
grassmannians of affine spacelike and lorentzian hyperplanes in
Minkowski spacetime, respectively, one might be forgiven for thinking
that $\zNi$ is the grassmannian of affine null hyperplanes. However
this is easily seen not to be the case. An affine null hyperplane is
of the form $p + W$ where $W \subset V$ is a null hyperplane. The
hyperplane is determined by the null line $W^\perp$, since
$W = (W^\perp)^\perp$. Unlike the case of spacelike or timelike
hyperplanes, $W^\perp$ is actually contained in $W$. The space of null
lines is the projectivised lightcone or, in other words, the celestial
sphere. Therefore the grassmannian of null hyperplanes (dually, the
projective space of null lines) is ($d-1$)-dimensional and since
affine null hyperplanes are obtained by translating a null hyperplane
by a vector $p$ with nonzero inner product with any generator of the
null line, the grassmannian of affine null hyperplanes is
$d$-dimensional. Of course, it is also a homogeneous space of the
Poincaré group, namely the identification of future and past null infinity,
$\scri$. This is straightforward to see using the dual
picture of light-like lines. The projectivised lightcone through
the origin maps to a sphere both at future and past null infinity.
Acting with a translation on the origin we can clearly reach any point
on both future and past null infinity.

The ($d+1$)-dimensional $\zNi_{d+1}$ is actually a
bundle over the grassmannian of affine null hyperplanes, i.e., a bundle over $\scri_{d}$. The
difference is that there is a Lorentz boost which preserves the null
line but rescales the points in the null line. Let us choose a null frame
$e_+,e_-,e_i$ for $V$. By a Lorentz transformation we can bring
the null line to be the line $\RR e_-$ spanned by $e_-$. The Lie
algebra of the stabiliser of $\RR e_-$ includes the boost $L_{+-}$,
which rescales $e_-$. Then $L_{+-}$ is in the stabiliser of the null
line, but not of any null vector generating that line.  The stabiliser
of $e_-$ in the Lorentz algebra consists of $L_{ij},L_{-i}$, whereas
the translations perpendicular to $e_-$ are spanned by $P_i$ and
$P_-$.  Hence the grassmannian of affine null hyperplanes in Minkowski
spacetime is described infinitesimally by the Klein pair $(\g,\k)$,
where $\k = \left<L_{ij},L_{-i},L_{+-},P_-, P_i\right>$, whereas that of
$\zNi_{d+1}$ is $(\g,\h)$ with
$\h = \left<L_{ij},L_{-i},P_-,P_i\right>$.

\begin{figure}[h!]
    \centering
    \begin{tikzpicture}[overlay]
\begin{pgfonlayer}{nodelayer}
		\node [style=ghost] (0) at (10.8, 4.7) {{$\zLC$}};
\end{pgfonlayer}
\end{tikzpicture}
    \includegraphics[width=0.8\textwidth]{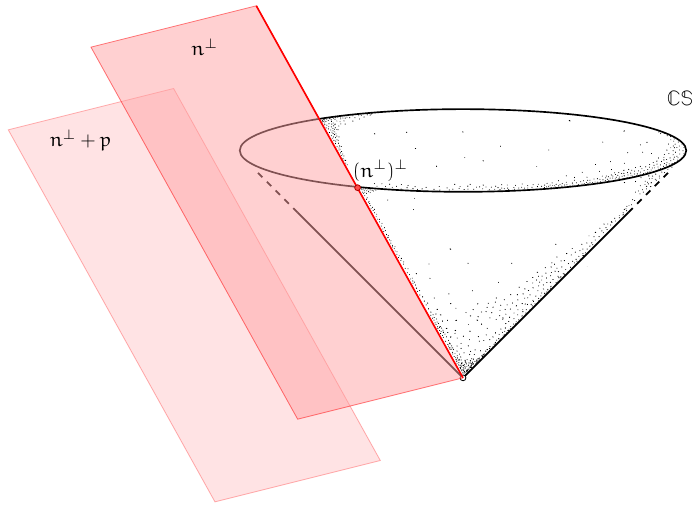}
\caption{Fibering of $\zNi$ over both the future-pointing lightcone and the space of affine null hyperplanes}
    \label{fig:LCCS}
\end{figure}

\subsection{$\zNi$ as the bundle of scales of the conformal carrollian structure on $\scri$}
\label{sec:znul}

In the following we want to further clarify the nature of the
homogeneous space $\zNi$, in addition to the geometric perspective
provided above. In order to do this, let us first consider the two
spaces on the right in the diagram \eqref{diag:relNul}, the light cone
$\zLC_d$ and the celestial sphere $\zCS^{d-1}$. The action of the
Poincaré group is not effective since the translations act trivially
on both $\zLC_d$ and $\zCS^{d-1}$. It is the quotient by the
translation ideal, isomorphic to the Lorentz group, which
acts effectively and it does so leaving invariant a carrollian
structure on $\zLC_d$ and a conformal riemannian structure on
$\zCS^{d-1}$. Indeed, it can be characterised as the symmetry group of
such structures; cf.~the last rows in Table
\ref{tab:overview-scriLCCS} which shows that they are isomorphic. The
fact that conformal symmetries of $\zCS^{d-1}$ correspond to
carrollian symmetries of $\zLC_d$ can be explained by the fact that
$\zLC_d$ is the total space of the bundle of scales of the conformal
manifold $\zCS^{d-1}$, as we will explain momentarily.

The Lorentz-invariant conformal structure on $\zCS^{d-1}$ consists of
all the metrics on $\zCS^{d-1}$ which are conformal to the round
metric $g$; that is, $[g] = \left\{\Omega^2 g \mid  
\Omega \in C^\infty(\zCS^{d-1})\right\}$.  Pick a point $\x \in
\zCS^{d-1}$, which we view as a unit-norm vector in $\RR^d$.
Evaluating the round metric at $\x$, we get $g_{\x} \in 
\odot^2 T_{\x}^*\zCS^{d-1}$ and hence a ray $Q_{\x} = \{ \lambda^2 g_{\x} \mid
\lambda \in \RR^+\} \subset \odot^2 T_{\x}^*\zCS^{d-1}$.  Let $Q =
\sqcup_{\x \in \zCS^{d-1}} Q_{\x}$ and define $\pi : Q \to \zCS^{d-1}$ by
sending $\lambda^2 g_{\x}$ to $\x$.  This then becomes a principal
$\RR^+$-bundle over $\zCS^{d-1}$ with $\sigma \in \RR^+$ acting on $Q$
via $\lambda^2 g_{\x} \mapsto \sigma^2 \lambda^2 g_{\x}$.  Since the
round metric defines a global section of $Q$, it is a trivial bundle,
so that $Q$ is diffeomorphic to $\RR^+ \times \zCS^{d-1}$.  By its
very definition, every section $\zCS^{d-1} \to Q$ defines a metric in
the conformal class $[g]$.

The (proper, orthochronous) Lorentz group $\SO(d-1,1)_0$ acts
transitively on $\zCS^{d-1}$ via conformal transformations and
therefore it acts on $Q$: if $A \in \SO(d-1,1)_0$, then
$A\cdot (\lambda^2 g_{\x}) = (\sigma(A,\x)^2 \lambda^2 g_{A\cdot
  \x})$, for some function
$\sigma : \SO(d-1,1)_0 \times \zCS^{d-1} \to \RR^+$ whose explicit
form is of no consequence. This action is also transitive and the
stabiliser of the point $g_{\x} \in Q$ is the subgroup of the
stabiliser of $\x \in \zCS^{d-1}$ which preserves $g_{\x}$; that is, a
subgroup isomorphic to $\ISO(d-1)$. This shows that $Q$ is isomorphic
to the future (deleted) lightcone $\zLC_d$ as a homogeneous space of
the (proper, orthochronous) Lorentz group: the diffeomorphism
$Q \to \zLC_d$ sends $\lambda^2 g_{\x} \in Q$ to
$(\lambda,\lambda \x) \in \zLC_d$.   This diffeomorphism is
equivariant under both the action of $\RR^+$ and that of
$\SO(d-1,1)_0$ and makes the following triangle commute:
\begin{equation}
  \begin{tikzcd}
    Q \arrow[rd] \arrow[rr,"\cong"]  & & \zLC_d \arrow[ld]\\
    & \zCS^{d-1} &
  \end{tikzcd}
\end{equation}
so that is both a bundle isomorphism and an isomorphism of homogeneous
spaces of $\SO(d-1,1)_0$. For more details and the relation of this
construction to tractor calculus we refer the reader
to~\cite{Curry:2014yoa}.

Returning to $\zNi$ we can mirror the above discussion and interpret
$\zNi$ as the \emph{bundle of scales} of the conformal carrollian
structure on $\scri$. Again the symmetries of the doubly-carrollian
structure of $\zNi_{d+1}$ and the conformal carrollian symmetries of
the carrollian structure of $\scri$ are isomorphic and given by
$BMS_{d+1}$; cf.~the last rows in Table~\ref{tab:overview} and
\ref{tab:overview-scriLCCS}. Let us consider $\scri$ with its
Poincaré-invariant conformal carrollian structure $[(\xi,h)]$
consisting of all carrollian structures
$(\Omega^{-1}\xi, \Omega^2 h)$, for $\Omega \in C^\infty_+(\scri)$ a
positive smooth function. Any carrollian structure in that class, say
$(\xi,h)$, defines a ray sub-bundle
$P \subset T\scri \oplus \odot^2T^*\scri$, where if $p \in \scri$,
$P_p = \{(\lambda^{-1} \xi_p, \lambda^2 h_p) \in T_p\scri \oplus
\odot^2T_p^*\scri \mid \lambda \in \RR^+\}$. Let $\pi: P \to \scri$ be
the restriction to $P$ of the projection
$T\scri \oplus \odot^2T^*\scri \to \scri$. Then $\pi: P \to \scri$ is
a principal $\RR^+$ bundle, where $\sigma \in \RR^+$ acts on $P$ via
$(\lambda^{-1} \xi_p, \lambda^2 h_p) \mapsto
(\sigma^{-1}\lambda^{-1}\xi_p, \sigma^2 \lambda^2 h_p)$. Since any
carrollian structure in the conformal class defines a section
$\scri \to P$, we see that this is a trivial bundle and hence
$P \cong \scri \times \RR^+$.

The (proper, orthochronous) Poincaré group $\ISO(d,1)_0$ acts on
$\scri$ preserving the conformal carrollian structure and hence it
acts on $P$: if $A \in \ISO(d,1)_0$ then
\begin{equation}
A \cdot (\lambda^{-1} \xi_p, \lambda^2 h_p) = (\sigma(A,p)^{-1}
\lambda^{-1} \xi_{A \cdot p}, \sigma(A,p)^2 \lambda^2 h_{A\cdot p}),
\end{equation}
for some function $\sigma : \ISO(d,1)_0 \times \scri \to \RR^+$ whose
explicit expression is not needed. This action is transitive and the
stabiliser of $(\lambda^{-1} \xi_p, \lambda^2 h_p)$ is the subgroup of
the stabiliser of $p \in \scri$ which leaves invariant the carrollian
structure, not just its conformal class.  We may choose $p$ to be the
point of $\scri$ with stabiliser subalgebra
$\h_\scri = \left<L_{ij}, L_{-i}, P_i, P_-, L_{+-}\right> =
\left<L'_{ij}, L'_{i}, B'_i, B'_-, P'_-\right> $, whose invariant
conformal carrollian structure is the one corresponding to
$P'_+ \mod \h_\scri$ and $\delta_{ij}\pi'^i\pi'^j$.  The subalgebra
$\left<L_{ij}, L_{-i}, P_i, P_-\right> = \left<L'_{ij}, L'_{i}, B'_i,
  B'_-\right>$ of $\h_\scri$ leaves the carrollian structure invariant
not just up to conformal rescaling and we see that it is isomorphic to
$\h_{\zNi}$.  Therefore $P$ is isomorphic to $\zNi$ as a homogeneous
space of the Poincaré group, which allows us to identify $\zNi$ as the
bundle of scales of the Poincaré-invariant conformal carrollian
structure of $\scri$.

The interpretation of $\zLC_d$ as the bundle of scales of $\zCS^{d-1}$
can be seen rather explicitly by embedding $\zLC_d$ in
$(d+1)$-dimensional Minkowski spacetime. In the same way, we can see
explicitly the relation between $\scri$ and $\zNi$ from the embedding
$\zNi_{d+1}\subset \EE^{d+1,2}$ and the projection down to $\scri_d
\subset \PP^{d+2}$ described in Section~\ref{sec:embedding-ni}.

\section{Conclusion and outlook}
\label{sec:conclusion}

Before we provide a few words concerning the relation to holography in
(anti-)de Sitter space and provide a list of some of the intriguing
questions for further studies, let us conclude and summarise the
results of this work.

\subsection{Conclusion}
\label{sec:conclusion-1}

With the aim to improve our understanding of flat holography we have studied
homogeneous spaces of the Poincaré group which are relevant to
asymptotically flat spacetime: see Figure~\ref{fig:Minkowski} for an
overview.

A concrete way to understand these spaces is provided in
Section~\ref{sec:embeddings}, where we embed Minkowski spacetime
$\MM_{d+1}$ together with $\zTi_{d+1}$ and $\zSpi_{d+1}$, the blow-ups of
timelike/spatial infinities~\cite{Ashtekar:1978zz}, and the novel space
$\zNi_{d+1}$ into the pseudo-euclidean space $\EE^{d+1,2}$.  We also
showed that $\zNi_{d+1}$ fibers over $\scri_{d}$ and the lightcone
$\zLC_{d}$ which in turn fiber over the celestial sphere $\zCS^{d-1}$
(see~\eqref{eq:double-fibration-1}).  The celestial sphere plays a 
distinguished rôle as the unique two-dimensional manifold admitting a
transitive (albeit non-effective) action of the Poincaré group in
$3+1$ dimensions.  This fact, however, also provides some challenges
for nonzero cosmological constant as we will discuss in
Section~\ref{sec:holography}.  Let us emphasise that $\zSpi_{d+1}$ 
and $\zTi_{d+1}$ are, in contradistinction to null infinity
$\scri_{d}$, of the same dimension as the bulk Minkowski spacetime
$\MM_{d+1}$ itself.  The novel space $\zNi_{d+1}$ is the natural $(d+1)$-dimensional
lift of $\scri_{d}$ that fills this gap.

In Section~\ref{sec:reconstruction} we reconstructed points of Minkowski
spacetime form intrinsic properties of these carrollian-like
geometries, roughly speaking from functions on $\zdS_{d}$, $\eH_{d}$,
$\zLC_{d}$ and $\zCS^{d-1}$. This can be seen as a form of
``holography''.  It also provides a generalisation of the ``good cut''
equation to generic dimension and is not necessarily tied to $\scri$.

In Section~\ref{sec:klein-pairs} we introduced these spaces as
homogeneous spaces of the Poincaré group.  This means that they are
characterised as quotients of the Poincaré group by different
subgroups, which implies that they have different geometrical and
physical interpretations, e.g., their invariants and, notably, their dimensions
differ.  We also studied their invariant structures: a carrollian
structure for $\zTi$, a pseudo-carrollian \cite{Gibbons:2019zfs}
structure for $\zSpi$, and a novel doubly-carrollian structure for
$\zNi$, intimately linked to the conformal carrollian structure of
$\scri$, in analogy to how the carrollian structure of the lightcone
$\zLC$ is linked to the conformal riemannian structure of the
celestial sphere $\zCS$.  See Tables~\ref{tab:overview},
\ref{tab:overview-HdS} and \ref{tab:overview-scriLCCS} for an overview
of the homogeneous spaces and their properties. Through the looking-glass, 
these spaces provide glimpses of an interesting and rich carrollian-like world that
lies beyond the by now well-studied flat carrollian space. The subgroups and
homogeneous spaces of this work have also appeared in various other
contexts: e.g., in relation to Dirac's form of relativistic
dynamics\footnote{More precisely $\zTi$ relates to the instant form,
  $\scri$ to the front form and $\MM$ to the point form
  of~\cite{Dirac:1949cp}.}, to the subalgebra of $\scri$ in the
context of the infinite momentum frame (see, e.g.,
\cite{Kogut:1972di}), and to induced representations as we will comment
on in Section~\ref{sec:outlook}.

As we also showed, the spaces $\zSpi_{d+1}$, $\zTi_{d+1}$ and $\zNi_{d+1}$ have the
following intriguing property: the symmetries of their invariant
structure match precisely the asymptotic symmetries one expects from
asymptotic flat spacetimes, e.g., BMS symmetries for $\zNi_{d+1}$. This
should be contrasted with the \emph{conformal} carrollian symmetries of
$\scri_d$, which are also given by the
very same BMS symmetries. The underlying reason is that $\zNi_{d+1}$
can be seen, as we prove in Section~\ref{sec:znul}, as the bundle of
scales of the conformal carrollian structure of $\scri_{d}$, in
complete analogy to how the lightcone is the bundle of scales of the
conformal structure on the celestial sphere.

Finally, in Section~\ref{sec:geometries}, we have developed a simple,
explicit and coordinate-independent geometric realisation of $\zTi$,
$\zSpi$, $\zNi$ and $\scri$ in terms of grassmannians of hyperplanes
in Minkowski spacetime.

\subsection{Relation to (anti-)de Sitter holography}
\label{sec:holography}

In this work we have focused on the Poincaré group and the asymptotic
infinities of flat space. Let us briefly discuss some aspects of the
relation to (anti-)de Sitter space. For simplicity we will restrict to
$3+1$ dimensions. These observations concerning the subalgebras are
based on~\cite{Patera:1976my}, of which we have summarised the
relevant details in Appendix~\ref{sec:homog-spac-poinc}.

Let us first observe that the asymptotic structure of asymptotically
flat space has more boundaries ($\scri^{\pm}$, $i^{\pm}$ (or
$\zTi^{\pm}$), $i^{0}$ (or $\zSpi$)) than their curved counterparts.
Anti-de Sitter space has one and de Sitter space has two: past and
future infinity. We have shown that the asymptotic geometries of flat
space are captured by homogeneous spaces of the Poincaré group.
Remarkably, this also generalises to the (anti-)de Sitter groups,
although the situation there is even simpler. Since the boundaries in these
cases are not singular we will focus on three-dimensional homogeneous
spaces, which can roughly be thought of as the boundaries of (anti-)de
Sitter space. In this sense they are close in spirit to $\scri$.

Upon inspection of the three-dimensional homogeneous spaces of the de
Sitter groups (see Appendix~\ref{sec:homog-spac-poinc} for details), we
obtain the conformal symmetries relevant for
AdS/CFT~\cite{Maldacena:1997re,Gubser:1998bc,Witten:1998qj} and the
(euclidean) conformal symmetries of
dS/(E)CFT~\cite{Strominger:2001pn}. For AdS there exists a second
homogeneous three dimensional space, but remarkably not more. For de
Sitter space the three dimensional geometry is unique. This
should be contrasted to the infinitely many three-dimensional spaces
of the Poincaré group~\cite{Patera:1975bw} (again, we refer to
Appendix~\ref{sec:homog-spac-poinc} for the details).

Reducing the homogeneous spaces by another dimension, i.e., looking at
putative holographic correspondences to  theories on a two-dimensional geometry
one finds that the de~Sitter groups do not possess homogeneous spaces
of the same dimension as the celestial sphere; only spaces of higher
dimension appear, where their putative dual three-dimensional (E)CFTs
live. This means there exists no homogeneous space with (A)dS symmetry
which could play the rôle of the celestial sphere in flat space
holography, and consequentially there is no flat limit. This is a
precise statement based only on symmetries and presumably quite
robust.

\subsection{Outlook}
\label{sec:outlook}

There are several interesting open questions which deserve further
exploration, some of which we list in the following.

\begin{description}
\item[Embedding formalism and holography]

  The embeddings we discuss in Section~\ref{sec:embeddings} can also
  be viewed as a generalisation of the embedding space formalism used
  in the AdS$_{d+1}$/CFT$_{d}$ correspondence (see, for
  example,~\cite{Dirac:1936fq,Costa:2011mg,Costa:2011dw}) to flat
  space.  This is a generalisation in the sense that both the bulk
  Minkowski spacetime as well as the ``boundaries'' embed in a higher
  dimensional space.  Let us emphasise that our space $\EE^{d+1,2}$ is
  one dimension higher than that commonly used for
  AdS$_{d+1}$/CFT$_{d}$, which opens the possibility to also embed
  these spaces and study their limits in $\EE^{d+1,2}$.

  A concrete way to do this would be to use the common embedding
  formalism and intersect it with a null hyperplane in the ambient
  space, as in Section~\ref{sec:embeddings}, in which case once could
  hope to obtain aspects of flat holography from AdS/CFT in one dimension
  higher by restricting to the null hyperplane.

\item[Reconstruction of Minkowski space] In Section
  \ref{sec:reconstruction} we employed our embedding space picture to
  show how points of Minkowski space can be related to certain
  sections of $\zTi,\zSpi,\zNi,$ and $\scri$. It would be interesting
  to relate this more explicitly to an intrinsically minkowskian
  construction that uses time-/space-/light-like curves to reconstruct
  a point in Minkowski space from a given section of the above spaces.
  We leave a more comprehensive discussion of this to future studies.

\item[Gauging and Cartan geometry]

  Homogeneous spaces are the flat models of Cartan
  geometries (see, e.g., \cite{sharpe2000differential}).
  The so-called gauging procedure may be re-interpreted as the
  construction of a Cartan geometry, with the gauge field defining a
  Cartan connection.  For the case of Minkowski spacetime this leads to
  pseudo-riemannian geometry and consequently to general relativity
  and for the de~Sitter spaces to MacDowell--Mansouri gravity
  \cite{Wise:2006sm}.

  The study of the Cartan geometries modelled on $\zSpi$, $\zTi$ and
  $\zNi$ via the gauging procedure will be the subject of a
  forthcoming paper. In \cite{Nguyen:2020hot} a Chern--Simons action
  for Cartan geometries based on $\zLC$ was written down, and it was
  shown that the so obtained geometries reproduced certain features of
  the asymptotic structure of asymptotically flat spacetimes. Cartan
  geometries modelled on $\scri$ have recently been discussed by
  Herfray in \cite{Herfray:2020rvq} and related to the geometry of
  asymptotically flat spacetimes~\cite{Herfray:2021xyp}
  (see~\cite{Herfray:2021qmp} for a review), and it would be
  interesting to extend those results to $\zSpi$, $\zTi$ and $\zNi$.
  
\item[Lower dimensional theories]

  In $2+1$ dimensions we can write down Chern--Simons theories for the
  homogeneous spaces discussed in this work. They are homogeneous
  spaces of the Poincaré group in $2+1$ dimensions, which admits a
  bi-invariant metric~\cite{Witten:1988hc}, equivalently the Poincaré
  Lie algebra admits an $\ad$-invariant scalar product. One can then
  generalise what was already done for $\zAdSC \cong \zTi$
  Chern--Simons theory~\cite{Matulich:2019cdo} (see~\cite{Ravera:2019ize} for the supergravity generalisation) and write
  down actions with an interpretation suited for the homogeneous
  spaces of this work.

  Similar remarks apply to $(1+1)$-dimensional generalisations of JT
  gravity, as well as their associated BF theory and dilaton gravity
  analogues~\cite{Grumiller:2020elf,Gomis:2020wxp}.
  
\item[Relation to novel (induced) representations of the Poincaré group]

  The homogeneous spaces we discuss have another interesting
  interpretation in the theory of induced representations, where one
  induces a representation of the Poincaré group using one of its subgroups; see
  \cite[Section 3]{Oblak:2016eij} for a review. The connection to our
  work comes from looking at the momentum orbits of the Poincaré
  particles that are given by homogeneous spaces of the Poincaré
  group~(see, e.g., \cite[Section 4.2]{Oblak:2016eij}). The momentum
  orbit of massive particles are related to~$\eH$, massless orbits to
  $\zLC$, and the tachyonic ones to $\zdS$.
  
  Besides these well-known Wigner momentum eigenstate representations,
  other interesting representations of the
  Poincaré group have recently been put forward~\cite{Pasterski:2016qvg,Pasterski:2017kqt,Banerjee:2018gce}.
  They have played a central role in advances in (celestial)
  holography, and it might be interesting to clarify if and how they
  are related to the homogeneous spaces described in this work. To our
  understanding, representations induced by $\scri$ have already been
  considered in~\cite{Banerjee:2018gce}, but we have discussed other
  interesting subgroups (see also Appendix~\ref{sec:homog-spac-poinc}
  for additional subalgebras).
\end{description}

\section*{Acknowledgments}
\label{sec:acknowledgments}

We are grateful to Tim Adamo, Andrew Beckett, Jelle Hartong, Yannick
Herfray, Sucheta Majumdar, and Alfredo Pérez for useful discussions.

We also thank the anonymous referee for various insightful comments
and suggestions.

The work of EH is supported by the Royal Society Research Grant for
Research Fellows 2017 “A Universal Theory for Fluid Dynamics” (grant
number RGF$\backslash$R1$\backslash$180017).

SP was supported by the Leverhulme Trust Research Project Grant
(RPG-2019-218) ``What is Non-Relativistic Quantum Gravity and is it
Holographic?''.

SP and JS acknowledge support of the Erwin Schrödinger Institute (ESI) in
Vienna where part of this work was conducted during the thematic
programme ``Geometry for Higher Spin Gravity: Conformal Structures,
PDEs, and Q-manifolds''.

The work of JS was supported by the Erwin-Schrödinger fellowship J
4135 of the Austrian Science Fund (FWF) and by the F.R.S.-FNRS Belgium
through the convention IISN 4.4503.

\appendix

\section{$\zSpi$ and $\zTi$ as blow-ups of spatial and timelike infinity}
\label{sec:AH-construction}

The following discussion is based on the original work of
Ashtekar--Hansen (AH)~\cite{Ashtekar:1978zz}.  Although a detailed
discussion of the AH construction lies beyond the scope of this work,
we will summarise the salient features in the following. In a
conformal compactification of Minkowski spacetime the conformal
boundary at spatial infinity is given by a single point $i^0$. This
remains true for more general asymptotically flat spacetimes in the
definition of AH. However, various physical fields, e.g., the
connection coefficients, admit only direction-dependent limits at
$i^0$. One therefore constructs a blow-up of $i^0$, such that fields
at $i^0$ can be regarded as smooth fields on a blow-up manifold
$\zSpi$.

The blow-up manifold is constructed using the behaviour of certain
inextensible spacelike curves approaching $i^0$. The AH definition
gives rise to a universal lorentzian metric at $i^0$ that is used to
demand that these curves have unit tangent vector at $i^0$. Such
tangent vectors form a hyperboloid in the tangent space of $i^0$, with
induced metric being $\zdS_d$. This defines the asymptotic geometry at
spatial infinity in the sense of~\cite{Geroch:1977jn}. However, the
differentiability conditions in the AH definition allow also to define
(direction-dependent) connection coefficients at $i^0$. Using these
one demands that the spacelike curves be geodesics of the original
asymptotically flat manifold. This requirement leaves undefined the
component of the acceleration along the tangent vector of the curve at
$i^0$. This additional parameter, taking values in the real numbers,
can therefore be used to distinguish between asymptotic spacelike
curves and thus becomes an additional coordinate on $\zSpi$.

From the above construction it is apparent that $\zSpi$ has the
structure of a fibre bundle. The base space is the (one-sheeted) unit
hyperboloid with fibre $\mathbb{R}$. There are two natural tensor
fields defined on $\zSpi$: a nowhere vanishing vector field
$n\in\eX(\zSpi)$ that generates diffeomorphism of the fibre and a
corank-one $\gamma\in\Gamma(\odot^2T^*\zSpi)$ of lorentzian signature
with constant positive curvature (the pullback of $\gamma$ to the base
space $\zdS$ is the metric on $\zdS$), which furthermore satisfies
$\gamma(n,-)=0$.  This is exactly the invariant structure that the
Klein pair of $\zSpi$ gives rise to, and we therefore recognise the AH
construction of $\zSpi$ as the (simply-connected) homogeneous space of
this Klein pair.  This observation was first made
in~\cite{Gibbons:2019zfs}.

This construction is applicable, mutatis mutandis, to future/past
timelike infinity. Here, the AH construction leads a universal
riemannian metric at $i^\pm$, which is now used to demand that
timelike curves approaching (or emanating from) $i^\pm$ have unit
tangent vector at $i^\pm$, with those tangent vectors now giving the
tangent space of $i^\pm$ the structure of hyperbolic space
$\eH_d$. Exactly as for $\zSpi$, the component of the acceleration
along the tangent vector of a curve at $i^\pm$ can be used to
distinguish between asymptotic timelike curves and is taken to be an
additional coordinate on $\zTi$. Hence $\zTi$ is a (trivial) line
bundle over hyperbolic space, whose invariant structure is a
carrollian structure, consisting of a nowhere vanishing
$\xi\in\eX(\zTi)$ and a corank-one positive semi-definite
$h\in\Gamma(\odot^2T^*\zTi)$ of constant negative curvature whose
kernel is spanned by $\xi$: $h(\xi,-)=0$.  This is precisely our space
$\zTi\cong \zAdSC$.

\section{Low-dimensional homogeneous spaces of $\ISO(3,1)$, $\SO(3,2)$
  and $\SO(4,1)$}
\label{sec:homog-spac-poinc}
  
As already discussed, up to coverings, a homogeneous space is
characterised by a Klein pair $(\g, \h)$ consisting of a Lie algebra
$\g$ and a Lie subalgebra $\h$.  This implies that the classification
of Lie subalgebras of the Poincaré algebra $\iso(3,1)$,
see~\cite{Patera:1974zd,Patera:1975bw} and references therein,
contains the classification of homogeneous spaces of the Poincaré
group.  (Not every Klein pair need be geometrically realisable, so it
could be that there are more subalgebras than homogeneous spaces.)
Although it might be interesting to study the geometry of all the
homogeneous spaces of the Poincaré group, in this work we concentrate
on $\MM$, $\zTi$, $\zSpi$ and $\zNi$ and their descendants as depicted
in \eqref{eq:null-fibrations}, which can be defined in any dimension,
and, in particular, capture the asymptotic structure of Minkowski
spacetime at infinity.

In this appendix we restrict to $3+1$ dimensions and we will comment
on homogeneous spaces of dimension four or lower, i.e., Lie
subalgebras of dimension six or higher. In the main part we were not
exhaustive with regard to three- and four-dimensional spaces (and
ignored the higher-dimensional ones). Here we want to provide
additional useful information, and relate and contextualise the
spaces of this work to the classification of~\cite{Patera:1975bw}. In
this way, we can read off from \cite{Patera:1975bw} the (generalised)
invariants of the Lie subalgebras.

A cursory glance at the classification of subalgebras of the Poincaré
Lie algebra in \cite[Table~VI]{Patera:1975bw} shows 10 six-dimensional
subalgebras of the Poincaré algebra, albeit that one of them has a
parameter: an angle $0<c<\pi$.  A slightly less cursory glance shows
that for three of the putative homogeneous four-dimensional spaces
($P_{8,1}$, $P_{9,1}$, $P_{10,1}$ in the notation of
\cite{Patera:1975bw}), the action of the Poincaré group is not
effective. The homogeneous spaces of this work are related in the
following way to the subalgebras~\cite[Table VI]{Patera:1975bw}
$\MM_{4} \leftrightarrow P_{1,2}$, $\zTi_{4} \leftrightarrow P_{3,2}$,
$\zSpi_{4} \leftrightarrow P_{4,2}$, $\zNi_{4} \leftrightarrow
P_{6,2}$. It is interesting to note that $\zNi_{4}$ can be seen as the
endpoints $c=0,\pi$ of the one-parameter family.

If we look at three-dimensional homogeneous spaces, i.e., subalgebras
of dimension seven~\cite[Table VII]{Patera:1975bw}, we find that there
exist $6$ cases of which one is a one-parameter family $0<c<\pi$. The
lightcone $\zLC_{3}$ corresponds to the endpoints $c=0,\pi$ of this
family. The subalgebras relate to the spaces of our work as
$\eH^{3} \leftrightarrow P_{3,1}$, $\zdS_{3} \leftrightarrow P_{4,1}$,
$\scri_{3} \leftrightarrow P_{2,2}$,
$\zLC_{3} \leftrightarrow P_{6,1}$. It is interesting to note that
$\scri_{3}$ is the unique effective three-dimensional homogeneous
space of the Poincaré group.

The unique two-dimensional Klein pair of the Poincaré algebra is not
effective and yields, upon reduction, a Klein pair for the celestial
sphere $\zCS_{2} \leftrightarrow P_{2,1}$.

This discussion of homogeneous spaces of the Poincaré group should be
contrasted with the case of the de~Sitter groups~\cite{Patera:1976my}.
We will again restrict to $3+1$ dimensions, but will now only discuss
subalgebras of dimension seven or higher.  Since the relevant Lie
algebras, $\so(3,2)$ and $\so(4,1)$, are simple, there are no
non-effective Klein pairs and because of dimension, it follows that
there are no two-dimensional spaces on which the corresponding groups
can act transitively.

Indeed, for  anti de-Sitter space the relevant Lie algebra is
$\so(3,2)$ and it follows from the classification in
\cite{Patera:1976my} that there are precisely two seven dimensional
subalgebras, $\a_{7,1}$ in Table IV and $\b_{7,1}$ in Table V, which are
maximal.  The Klein pair $(\so(3,2),\a_{7,1})$ is conformally
compactified Minkowski space and hence of relevance in the AdS/CFT
correspondence.  The Klein pair $(\so(3,2),\b_{7,1})$ can be
interpreted as the grassmannian of maximally isotropic planes in
$\RR^{3,2}$ and has still to find relevance in holography.  It admits
a (non-metric) conformal structure whose bundle of scales is the null
quadric $\eQ_0 \subset \EE^{3,2}$ introduced (in general dimension) in
Section~\ref{sec:poincare-orbits}.  This interpretation of the null
quadric reveals that it has an $\Ort(3,2)$-invariant pseudo-carrollian
structure.  There are no (proper) subalgebras of $\so(3,2)$ of
dimension higher than $7$.  In other words, there are no
two-dimensional spaces on which $\Ort(3,2)$ acts transitively.

The relevant Lie algebra for de~Sitter space is now $\so(4,1)$, and
it follows from the classification in \cite[Table XI]{Patera:1976my}
there is now a unique $7$-dimensional subalgebra $\a_{7,1}$ and again no
subalgebra of higher dimension. This implies that there is a unique
three-dimensional homogeneous space of $\Ort(4,1)$ with Klein pair
$(\so(4,1),\a_{7,1})$ corresponding to the celestial $3$-sphere with
its invariant conformal structure.  Its bundle of scales is the
four-dimensional lightcone, with Klein pair $(\so(4,1), \a_{6,1})$ in
the notation of \cite[Table  XI]{Patera:1976my}.  No lower-dimensional
homogeneous spaces (or even Klein pairs) exist.

\section{Symmetries of the doubly-carrollian structure of $\zNi$}
\label{sec:symmetries-znul}

In this appendix we work out the Lie algebra of vector fields
preserving the ``doubly-carrollian'' structure of $\zNi$.  We recall
from Section~\ref{sec:null} that $\zNi$ is one of the two
smooth components of the intersection of the null quadric $\eQ_0$ with
the null hyperplane $\eN_0$ in $\EE^{d+1,2}$, so that it consists of the
points $(r,x^a,x^+,0) \in \EE^{d+1,2}$, where
$r = \sqrt{\sum_{a=1}^d (x^a)^2} > 0$. This shows that $\zNi$ is
diffeomorphic to $N:= (\RR^d \setminus\{0\}) \times \RR$ and the
explicit embedding $j : N \to \EE^{d+1,2}$ is defined by
$j(x^a,x^+) = (r,x^a,x^+,0)$.

The Poincaré generators on $\EE^{d+1,2}$ restricted to $\zNi$ are the
image under the embedding of the following vector fields on $N$:
\begin{equation}
  L_{ab} := x^a \d_b - x^b \d_a, \quad B_a = - r \d_a, \quad P_a = x^a
  \d_+ \quad\text{and}\quad H = -r \d_+.
\end{equation}
The vector fields on $N$ which commute with the Poincaré generators
form a two-dimensional nonabelian Lie algebra with basis
\begin{equation}
  \xi_+ = \d_+ \qquad\text{and}\qquad \xi_- = r \d_r + x^+ \d_+,
\end{equation}
with Lie bracket $[\xi_+,\xi_-] = \xi_+$.  The Poincaré-invariant
$(0,2)$-tensor is the pull-back to $N$ of the pseudo-euclidean
metric on $\EE^{d+1,2}$:
\begin{equation}
  j^* g_{\EE} = -dr^2 + \sum_{a=1}^d (dx^a)^2.
\end{equation}
Using spherical polar coordinates in $\RR^d\setminus\{0\}$,
\begin{equation}
  \sum_{a=1}^d (dx^a)^2 = dr^2 + r^2 g_S,
\end{equation}
where $g_S$ is the round metric on the
unit sphere in $\RR^{d+1}$, so that $j^*g_{\EE} = r^2 g_S$.
In summary, the doubly-carrollian structure on $N$ is given by the
data $(\d_+, r\d_r + x^+\d_+, r^2 g_S)$.

We now determine the Lie algebra of symmetries of the
doubly-carrollian structure of $\zNi$.  Let $\zeta \in \eX(N)$ be a
vector field on $N$, which we choose to decompose as
\begin{equation}
  \zeta = \zeta^r \d_r + \zeta^+ \d_+ + \zeta^\perp,
\end{equation}
where the vector field $\zeta^\perp$ is tangent to the spheres (so in
the kernel of $dr$ and $dx^+$), but depends a priori on all the
coordinates.  Demanding that $[\zeta,\xi_+] = 0$ says that the
functions $\zeta^r$, $\zeta^+$ and the vector field $\zeta^\perp$ do
not depend on $x^+$.  The Lie bracket $[\xi_-,\zeta]$ is given by
\begin{equation}
  [\xi_-,\zeta] = (r\d_r \zeta^r - \zeta^r) \d_r + (r \d_r \zeta^+ -
  \zeta^+) \d_+ + r\d_r \zeta^\perp,
\end{equation}
which vanishes provided that
\begin{equation}
  \zeta^r = r f^r,\qquad \zeta^+ = r f^+ \qquad\text{and}\qquad
  \zeta^\perp \in \eX(S^{d-1}),
\end{equation}
where $f^r,f^+ \in C^\infty(S^{d-1})$.  Demanding that
\begin{equation}
  \zeta = r f^r \d_r + r f^+ \d_+ + \zeta^\perp
\end{equation}
leaves invariant $r^2 g_S$ results in
\begin{equation}
  \eL_{\zeta^\perp} g_S = -2 f^r g_S,
\end{equation}
so that $\zeta^\perp$ is a conformal Killing vector on $S^{d-1}$ and
$f^ r$ is related to its divergence by
\begin{equation}
f^r = \tfrac{1}{1-d} \div \zeta^\perp.
\end{equation}
The function $f^+$ is unconstrained and hence we find that as a vector
space, the symmetry Lie algebra of $\zNi$ (as a doubly-carrollian
manifold) is $\ckv(S^{d-1}) \oplus C^\infty(S^{d-1})$, with
$\ckv(S^{d-1})$ the Lie algebra of conformal Killing vectors for the
round metric on $S^{d-1}$.  For $d\geq 3$, $\ckv(S^{d-1}) \cong
\so(d,1)$, whereas for $d=2$, $\ckv(S^1) = \eX(S^1)$ since every
smooth vector field on the circle is conformal Killing.

To understand the Lie algebra structure, let us write for $(X,f) \in
\ckv(S^{d-1}) \oplus C^\infty(S^{d-1})$, the corresponding vector 
field as
\begin{equation}
  \zeta_{(X,f)} = X - \tfrac{\div X}{d-1} r \d_r + r f \d_+
\end{equation}
and we calculate
\begin{equation}
  \label{eq:semi-direct}
  \begin{split}
    [\zeta_{(X,0)}, \zeta_{(Y,0)}] &= \zeta_{([X,Y],0)}\\
    [\zeta_{(X,0)}, \zeta_{(0,f)}] &= \zeta_{(0,X\cdot f)}\\
    [\zeta_{(0,f)}, \zeta_{(0,g)}] &= 0,
  \end{split}
\end{equation}
where $[X,Y]$ is the Lie bracket in $\ckv(S^{d-1})$ and
\begin{equation}
\label{eq:action-of-X-on-f}
  X \cdot f = X^i \tfrac{\d f}{\d x^i} - \tfrac{\div X}{d-1} f.
\end{equation}
So the Lie algebra of symmetries of $\zNi$ is a semidirect product
with $C^\infty(S^{d-1})$ an abelian ideal and the action of
$X \in \ckv(S^{d-1})$ on $f \in C^\infty(S^{d-1})$ is such that
$f$ does not transform as a function, but as a section of the density
line bundle in a conformal geometry.

We should contrast these results with those in
\cite[Section~10]{Figueroa-OFarrill:2019sex} for the conformal
symmetries of carrollian spacetimes at level $N=2$.  We find that the
symmetry algebra of $\zNi_{d+1}$ is isomorphic to the conformal
symmetry algebra of $\zdSC_d$ which, as shown there, is itself
isomorphic to the conformal symmetry algebra of $\zLC_d$.

\section{Another choice of sections for reconstruction}
\label{sec:another-choice-section}

In Section~\ref{sec:-m-from-SpiNiTi} we discussed how to interpret
Minkowski spacetime as the parameter space of certain hypersurfaces of
$\zSpi$, $\zTi$ and $\zNi$ which arise as sections of the fibrations
$\zSpi_{d+1} \to \zdS_d$, $\zTi_{d+1} \to \eH^{d}$ and
$\zNi_{d+1} \to \zLC_d$.  More concretely we showed that once we pick
one such section, any other such section is related to it by a
Poincaré translation.  The choice of the initial section, and hence
all the sections which correspond to points in $\MM$, does not follow
from the formalism, but we presented a geometric construction,
analogous to the interpretation of the good cuts (sections of
$\scri \to \zCS$) in \cite{Kozameh:1983yu}, which exhibits the desired
sections as intersections with (generalised) lightcones in the
pseudo-euclidean space $\EE$ based at the points of the embedded
Minkowski spacetime.  In this appendix we give an alternative
construction which results in another choice of section; although both
constructions agree for the case of $\zNi$.

In the construction in this appendix, the choice $x^+=0$ would seem to
be preferred by the fact that the resulting linear functions defining
the sections are eigenfunctions of the second Casimir of the Lorentz
algebra.  We assume that $d>1$ in what follows, since only for $d>1$
is the Lorentz algebra $\so(d,1)$ semisimple.

Let $L_{\mu\nu} = - L_{\nu\mu}$ be generators of the Lorentz algebra,
thought of as vector fields on $\RR^{d,1}$.  Relative to the cartesian
coordinates,
\begin{equation}
  L_{\mu\nu} = \bar\eta_{\mu\rho} x^\rho \d_\nu - \bar\eta_{\nu\rho}
  x^\rho \d_\mu.
\end{equation}
The Lorentz algebra is semisimple and hence the Killing form
\begin{equation}
  K_{\mu\nu,\rho\sigma} = \Tr (\ad_{L_{\mu\nu}} \circ \ad_{L_{\rho\sigma}})
\end{equation}
is non-degenerate.  Up to a dimension-dependent proportionality
constant, it is given by
\begin{equation}
  K_{\mu\nu,\rho\sigma} = \bar\eta_{\mu\rho}\bar\eta_{\nu\sigma} -
  \bar\eta_{\nu\rho}\bar\eta_{\mu\sigma},
\end{equation}
with inverse (again up to a dimension-dependent multiplicative factor)
\begin{equation}
  K^{\mu\nu,\rho\sigma} = \bar\eta^{\mu\rho}\bar\eta^{\nu\sigma} -
  \bar\eta^{\nu\rho}\bar\eta^{\mu\sigma}.
\end{equation}
The second Casimir element is then given (up to normalisation) by
\begin{equation}
  C_2 = \tfrac14 K^{\mu\nu,\rho\sigma} L_{\mu\nu} L_{\rho\sigma},
\end{equation}
which becomes the following second-order differential operator on
$\RR^{d,1}$:
\begin{equation}
  C_2 = \x^2 \Box + (2-d) E - E^2,
\end{equation}
where $\x^2 = \bar\eta_{\mu\nu} x^\mu x^\nu$, $E = x^\mu\d_\mu$ is the
Euler vector field and $\Box = \bar\eta^{\mu\nu}\d_\mu \d_\nu$ is
the D'Alembertian.

Acting on an affine function $f(\x) = x^+ - \bar\eta(\bv,\x)$, we find
\begin{equation}
  C_2 f = (d-1) \bar\eta(\bv,\x),
\end{equation}
so that if $x^+=0$ then the resulting linear function $f(\x) = -
\bar\eta(\bv,\x)$ is an eigenfunction of $C_2$ with nonzero
eigenvalue, since we assumed that $d>1$.

\section{Other signatures}
\label{sec:other-signatures}

In this appendix we indicate how the embedding formalism and results
of Section~\ref{sec:embeddings} extend to other signatures. Due to
their relevance for scattering amplitudes in quantum field theory we
put particular emphasis on the case of the euclidean $(4,0)$-signature
and split $(2,2)$-signature version of Minkowski spacetime. The
conformal compactified split signature case was already discussed in
\cite{Mason:2005qu} and is called Klein space in
\cite{Atanasov:2021oyu}.

Let us consider $\EE := \EE^{p+1,q+1}$, where $p,q \geq 0$, with global
coordinates $x^A = (x^\mu,x^+,x^-)$ where $x^\mu$ are coordinates on
$\RR^{p,q}$, relative to which the flat pseudo-euclidean metric is
given by
\begin{equation}
  \label{eq:metric-pq}
  g = \eta_{AB} dx^A dx^B = \eta_{\mu\nu} dx^\mu dx^\nu + 2 dx^+ dx^-,
\end{equation}
with $\eta_{\mu\nu}$ of signature $(p,q)$.  We define quadrics for
$\epsilon \in \RR$ by
\begin{equation}
  \label{eq:quadrics-pq}
  \eQ_\epsilon = \left\{ x \in \EE ~ \middle | ~  \eta_{AB} x^A x^B =
    \epsilon \right\}
\end{equation}
and hyperplanes for $\sigma \in \RR$ by
\begin{equation}
  \label{eq:hyper-pq}
  \eN_\sigma = \left\{x \in \EE ~ \middle | ~ x^- = \sigma\right\}.
\end{equation}
A subgroup $\Ort(p+1,q+1) \subset GL(p+q+2,\RR)$ preserves every
$\eQ_\epsilon$ and acts transitively on $\eQ_\epsilon$ for $\epsilon
\neq 0$.  If $\epsilon = 0$, $\eQ_0$ contains the origin ($x=0$),
which is a point-like orbit and $\Ort(p+1,q+1)$ acts transitively on
the complement.

The subgroup $\widetilde G \subset \Ort(p+1,q+1)$ which preserves
$\eN_\sigma$ (for $\sigma \neq 0$)  consists of matrices formally
identical to those in equation~\eqref{eq:poincare-subgroup-ort} except
that $\bv \in \RR^{p+q}$ and $A \in \Ort(p,q)$.  It follows that
$\widetilde G \cong \Ort(p,q) \ltimes \RR^{p+q}$.  If $\sigma =0$
there is an enhancement to a $\CO(p,q) \ltimes \RR^{p+q}$ subgroup of
$\Ort(p+1,q+1)$.  The formulae are \emph{mutatis mutandis} as in
Section~\ref{sec:embeddings}.

Let $G$ denote the identity component of $\widetilde G$ and let us
decompose $\EE$ into $G$-orbits.  By construction, $G$ preserves every
$\eM_{\epsilon,\sigma} = \eQ_\epsilon \cap \eN_\sigma$.

If $\sigma \neq 0$, then we may solve for $x^+$ as in
equation~\eqref{eq:solving-x+} and we find that
$\eM_{\epsilon,\sigma\neq 0}$ is an embedding of $\RR^{p,q}$
into $\EE$:
\begin{equation}
  \begin{tikzcd}
    \RR^{p,q} \arrow[r] & \eM_{\epsilon,\sigma\neq 0}\\
    \x \arrow[r,mapsto] &
    \begin{pmatrix}
      \x \\ x^+(\x) \\ \sigma
    \end{pmatrix}
  \end{tikzcd}
\end{equation}
which is $G$-equivariant under the $G$-action $\x \mapsto A \x + \bv$
on $\RR^{p,q}$ and the linear action of $G$ on $\EE$.  The pull-back
to $\RR^{p,q}$ of the metric \eqref{eq:metric-pq} on $\EE$ is
$\eta_{\mu\nu}dx^\mu dx^\nu$, so that the embedding is isometric
relative to the pseudo-euclidean metric on $\RR^{p,q}$.

Pick as the origin of $\eM_{\epsilon,\sigma\neq 0}$ the point with
coordinates $(\bzero,\frac{\epsilon}{2\sigma},\sigma)$.  Its
stabiliser is the copy of the identity of component of $\Ort(p,q)$
which is formally the same as the subgroup $H$ in
equation~\eqref{eq:lorentz-subgroup-ort}, except that $A \in \SO(p,q)_0$.

So far this is \emph{mutatis mutandis} as in
Section~\ref{sec:embeddings}.  The only changes, albeit minor, arise
when $\sigma = 0$.  In this case we have $\eM_{\epsilon,0}$ and we
consider three cases depending on whether $\epsilon >0$, $\epsilon =
0$ or $\epsilon < 0$.

\subsubsection*{\textbf{The case $\epsilon = \rho^2 > 0$}}
\label{sec:epsilon-positive}

Here,
\begin{equation}
  \eM_{\rho^2,0} = \left\{
    \begin{pmatrix}
      \x \\  x^+ \\ 0
    \end{pmatrix}
~ \middle | ~ \eta(\x,\x) = \rho^2\quad\text{and}\quad x^+ \in \RR\right\}.
\end{equation}
Let us break up $\x = (\y,\zz) \in \RR^p \oplus \RR^q$, so that
$\eta(\x,\x) = |\y|^2 - |\zz|^2 = \rho^2$, so that $|\y|^2 = |\zz|^2 +
\rho^2$.  We have several cases to consider:
\begin{itemize}
\item If $p=0$, $\y=0$ and there are no solutions: $\eM_{\rho^2,0} = \varnothing$.
\item If $p=1$, $\y = y \in \RR$ and $y^2 = \rho^2 + |\zz|^2$, so $y =
  \pm \sqrt{\rho^2 + |\zz|^2}$.  This breaks up into two subcases
  depending on whether $q >0$ or $q=0$:
  \begin{itemize}[label=$\circ$]
  \item if $q>0$, then the equation $y = \pm \sqrt{\rho^2 + |\zz|^2}$
    defines a two-sheeted hyperboloid $\eH_\rho = \eH^+_\rho \sqcup
    \eH^-_\rho$ and, since $G$ is connected,
    $\eM_{\rho^2,0}$ decomposes into two $G$-orbits:
    \begin{equation}
      \eM_{\rho^2,0}= (\eH^+_\rho \times \RR) \sqcup (\eH^-_\rho \times \RR);
    \end{equation}
  \item whereas if $q=0$, then we have two points $y = \pm \rho$ and
    hence
    $\eM_{\rho^2,0}$ decomposes into two $G$-orbits:
    \begin{equation}
      \eM_{\rho^2,0} = (\{\rho\} \times \RR) \sqcup (\{-\rho\} \times \RR).
    \end{equation}
  \end{itemize}
\item Finally, if $p>1$ then $|\y^2| = \rho^2 + |\zz|^2$ is connected
  and $\eM_{\rho^2,0}$ is its own $G$-orbit.
\end{itemize}

\subsubsection*{\textbf{The case $\epsilon = -\rho^2 < 0$}}
\label{sec:epsilon-negative}

This case is virtually identical to the previous case interchanging
$p \leftrightarrow q$ and $\y \leftrightarrow \zz$.

\subsubsection*{\textbf{The case $\epsilon = 0$}}
\label{sec:epsilon-zero}

Now
\begin{equation}
  \eM_{0,0} = \left\{
    \begin{pmatrix}
      \x \\ x^+ \\ 0
    \end{pmatrix}
    ~ \middle | ~  \x \in \RR^{p,q},\quad \eta(\x,\x) = 0 \quad\text{and}\quad x^+ \in \RR \right\}.
\end{equation}
We again decompose $\x = (\y,\zz)$ with $\y \in \RR^p$ and $\zz \in
\RR^q$ and now $|\y|^2 = |\zz|^2$.  We have several cases:
\begin{itemize}
\item If either $p=0$ or $q=0$, then $\x=0$ and $\eM_{0,0}$ is
  a line of point-like orbits:
  \begin{equation}
    \eM_{0,0} = \bigsqcup_{x^+\in\RR}\left\{
      \begin{pmatrix}
        \bzero \\ x^+ \\ 0
      \end{pmatrix}\right\}.
  \end{equation}
\item If either $p=1$ or $q=1$ we are in the situation discussed in
  Section~\ref{sec:embeddings}:
  \begin{equation}
    \eM_{0,0} = \eM_{0,0}^+ \sqcup \bigsqcup_{x^+\in\RR} \left\{
      \begin{pmatrix}
        \bzero \\ x^+ \\ 0
      \end{pmatrix}\right\} \sqcup \eM_{0,0}^-
  \end{equation}
  with
  \begin{equation}
    \eM_{0,0}^\pm = \left\{ \begin{pmatrix}
        \x \\ x^+ \\ 0
      \end{pmatrix} ~ \middle | ~ \x \in \eL^\pm \quad \text{and}
      \quad x^+ \in \RR\right\}.
  \end{equation}
\item Finally if $p>1$ and $q>1$, we have that
  \begin{equation}
    \eM_{0,0} = \eM'_{0,0} \sqcup  \bigsqcup_{x^+ \in \RR} \left\{
      \begin{pmatrix}
        \bzero \\ x^+ \\ 0
      \end{pmatrix}\right\},
  \end{equation}
  where
  \begin{equation}
    \eM'_{0,0} = \left\{
      \begin{pmatrix}
        \x \\ x^+ \\ 0
      \end{pmatrix} ~ \middle | ~ \eta(\x,\x) = 0,\quad \x \neq 0
      \quad\text{and}\quad x^+ \in \RR\right\}.
  \end{equation}
\end{itemize}

Let us contrast the three cases $(p,q) \in \{(4,0), (3,1), (2,2)\}$.
The case $(p,q)=(3,1)$ is as in Section~\ref{sec:embeddings}:
\begin{equation}
  \EE^{4,2} = \bigsqcup_{\substack{\epsilon,\sigma\in \RR\\ \sigma \neq 0}} \eM_{\epsilon,\sigma} \sqcup \bigsqcup_{\epsilon>0} \eM_{\epsilon,0} \sqcup \bigsqcup_{\epsilon<0} \left( \eM_{\epsilon,0}^+  \sqcup \eM_{\epsilon,0}^-\right) \sqcup \eM_{0,0}^+ \sqcup \eM_{0,0}^- \sqcup \bigsqcup_{x^+ \in \RR} \left\{
    \begin{pmatrix}
      \bzero \\ x^+ \\ 0
    \end{pmatrix}
  \right\},
\end{equation}
giving, in order of appearance, embeddings of Minkowski spacetime, $\zSpi$, $\zTi^+$,
$\zTi^-$, $\zNi^+$, $\zNi^-$ and the line of fixed points
$(\bzero,x^+,0)$.
  
The case $(p,q)=(4,0)$ gives
\begin{equation}
  \EE^{5,1} = \bigsqcup_{\substack{\epsilon,\sigma\in \RR\\ \sigma \neq 0}} \eM_{\epsilon,\sigma} \sqcup \bigsqcup_{\epsilon>0} \eM_{\epsilon,0} \sqcup \bigsqcup_{x^+ \in \RR} \left\{
    \begin{pmatrix}
      \bzero \\ x^+ \\ 0
    \end{pmatrix}
  \right\},
\end{equation}
giving, in order of appearance, embeddings of euclidean space,
cylinders and the line of fixed points $(\bzero,x^+,0)$.  We may think
of each cylinder as the euclidean version of $\zSpi$: the blow-up of
the point at infinity where all geodesics end.  Of course in euclidean
signature there are no timelike nor null infinities, which explains
the absence of orbits corresponding to $\zTi^\pm$ or $\zNi$.

Finally, if $(p,q)=(2,2)$ we have
\begin{equation}
  \EE^{3,3} = \bigsqcup_{\substack{\epsilon,\sigma\in \RR\\ \sigma
      \neq 0}} \eM_{\epsilon,\sigma} \sqcup \bigsqcup_{\epsilon\neq 0} \eM_{\epsilon,0} \sqcup \eM'_{0,0} \sqcup \bigsqcup_{x^+ \in \RR} \left\{
    \begin{pmatrix}
      \bzero \\ x^+ \\ 0
    \end{pmatrix}
  \right\},
\end{equation}
giving, in order or appearance, embeddings of the Klein space
$\RR^{2,2}$, a pseudo-carrollian three-dimensional manifold which is
the blow-up of either timelike ($\epsilon<0$) or spacelike
($\epsilon>0$) infinities, as described in
\cite[Section~2]{Atanasov:2021oyu}, and $\eM'_{0,0}$ which, just like
$\zNi^\pm$ to $\scri^\pm$, can be interpreted as the bundle of scales
of the null infinity of the Klein space which is connected, unlike for
Minkowski spacetime. The final expression is again the line of fixed points $(\bzero,x^+,0)$.

Under the projection $\EE^{p+1,q+1}\setminus\{0\} \to \PP^{p+q+1}$,
what happens to the null quadric $\eQ_0\setminus\{0\}$ now?  Let us
again contrast $(p,q) \in \{(4,0),(3,1),(2,2)\}$.

The case $(p,q) = (3,1)$ is as in Section~\ref{sec:embeddings} and
gives a conformal compactification of Minkowski spacetime
$\MM^\sharp = \MM \sqcup \scri \sqcup \{I\}$. Here $\scri$ is
antipodally identified $\scri^{\pm}$ and $\{I\}$ is a point where
$i^{\pm}$ and $i^{0}$ are identified, see~\cite[Section 9.2]{MR838301}
for more details. This compactification has topology
$S^{3} \times S^{1}$ with boundary $S^{2}\times S^{1}$ where $S^{2}$
is the celestial sphere.

For $(p,q)=(4,0)$ we get the one-point compactification of euclidean
space: namely, $S^4 = \RR^4 \sqcup \{\infty\}$, where $\infty$ is the
projective image of the line of fixed points (minus the origin). The
boundary consists of one point.

Finally, for $(p,q)=(2,2)$ the conformal compactification of the Klein
space has topology $S^2\times S^{2}/\ZZ^{2}$~\cite{Mason:2005qu} with
a boundary of topology $S^{3}$~\cite{Atanasov:2021oyu}. It might be
interesting to see if the celestial tori can be understood from the
point of view of the Reeb foliation of $S^{3}$.

\providecommand{\href}[2]{#2}\begingroup\raggedright\endgroup


\end{document}